\documentclass[]{aastex631}

\usepackage[flushleft]{threeparttable}

\newcommand{\msun}{M$_{\odot}$}
\newcommand{\sfr}{M$_{\odot}$ yr$^{-1}$}
\defcitealias{bruzual_stellar_2003}{BC03}

\received{November 22}
\revised{--}
\accepted{--}

\submitjournal{ApJ}

\shorttitle{PHANGS-HST Star Clusters and Compact Associations I: Observed Properties}
\shortauthors{Maschmann et al.}

\begin{document}
\title{PHANGS-HST catalogs for $\sim$100,000 star clusters and compact associations in 38 galaxies: I. Observed properties}
\newcommand{\Arizona}{\affil{Steward Observatory, University of Arizona, Tucson, AZ 85721, USA
}}

\newcommand{\GEMINI}{\affil{Gemini Observatory/NSF’s NOIRLab, 950 N. Cherry Avenue, Tucson, AZ, 85719, USA
}}
\newcommand{\ASCL}{\affil{Astrophysics Source Code Librar
y, Michigan Technological University, 1400 Townsend Drive, Houghton, MI 49931}}

\newcommand{\OSU}{\affil{Department of Astronomy, The Ohio State University, 140 West 18th Avenue, Columbus, Ohio 43210, USA}}

\newcommand{\Alberta}{\affil{Department of Physics, University of Alberta, Edmonton, AB T6G 2E1, Canada}}

\newcommand{\ANU}{\affil{Research School of Astronomy and Astrophysics, Australian National University, Canberra, ACT 2611, Australia}}

\newcommand{\IPARCOS}{\affil{Instituto de F\'{\i}sica de Part\'{\i}culas y del Cosmos, Universidad Complutense de Madrid, E-28040 Madrid, Spain}}

\newcommand{\IPAC}{\affil{Caltech-IPAC, 1200 E. California Blvd. Pasadena, CA 91125, USA}}

\newcommand{\Caltech}{\affil{Caltech, 1200 E. California Blvd. Pasadena, CA 91125, USA}}

\newcommand{\Carnegie}{\affil{Observatories of the Carnegie Institution for Science, 813 Santa Barbara Street, Pasadena, CA 91101, USA}}

\newcommand{\CCAPP}{\affil{Center for Cosmology and Astroparticle Physics, 191 West Woodruff Avenue, Columbus, OH 43210, USA}}

\newcommand{\CfA}{\affil{Harvard-Smithsonian Center for Astrophysics, 60 Garden Street, Cambridge, MA 02138, USA}}

\newcommand{\CITEVA}{\affil{Centro de Astronomía (CITEVA), Universidad de Antofagasta, Avenida Angamos 601, Antofagasta, Chile}}

\newcommand{\CNRS}{\affil{CNRS, IRAP, 9 Av. du Colonel Roche, BP 44346, F-31028 Toulouse cedex 4, France}}

\newcommand{\ESO}{\affil{European Southern Observatory, Karl-Schwarzschild Stra{\ss}e 2, D-85748 Garching bei M\"{u}nchen, Germany}}

\newcommand{\Heidelberg}{\affil{Astronomisches Rechen-Institut, Zentrum f\"{u}r Astronomie der Universit\"{a}t Heidelberg, M\"{o}nchhofstra\ss e 12-14, D-69120 Heidelberg, Germany}}

\newcommand{\ICRAR}{\affil{International Centre for Radio Astronomy Research, University of Western Australia, 35 Stirling Highway, Crawley, WA 6009, Australia}}

\newcommand{\IRAM}{\affil{Institut de Radioastronomie Millim\'{e}trique (IRAM), 300 Rue de la Piscine, F-38406 Saint Martin d'H\`{e}res, France}}

\newcommand{\IRAP}{\affil{CNRS, IRAP, 9 Av. du Colonel Roche, BP 44346, F-31028 Toulouse cedex 4, France}}

\newcommand{\UPS}{\affil{Universit\'{e} de Toulouse, UPS-OMP, IRAP, F-31028 Toulouse cedex 4, France}}

\newcommand{\ITA}{\affil{Universit\"{a}t Heidelberg, Zentrum f\"{u}r Astronomie, Institut f\"{u}r Theoretische Astrophysik, Albert-Ueberle-Str 2, D-69120 Heidelberg, Germany}}

\newcommand{\IWR}{\affil{Universit\"{a}t Heidelberg, Interdisziplin\"{a}res Zentrum f\"{u}r Wissenschaftliches Rechnen, Im Neuenheimer Feld 205, D-69120 Heidelberg, Germany}}

\newcommand{\JHU}{\affil{Department of Physics and Astronomy, The Johns Hopkins University, Baltimore, MD 21218, USA}}

\newcommand{\Leiden}{\affil{Leiden Observatory, Leiden University, P.O. Box 9513, 2300 RA Leiden, The Netherlands}}

\newcommand{\Maryland}{\affil{Department of Astronomy, University of Maryland, College Park, MD 20742, USA}}

\newcommand{\MPE}{\affil{Max-Planck-Institut f\"{u}r extraterrestrische Physik, Giessenbachstra{\ss}e 1, D-85748 Garching, Germany}}

\newcommand{\MPIA}{\affil{Max-Planck-Institut f\"{u}r Astronomie, K\"{o}nigstuhl 17, D-69117, Heidelberg, Germany}}

\newcommand{\Nagoya}{\affil{Department of Physics, Nagoya University, Furo-cho, Chikusa-ku, Nagoya, Aichi 464-8602, Japan}}

\newcommand{\NRAO}{\affil{National Radio Astronomy Observatory, 520 Edgemont Road, Charlottesville, VA 22903-2475, USA}}

\newcommand{\OAN}{\affil{Observatorio Astron\'{o}mico Nacional (IGN), C/Alfonso XII, 3, E-28014 Madrid, Spain}}

\newcommand{\ObsParis}{\affil{Sorbonne Universit\'{e}, Observatoire de Paris, Universit\'{e} PSL, CNRS, LERMA, F-75014, Paris, France}}

\newcommand{\Princeton}{\affil{Department of Astrophysical Sciences, Princeton University, Princeton, NJ 08544 USA}}

\newcommand{\UToledo}{\affil{University of Toledo, 2801 W. Bancroft St., Mail Stop 111, Toledo, OH, 43606}}

\newcommand{\Toulouse}{\affil{Universit\'{e} de Toulouse, UPS-OMP, IRAP, F-31028 Toulouse cedex 4, France}}

\newcommand{\UBonn}{\affil{Argelander-Institut f\"ur Astronomie, Universit\"at Bonn, Auf dem H\"ugel 71, 53121 Bonn, Germany}}

\newcommand{\UChile}{\affil{Departamento de Astronom\'{i}a, Universidad de Chile, Camino del Observatorio 1515, Las Condes, Santiago, Chile}}

\newcommand{\UCM}{\affil{Departamento de F\'{\i}sica de la Tierra y Astrof\'{\i}sica, Universidad Complutense de Madrid, E-28040 Madrid, Spain}}

\newcommand{\UCSD}{\affil{Department of Astronomy and Astrophysics,  University of California,\\ San Diego, 9500 Gilman Drive, La Jolla, CA 92093, USA}}

\newcommand{\ULyon}{\affil{Univ Lyon, Univ Lyon 1, ENS de Lyon, CNRS, Centre de Recherche Astrophysique de Lyon UMR5574,\\ F-69230 Saint-Genis-Laval, France}}

\newcommand{\UMass}{\affil{University of Massachusetts—Amherst, 710 N. Pleasant Street, Amherst, MA 01003, USA}}

\newcommand{\UWyoming}{\affil{Department of Physics and Astronomy, University of Wyoming, Laramie, WY 82071, USA}}

\newcommand{\LAM}{\affil{
Aix Marseille Univ, CNRS, CNES, LAM (Laboratoire d’Astrophysique de Marseille),  F-13388 Marseille,
France}}

\newcommand{\UHawaii}{\affil{Institute for Astronomy, University of Hawaii, 2680 Woodlawn Drive, Honolulu, HI 96822, USA}}

\newcommand{\UGent}{\affil{Sterrenkundig Observatorium, Universiteit Gent, Krijgslaan 281 S9, B-9000 Gent, Belgium}}

\newcommand{\IPARC}{\affil{Instituto de F\'{\i}sica de Part\'{\i}culas y del Cosmos IPARCOS, Facultad de Ciencias F\'{\i}sicas, Universidad Complutense de Madrid, E-28040, Spain}}

\newcommand{\STScI}{\affil{Space Telescope Science Institute, 3700 San Martin Drive, Baltimore, MD 21218, USA}}

\newcommand{\STScIESA}{\affiliation{AURA for the European Space Agency (ESA), Space Telescope Science Institute, 3700 San Martin Drive, Baltimore, MD 21218, USA}}

\newcommand{\McMaster}{\affil{Department of Physics and Astronomy, McMaster University, Hamilton, ON L8S 4M1, Canada}}

\newcommand{\INAF}{\affil{INAF -- Osservatorio Astrofisico di Arcetri, Largo E. Fermi 5, I-50157, Firenze, Italy}}

\newcommand{\Sydney}{\affil{Sydney Institute for Astronomy, School of Physics A28, The University of Sydney, NSW 2006, Australia}}

\newcommand{\UA}{\affil{Centro de Astronomía (CITEVA), Universidad de Antofagasta, Avenida Angamos 601, Antofagasta, Chile}}

\newcommand{\LERMA}{\affil{Observatoire de Paris, PSL Research University, CNRS, Sorbonne Universit\'es, 75014 Paris}}

\newcommand{\SAIMSU}{\affil{Sternberg Astronomical Institute, Lomonosov Moscow State University, Universitetsky pr. 13, 119234 Moscow, Russia}}

\newcommand{\UTA}{\affil{Instituto de Alta Investigación, Universidad de Tarapacá, Casilla 7D, Arica, Chile}}

\newcommand{\IAC}{\affil{Instituto de Astrof\'isica de Canarias, C/ V\'ia L\'actea s/n, E-38205, La Laguna, Spain}}

\newcommand{\UNAM}{\affil{Instituto de Astronom\'ia, Universidad Nacional Aut\'onoma de M\'exico, Unidad Acad\'emica en Ensenada, Km 103 Carr. Tijuana-Ensenada, Ensenada, B.C.,
C.P. 22860, M\'exico}}

\newcommand{\ULL}{\affil{Departamento de Astrof\'isica, Universidad de La Laguna, Av. del Astrof\'isico Francisco S\'anchez s/n, E-38206, La Laguna, Spain}}

\newcommand{\AAPF}{\altaffiliation{NSF Astronomy and Astrophysics Postdoctoral Fellow}}

\newcommand{\DECRA}{\altaffiliation{ARC DECRA Fellow}}

\newcommand{\Oxford}{\affil{Sub-department of Astrophysics, Department of Physics, University of Oxford, Keble Road, Oxford OX1 3RH, UK}}

\newcommand{\wesleyan}{\affil{Astronomy Department and Van Vleck Observatory, Wesleyan University, 96 Foss Hill Drive, Middletown, CT 06459, USA}}

\newcommand{\PLATA}{\affil{Instituto de Astrof\'{\i}sica de La Plata, CONICET--UNLP, Paseo del Bosque S/N, B1900FWA La Plata, Argentina }}

\newcommand{\ARC}{\affil{ARC Centre of Excellence for All Sky Astrophysics in 3 Dimensions (ASTRO 3D), Australia}}

\newcommand{\UVirginia}{\affil{University of Virginia Astronomy Department, 530 McCormick Road, Charlottesville, VA 22904, USA}}

\newcommand{\UniCA}{\affil{Université Côte d'Azur, Observatoire de la Côte d'Azur, CNRS, Laboratoire Lagrange, 06000, Nice, France}}

\newcommand{\CamIoA}{\affil{Institute of Astronomy, University of Cambridge, Madingley Road, Cambridge CB3 0HA, UK}}

\newcommand{\KICC}{\affil{Kavli Institute for Cosmology Cambridge, Madingley Road, Cambridge CB3 0HA, UK}}

\hyphenation{Cosmic-Flows}
\hyphenation{Hyper-LEDA}
\hyphenation{HERA-CLES}

\correspondingauthor{Daniel Maschmann}
\email{danielmaschmann@arizona.edu}

\author[0000-0001-6038-9511]{Daniel Maschmann}
\Arizona

\author[0000-0002-2278-9407]{Janice C. Lee}
\STScI
\Arizona
\GEMINI

\author[0000-0002-8528-7340]{David A. Thilker}
\JHU

\author[0000-0002-3784-7032]{Bradley C. Whitmore}
\STScI

\author[0000-0003-1943-723X]{Sinan Deger}
\CamIoA
\KICC

\author[0000-0003-0946-6176]{M\'ed\'eric Boquien}
\UniCA

\author[0000-0003-0085-4623]{Rupali Chandar}
\UToledo

\author[0000-0002-5782-9093]{Daniel A. Dale}
\UWyoming

\author[0000-0001-8289-3428]{Aida Wofford}
\UNAM
\UCSD

\author{Stephen Hannon}
\MPIA

\author[0000-0003-3917-6460]{Kirsten L. Larson}
\STScIESA

\author[0000-0002-2545-1700]{Adam K. Leroy}
\OSU

\author[0000-0002-3933-7677]{Eva Schinnerer}
\MPIA

\author[0000-0002-5204-2259]{Erik Rosolowsky}
\Alberta

\author[0000-0001-7130-2880]{Leonardo \'Ubeda}
\STScI

\author[0000-0003-0410-4504]{Ashley T. Barnes}
\ESO

\author[0000-0002-6155-7166]{Eric Emsellem}
\ESO
\ULyon

\author{Kathryn Grasha}
\DECRA
\ANU
\ARC

\author[0000-0002-9768-0246]{Brent~Groves}
\ICRAR

\author[0000-0002-4663-6827]
{R\'emy Indebetouw}
\UVirginia
\NRAO

\author[0000-0003-4770-688X]{Hwihyun Kim}
\GEMINI

\author[0000-0002-0560-3172]{Ralf S. Klessen}
\ITA
\IWR

\author[0000-0001-6551-3091]{Kathryn Kreckel}
\Heidelberg

\author[0000-0003-2508-2586]{Rebecca C. Levy}
\AAPF
\Arizona

\author[0000-0001-5965-3530]{Francesca Pinna}
\IAC
\ULL
\MPIA

\author[0000-0002-0579-6613]{M. Jimena Rodr\'{\i}guez}
\Arizona
\PLATA

\author[0009-0009-9148-2159]{Qiushi Tian}
\wesleyan

\author[0000-0002-0012-2142]{Thomas G. Williams}
\Oxford

\begin{abstract}

We present the largest catalog to-date of star clusters and compact associations in nearby galaxies.  We have performed a V-band-selected census of clusters across the 38 spiral galaxies of the PHANGS-HST Treasury Survey, and measured integrated, aperture-corrected NUV-U-B-V-I photometry.
This work has resulted in uniform catalogs that contain $\sim$20,000 clusters and compact associations which have passed human inspection and morphological classification, and a larger sample of $\sim$100,000 classified by neural network models. 
Here, we report on the observed properties of these samples, and demonstrate that tremendous insight can be gained from just the observed properties of clusters, even in the absence of their transformation into physical quantities. 
In particular, we show the utility of the UBVI color-color diagram, and the three principal features revealed by the PHANGS-HST cluster sample: the young cluster locus, the middle-age plume, and the old globular cluster clump.  
We present an atlas of maps of the 2D spatial distribution of clusters and compact associations in the context of the molecular clouds from PHANGS-ALMA.  We explore new ways of understanding this large dataset in a multi-scale context by bringing together once-separate techniques for the characterization of clusters (color-color diagrams and spatial distributions) and their parent galaxies (galaxy morphology and location relative to the galaxy main sequence). 
A companion paper presents the physical properties: ages, masses, and dust reddenings derived using improved spectral energy distribution (SED) fitting techniques.

\end{abstract}


\section{Introduction} \label{sec:intro}
Decades of research on star formation have taught us that systematic observations - spanning key spatial scales and phases of the star formation cycle, over a full set of galactic environments - are essential for development of a robust, unified model of star formation and galaxy evolution \citep[e.g.][]{kennicutt_star_2012}.  To enable such an integrated multi-phase, multi-scale study of star formation, the PHANGS (Physics at High Angular resolution in Nearby GalaxieS) collaboration \citep{schinnerer_physics_2019} has conducted large surveys with ALMA \citep{leroy_phangs-alma_2021}, VLT/MUSE \citep{emsellem_phangs-muse_2022}, HST \citep{lee_phangs-hst_2022}, and JWST \citep{lee_phangs-jwst_2023}, and is studying the relationships between molecular clouds, HII regions, dust, and star clusters across the large diversity of environments found in nearby galaxies. Beyond these four principal surveys, a wealth of additional supporting data is available and continues to be obtained by PHANGS including Astrosat far-/near-ultraviolet imaging (PI: E. Rosolowsky; Hassani 2023, submitted), HST H$\alpha$ narrowband imaging (PIs: R. Chandar, D. Thilker, F. Belfiore), ground-based wide-field H$\alpha$ narrow-band imaging (PIs: G. Blanc, I.-T. Ho), and H I 21 cm observations with the VLA and MeerKAT.  To support science analysis with this wealth of data, PHANGS has been producing and publicly releasing an extensive set of ``higher level science products.'' \footnote{\url{https://sites.google.com/view/phangs/home/data}}

In the context of this comprehensive effort, NUV-U-B-V-I imaging for 38 spiral galaxies was obtained from 2019-21 through a HST Cycle 26 Treasury program.  The galaxies were drawn from the PHANGS-ALMA parent sample and thus have $^{12}$CO(J = 2$\rightarrow$1) observations at $\sim1^{\prime\prime}$ resolution.  Half of the sample (19 galaxies) are covered by all four principal surveys of PHANGS; i.e., in addition to the HST and ALMA observations, integral field spectroscopic mapping from 4800-9300\AA\, has been performed with VLT-MUSE, and imaging in eight bands from 2-21 $\mu$m is being obtained through a JWST Cycle 1 Treasury program.  Details on the design of the PHANGS-ALMA, PHANGS-HST, PHANGS-MUSE, and PHANGS-JWST foundational surveys are provided in the papers cited above.  New large HST and JWST surveys to expand the number of galaxies with HST-JWST-ALMA data to 74 have been recently approved in 2023 (JWST Cycle 2, GO-3707, PI A. Leroy; HST Cycle 31, GO-17502, PI D. Thilker).

As discussed in \cite{lee_phangs-hst_2022}, one of the main goals of the PHANGS-HST Treasury Survey is to conduct a uniform census of star clusters and stellar associations in 38 nearby spiral galaxies (${\rm d \lesssim 20 Mpc}$) to probe cluster formation and evolution, and to utilize these effectively single-age stellar populations as `clocks' to time star formation and ISM processes.  Here, we present the result of this census: catalogs providing the photometric properties of $\sim$100,000 star clusters and compact associations, the largest such sample to-date.  These catalogs are the culmination of technical efforts as summarized in \cite{lee_phangs-hst_2022} to establish improved techniques for cluster candidate detection and selection \citep{whitmore_star_2021, thilker_phangs-hst_2022}, photometry \citep{deger_bright_2022}, and automated morphological classification using machine learning techniques \citep{wei_deep_2020, whitmore_star_2021, hannon_star_2023}.  

A companion paper \citep[][hereafter Paper II]{thilker23sed} presents the physical properties of the sample (specifically, age, mass, and reddening) derived using improved strategies for spectral energy distribution (SED) fitting, which were initially explored by \citet{whitmore_improving_2023}.  The improvements seek to mitigate the age-reddening-metallicity degeneracy by building upon conventional SED fitting techniques for star clusters which were adopted at the outset of the PHANGS-HST survey \citep{turner_phangs-hst_2021}.  All of the catalogs described here can be accessed through the Mikulski Archive for Space Telescopes (MAST)\footnote{\url{https://archive.stsci.edu/hlsp/phangs/phangs-cat}}.

The PHANGS-HST star cluster catalogs enable a wide range of science investigations. Many of the studies by the PHANGS team which have utilized these catalogs so far have focused on star formation feedback and timescales, but investigations of the old stellar populations (globular clusters) has also begun \citep{floyd}. We briefly describe some of these studies below.

\citet{barnes_linking_2022} examine the clusters and associations within isolated, compact H II regions in NGC1672, identified through HST narrowband imaging.  They find higher pressures (as measured from PHANGS-MUSE) within more compact H II regions, though with significant scatter which is presumably introduced by variation in the stellar population properties (e.g. mass, age, metallicity). 

By cross matching star clusters and multi-scale stellar associations with H\,II regions from PHANGS-MUSE across the full set of 19 galaxies with PHANGS-HST$+$MUSE data, \citet{scheuermann_stellar_2023} study how H II regions evolve over time.   They find that younger nebulae are more attenuated by dust and closer to giant molecular clouds, as expected by feedback-regulated models of star formation.  They also report strong correlations with local metallicity variations and age, suggesting that star formation preferentially occurs in locations of locally enhanced metallicity. 

Across this same set of 19 galaxies, \citet{egorov_quantifying_2023} study the star clusters and associations within nebular regions of locally elevated velocity dispersion, including expanding superbubbles, identified with PHANGS-MUSE.  They find that the kinetic energy of the ionised gas is correlated with the inferred mechanical energy input from supernovae and stellar winds which can be interpreted as a coupling efficiency of $10-20\,\%$. They also find that young clusters and associations are preferentially located along the rims of superbubbles, which provides possible evidence for star formation propagation or triggering. 

\citet{watkins_quantifying_2023} perform a similar analysis, but starting with molecular gas superbubbles in PHANGS-ALMA.  They measure radii and expansion velocities, and dynamically derive bubble ages and the mechanical power from young stars required to drive the bubbles. They find that the masses and ages of the PHANGS-HST clusters and associations are consistent with the required power, if a supernova (SN) model that injects energy with a coupling efficiency of $\sim10\,\%$ is assumed.  

Joint HST-JWST analysis with the IR imaging from the PHANGS-JWST Cycle 1 Treasury has also begun, and first results have been published in a collection of papers for a PHANGS-JWST ApJL focus issue.\footnote{\url{https://iopscience.iop.org/collections/2041-8205_PHANGS-JWST-First-Results}}  

One of the most striking features of the PHANGS-JWST MIRI imaging is the ubiquitous bubble structure \citep{lee_phangs-jwst_2023, williams_phangs-jwst_2024}. \citet{watkins_phangs-jwst_2023} and \citet{barnes_phangs-jwst_2023} demonstrate star formation feedback are likely to be the origin of these bubbles, based on analysis of the PHANGS-HST star cluster and associations catalogs for NGC~628.  \citet{thilker_phangs-jwst_2023} study the dust filament network in NGC~628 and its relation to sites of star formation, finding that $>$60\% optically-selected young clusters ($<$5 Myr) occurs within $\sim$25~pc dust filaments. \citet{rodriguez_phangs-jwst_2023} and \citet{whitmore_phangs-jwst_2023} present first results on dust embedded star clusters, which trace the youngest sites of star formation, with the PHANGS-HST clusters and associations serving as a essential reference for computing constraints on the timescales for dust clearing and the embedded phase.

In this paper, we describe the PHANGS-HST catalogs of star clusters and compact associations with the aim of supporting further science with this extensive dataset.  The characterization of the observed properties presented in this paper provide a starting point for the utilization of the full census of star clusters and compact associations across the PHANGS-HST 38 galaxy sample to realize the aim of PHANGS to understand the interplay of the small-scale physics of gas and star formation with galactic structure and galaxy evolution.

The remainder of this paper is organized as follows. In Section~\ref{sect:catalog_content}, we provide an overview of the PHANGS-HST galaxy sample, HST observations, star cluster and compact association catalog production pipeline, and describe the publicly-released catalog structure and contents.  In Section~\ref{sect:catalog_properties}, we examine the size and photometric depths of the samples detected in each galaxy.  In Section~\ref{sect:color_color}, we continue to develop the ideas introduced in \cite{lee23ubvi} on using UV-optical color-color diagrams to gain insight into star cluster formation and evolution.  We explore new ways of understanding the data in a multi-scale context by studying the features of the UBVI star cluster color-color diagrams for each galaxy in relation to its position relative to the star-forming galaxy main sequence in star formation rate (SFR) and stellar mass (M$_*$).  This composite diagram provides a framework for understanding cluster formation, evolution, and destruction in the context of the global properties of their host galaxies.  In Section~\ref{sect:spatialdist}, we present an atlas of maps illustrating the 2D spatial distributions of the clusters and compact associations relative to giant molecular clouds from the PHANGS-ALMA CO(2-1) catalogs.  We bring together characteristics of the cluster spatial distributions and color-color diagrams, with galaxy morphology and position along the main sequence to gain qualitative insight into the global drivers of cluster formation and evolution.  In Section~\ref{sect:discussion}, we discuss issues related to sample completeness to outline future work and to provide advice to users of the catalog.  Key conclusions are summarized in Section~\ref{sect:summary}.

Vega magnitudes are used in this paper to facilitate comparison to prior work.

\section{Star Cluster Catalogs}\label{sect:catalog_content}
As mentioned in the introduction, the PHANGS-HST catalogs of star clusters and compact associations are the end-product of an extensive processing pipeline.  In this section, we provide a brief overview of the HST observations and this pipeline.  The reader is referred to the corresponding technical papers, as cited in the Introduction and below, for a full discussion.  Documentation of the PHANGS-HST imaging filters and exposure times for individual galaxies are provided in the next section as these are needed to understand the depth of the cluster catalogs.

\subsection{Galaxy sample and observations}
Galaxies for the PHANGS-ALMA parent sample were selected to be nearby (D$\lesssim$20~Mpc), massive ($M_{*} \gtrsim 10^{9.75}$~\msun), on the star-forming galaxy main sequence, and relatively face-on \citep{leroy_phangs-alma_2021}.  A subset of these were chosen for observation with HST (GO-15654) as discussed in \citet{lee_phangs-hst_2022}. The resulting PHANGS-HST sample is comprised of 38 spiral galaxies with morphological types of Sa through Sd, specific SFR (sSFR) from ${\sim}10^{-10.5} {-} 10^{-9}$~yr$^{-1}$, SFR from ${\sim}0.2{-}17$ \sfr, and molecular gas surface density ($\Sigma_{\mbox{mol}}$) from ${\sim}10^{0.5}{-}10^{2.7}$ \msun~pc$^{-2}$ \citep[see ][Table 1 \& Figure 1]{lee_phangs-hst_2022}.

PHANGS-HST imaging targeted the star-forming galaxy disk, and includes a combination of new and archival observations in five filters: F275W (NUV), F336W (U), F438W or F435W (B), F555W (V), and F814W (I).\footnote{Parallel imaging with ACS targeting the galaxy halo was also performed to constrain distances by measuring the brightness of the tip of the red giant branch (TRGB) (see \citealt{anand_distances_2021} and Section 3.2 of \citealt{lee_phangs-hst_2022})} We obtained new imaging of 34 galaxies with 43 WFC3 pointings using an allocation of 122 orbits.  Archival NUV-U-B-V-I observations from the LEGUS survey \citep{calzetti_legacy_2015}\footnote{LEGUS data products: \url{https://archive.stsci.edu/prepds/legus/dataproducts-public.html}} were used for seven galaxies (NGC~0628, NGC~1433, NGC~1512, NGC~1566, NGC~3351, NGC~3627, NGC~6744; for the latter three, we obtained additional imaging to increase the HST footprint and match PHANGS-ALMA coverage of the disk).  Suitable archival imaging in selected bands was available for 16 other galaxies.

Table\,\ref{tab:exp_time} summarizes all HST observations, specifies the cameras used and the exposure times. The new data obtained for PHANGS-HST and the archival data were processed together using the same data reduction pipeline \citep[as summarized by][]{lee_phangs-hst_2022} to ensure homogeneity in the data products to the extent possible.  All of the PHANGS-HST science-ready drizzled images and co-aligned single exposures are available for download at MAST\footnote{\url{https://archive.stsci.edu/hlsp/phangs/phangs-hst}}. 
\begin{table*}
\begin{center}
\caption{Exposure time and detector type for each band. This table presents for each PHANGS-HST galaxy the Proposal ID (HST-GO-PID), the exposure time ($t_{\rm exp}$) and number of pointings ($n_{p}$) for each band. We also specify the HST instrument/detector used (Det) except for the bands F275W and F336W as they were all observed with the UVIS detector. We abbreviate ACS/WFC as WFC, and WFC3/UVIS as UVIS. For the B band all observations taken with the UVIS (WFC) detector are performed with the filter F438W (F435W). 
}  
\label{tab:exp_time}
\begin{tabular}{lcccccccccc}
\multicolumn{1}{c}{Galaxy} & \multicolumn{1}{c}{HST-GO-PID} & \multicolumn{1}{c}{${n_{\rm p}}$} & \multicolumn{1}{c}{F275W} & \multicolumn{1}{c}{F336W} & \multicolumn{2}{c}{F435W/F438W} & \multicolumn{2}{c}{F555W} & \multicolumn{2}{c}{F814W} \\ 
\hline
\multicolumn{1}{c}{} & \multicolumn{1}{c}{} & \multicolumn{1}{c}{} & \multicolumn{1}{c}{${t_{\rm exp}}$} & \multicolumn{1}{c}{${t_{\rm exp}}$} & \multicolumn{1}{c}{Det} & \multicolumn{1}{c}{${t_{\rm exp}}$} & \multicolumn{1}{c}{Det} & \multicolumn{1}{c}{${t_{\rm exp}}$} & \multicolumn{1}{c}{Det} & \multicolumn{1}{c}{${t_{\rm exp}}$} \\ 
\hline
\multicolumn{1}{c}{} & \multicolumn{1}{c}{} & \multicolumn{1}{c}{} & \multicolumn{1}{c}{[s]} & \multicolumn{1}{c}{[s]} & \multicolumn{1}{c}{} & \multicolumn{1}{c}{[s]} & \multicolumn{1}{c}{} & \multicolumn{1}{c}{[s]} \\ 
\hline
IC\,1954 & 15654 & 1 & 2083 & 1059 & UVIS & 1006 & UVIS & 649 & UVIS & 844\\
IC\,5332 & 15654 & 1 & 2089 & 1061 & UVIS & 1011 & UVIS & 650 & UVIS & 804\\
NGC\,628C & 10402, 13364 & 1 & 2434 & 2323 & WFC & 864 & WFC & 546 & WFC & 587\\
NGC\,628E & 9796, 13364 & 1 & 2311 & 1102 & WFC & 2967 & UVIS & 947 & WFC & 986\\
NGC\,685 & 15654 & 1 & 1421 & 712 & UVIS & 683 & UVIS & 464 & UVIS & 554\\
NGC\,1087 & 15654 & 1 & 2095 & 1067 & UVIS & 1014 & UVIS & 649 & UVIS & 778\\
NGC\,1097 & 13413, 15654 & 2 & 2220 & 1236 & UVIS & 805 & UVIS & 2229 & UVIS & 697\\
NGC\,1300 & 10342, 15654 & 2 & 2239 & 2202 & WFC & 1710 & WFC & 858 & WFC & 858\\
NGC\,1317 & 15654 & 1 & 2083 & 1063 & UVIS & 1014 & UVIS & 649 & UVIS & 805\\
NGC\,1365 & 15654 & 1 & 2101 & 1071 & UVIS & 1020 & UVIS & 646 & UVIS & 812\\
NGC\,1385 & 15654 & 1 & 2091 & 1066 & UVIS & 1015 & UVIS & 649 & UVIS & 809\\
NGC\,1433 & 13364 & 1 & 2321 & 1097 & UVIS & 950 & UVIS & 1120 & UVIS & 970\\
NGC\,1512 & 13364 & 3 & 2315 & 1095 & UVIS & 944 & UVIS & 1119 & UVIS & 966\\
NGC\,1559 & 14253, 15145, 15654 & 1 & 4330 & 1062 & UVIS & 1196 & UVIS & 1833 & UVIS & 3514\\
NGC\,1566 & 13364 & 1 & 2329 & 1102 & UVIS & 950 & UVIS & 1127 & UVIS & 973\\
NGC\,1672 & 10354, 15654 & 2 & 2730 & 2392 & WFC & 811 & UVIS & 1466 & WFC & 814\\
NGC\,1792 & 15654 & 1 & 2096 & 1071 & UVIS & 1018 & UVIS & 649 & UVIS & 805\\
NGC\,2775 & 15654 & 1 & 2083 & 1061 & UVIS & 1018 & UVIS & 650 & UVIS & 792\\
NGC\,2835 & 15654 & 1 & 2095 & 1064 & UVIS & 1015 & UVIS & 648 & UVIS & 813\\
NGC\,2903 & 15654 & 2 & 2158 & 1096 & UVIS & 1039 & UVIS & 665 & UVIS & 829\\
NGC\,3351 & 13364 & 2 & 2268 & 1092 & UVIS & 1023 & UVIS & 1421 & UVIS & 1550\\
NGC\,3621 & 9492, 15654 & 2 & 2237 & 2210 & WFC & 687 & WFC & 687 & WFC & 917\\
NGC\,3627 & 13364 & 2 & 2200 & 1092 & UVIS & 971 & UVIS & 847 & UVIS & 861\\
NGC\,4254 & 12118, 15654 & 2 & 2126 & 1167 & UVIS & 1023 & UVIS & 696 & UVIS & 758\\
NGC\,4298 & 14913, 15654 & 1 & 2136 & 1867 & UVIS & 1024 & UVIS & 2037 & UVIS & 1026\\
NGC\,4303 & 15654 & 1 & 2097 & 1070 & UVIS & 1016 & UVIS & 651 & UVIS & 780\\
NGC\,4321 & 15654 & 2 & 2306 & 1170 & UVIS & 1108 & UVIS & 708 & UVIS & 891\\
NGC\,4535 & 15654 & 1 & 2088 & 1066 & UVIS & 1014 & UVIS & 646 & UVIS & 789\\
NGC\,4536 & 11570, 15654 & 2 & 2231 & 1158 & UVIS & 1080 & UVIS & 722 & UVIS & 848\\
NGC\,4548 & 15654 & 1 & 2089 & 1066 & UVIS & 1016 & UVIS & 650 & UVIS & 804\\
NGC\,4569 & 15654 & 1 & 2088 & 1064 & UVIS & 1013 & UVIS & 648 & UVIS & 803\\
NGC\,4571 & 15654 & 1 & 2087 & 1064 & UVIS & 1015 & UVIS & 649 & UVIS & 803\\
NGC\,4654 & 15654 & 1 & 2089 & 1067 & UVIS & 1015 & UVIS & 648 & UVIS & 803\\
NGC\,4689 & 15654 & 1 & 2077 & 1062 & UVIS & 1013 & UVIS & 647 & UVIS & 803\\
NGC\,4826 & 15654 & 1 & 2085 & 1069 & UVIS & 1015 & UVIS & 650 & UVIS & 812\\
NGC\,5068 & 15654 & 2 & 1572 & 802 & UVIS & 1023 & UVIS & 655 & UVIS & 817\\
NGC\,5248 & 15654 & 1 & 2096 & 1069 & UVIS & 1016 & UVIS & 651 & UVIS & 792\\
NGC\,6744 & 13364 & 2 & 2250 & 1099 & UVIS & 977 & UVIS & 1099 & UVIS & 957\\
NGC\,7496 & 15654 & 1 & 2078 & 1058 & UVIS & 1008 & UVIS & 646 & UVIS & 807\\
\hline
\hline
\end{tabular} 
\end{center}
\end{table*}

\subsection{Candidate star cluster selection and photometry}\label{ssect:select_photo}

Initial source detection \citep{thilker_phangs-hst_2022} on the HST imaging was performed with a combination of the PSF-fitting photometry software \textsc{DOLPHOT}\footnote{\url{http://americano.dolphinsim.com/dolphot/}} \citep[v2.0,][]{dolphin_dolphot_2016} and \textsc{photutils/DAOStarFinder}\footnote{\url{https://photutils.readthedocs.io/en/stable/api/photutils.detection.DAOStarFinder.html}}\citep[][]{bradley_astropyphotutils_2023}, a python implementation of \textsc{DAOPHOT}\footnote{\url{https://www.star.bris.ac.uk/~mbt/daophot/}}\citep[v1.3-2][]{stetson_daophot_1987}. A combined ``all-source" V-band detection catalog was created as described in \citet{thilker_phangs-hst_2022}. The number of sources detected in each galaxy ranges from 200K to 1.2M, with a median of 300K.

Star clusters have effective radii between 0.5 and 10~pc \citep{portegies_zwart_young_2010,ryon_effective_2017,krumholz_star_2019,brown_radii_2021}. At the distance of our targets, they appear single-peaked and are marginally-resolved in HST WFC3 images, which have a pixel scale of 0\farcs04.  To distinguish point sources from cluster candidates, multiple concentration indices \citep[MCI, ][]{thilker_phangs-hst_2022} are computed using V-band photometry measured in series of circular apertures with radii from 1-5 pixels.  Across all 38 galaxies, a total of $\sim$190K cluster candidates are found.  The candidates are inspected and morphologically classified as described in the next section.  

Fluxes are computed using photometry in circular apertures with radii of 4 pixels \citep[$\sim$ 0\farcs16; which subtends 3.4 to 18 pc for galaxy distances 5 to 23 Mpc spanned by the PHANGS-HST sample, see][Table 1]{lee_phangs-hst_2022}. Sky background at the position of each object is estimated in a sky annulus between 7--8 pixels radius. To compute total fluxes, we apply a correction for the light outside the aperture, carefully derived for each filter and for each galaxy as described in \cite{deger_bright_2022}.  The details of the aperture correction can introduce important differences in the colors and derived physical properties of the sources as discussed by \cite{deger_bright_2022}.

\subsection{Human and Machine Learning (ML) morphological classification}\label{ssec:classification}
Cluster candidates are inspected to eliminate spurious sources, and to place them into three morphological classes associated
with the likelihood of gravitational boundedness for clusters older than the crossing time $\sim$10 Myr \citep{whitmore_antennae_2010, gieles_distinction_2011, bastian_stellar_2012, fall_similarities_2012, chandar_star-cluster_2014, grasha_spatial_2015, adamo_legacy_2017, krumholz_star_2019, cook_star_2019, wei_deep_2020}.  We use the following commonly adopted classes:
\begin{itemize}
    \item Class~1 (C1): star cluster -- single peak, circularly symmetric, with radial profile more extended relative to point source
    \item Class~2 (C2): star cluster -- similar to Class~1, but elongated or asymmetric 
    \item Class~3 (C3): compact stellar association -- asymmetric, multiple peaks 
    \item Class~4 (C4): not a star cluster or compact stellar association (e.g. image artifacts, background galaxies, individual stars or pairs of stars) 
\end{itemize}
The reader is referred to \citet{whitmore_star_2021} for a full description of the PHANGS-HST classification process, and discussion of differences from the LEGUS cluster classification of \citet{adamo_legacy_2017}.  Figures with examples of each of these morphological classes can be found in \citet{wei_deep_2020} Figure 1, \citet{whitmore_star_2021} Figures 1-4, \citet{lee_phangs-hst_2022} Figure 9, \citet{deger_bright_2022} Figures 11-12, \citet{hannon_star_2023} Figure 1.

Historically, a bottleneck in the production of extragalactic cluster catalogs has been the visual inspection of candidates.
To address this bottleneck, the classification of the $\sim190,000$ PHANGS-HST cluster candidates was automated using convolutional neural networks (CNNs).  CNNs were trained using ``deep transfer" machine learning (ML) techniques and samples of human classified candidates, as discussed in detail in \cite{wei_deep_2020, whitmore_star_2021, hannon_star_2023}. \footnote{\textsc{VGG19-BN} \citep{simonyan_very_2015} and \textsc{ResNet18} \citep{he_deep_2015} network architectures were both explored, although we adopt \textsc{VGG19-BN} for the present work.}

To produce the training sets, human classification was performed for the brightest $\sim$1000 candidates in each galaxy by co-author BCW. As a result, the brightest clusters appear in both the human and ML catalogs, but in galaxies with larger candidate samples, fainter clusters are missing from the human catalog.  The ML samples are $\sim$ 1 mag (median) deeper in the V-band \citep[and Section~\ref{ssect:v_mag}]{whitmore_star_2021}, with a range of 16-26\,mag. 
This is an aspect of the human classified cluster samples that users of the PHANGS-HST catalogs should keep in mind, and leads to a number of key characteristics of the catalogs as discussed in Sec.~\ref{sect:catalog_properties}.   

As reported in \citet{hannon_star_2023}, the PHANGS-HST ML and human classifications agreement rates are 74, 60 and 71\,\% for class 1, 2 and 3, respectively.  
The model accuracy slightly decreases as the galaxy distance increases ($\lesssim$ 10\% from 10 to 23 Mpc), and as the clusters become fainter ($\sim$10\% for $m_v>$ 23.5 mag). \citet{whitmore_star_2021} demonstrated that analysis of mass and age functions are robust to the uncertainties in machine learning classifications, and also provided essential advice on science analysis of catalogs based on machine classifications.  Differences in the observed properties of the PHANGS-HST catalogs based on human and machine classifications are explored further in later in this paper.

Overall, the performance of our neural network models is comparable to the consistency between human classifiers \citep{wei_deep_2020, whitmore_star_2021}, as well as the \textsc{STARCNET} models of \citet{perez_starcnet_2021}, developed for classification of star clusters in the LEGUS survey \citep{calzetti_legacy_2015, linden_star_2022}; i.e., 78, 55 and 45\,\%.  It is important to be aware that there is still significant variation in the classification of C2 and C3 objects among different studies and classifiers \citep[e.g., discussion in section 6.3.3 of][]{whitmore_star_2021}.  Part of the issue is that the characteristics of the classes have not been documented with detail much beyond the descriptions at the beginning of this section \citep[e.g., see section 2 in both][]{adamo_legacy_2017, perez_starcnet_2021}.  To help make progress, in \citet{whitmore_star_2021} we provide a full description of the methodology and criteria underlying the BCW classification scheme.  However, further improvement in classification consistency still requires agreement on the criteria among a full range of experts in the field, and the development of a standardized reference set of human-labelled star clusters, as we discuss in \citet{wei_deep_2020}.  

\subsection{Catalog structure and contents}\label{ssect:cat_content}
The observed properties of our census of star clusters and compact associations throughout the PHANGS-HST 38 galaxy sample are provided as part of PHANGS-HST Data Release~4 / Catalog Release~2 (DR4/CR2) hosted at MAST
\footnote{\url{https://archive.stsci.edu/hlsp/phangs/phangs-cat}}.  Four separate catalogs are provided for each galaxy: 
\begin{itemize}
    \item human-classified clusters (Human C1+C2)
    \item ML-classified clusters (ML C1+C2)
    \item human-classified compact associations (Human C3)
    \item ML-classified compact associations (ML C3)
\end{itemize}
The corresponding physical quantities (ages, masses, reddenings) derived through SED fitting are provided in companion catalogs, as described in Paper II.  This catalog structure is motivated by the expectation that the physical quantities may continue to evolve, in particular with the addition of JWST photometry, while the observed properties (from HST) will remain fixed with this release. Thus, overall, 38 (galaxies) $\times$ 4 (morphological classification subsets) $ \times$2 (observed or physical properties) catalogs are available.

The C1+C2 clusters are provided separately from the C3 compact associations for two main reasons.  First, in studies of star cluster evolution, particularly those which seek to constrain cluster disruption, analysis is often performed with only C1+C2 single-peaked, centrally concentrated objects, which are thought to have a higher probability of being gravitationally bound, and exclude C3 multi-peaked objects \citep{bastian_stellar_2012, chandar_star-cluster_2014}.  Terminology was introduced by \citet{krumholz_star_2019} to facilitate discussion of the differences in the approaches taken by various groups: C1+C2 samples are referred to as ``exclusive" samples, while C1+C2+C3 are referred to as ``inclusive samples."  This delineation is explicitly reflected in our catalog structure. Second, the selection methods implemented in the pipeline described above are optimized for the detection of single-peaked clusters, and yield a highly incomplete inventory for multi-peaked stellar associations.  

Science applications requiring a more complete sampling of the young stellar population should not rely on the C1+C2+C3 catalogs alone.  We have developed a second PHANGS-HST pipeline focused on the identification of multi-scale stellar associations to address this issue, by deploying a watershed algorithm to segment point source catalogs into hierarchically nested structures spanning physical scales from 8~pc to 64~pc \citep{larson_multiscale_2023}.  We find that the majority of C3 objects have a position located within these watershed-identified multi-scale stellar associations. Preliminary products from both the PHANGS-HST multi-scale stellar association pipeline and the cluster pipeline have been released for five galaxies as part of PHANGS-HST DR3/CR1.  The current DR4/CR2 for the full 38 galaxy sample supersedes the preliminary DR3/CR1 cluster catalogs.  A complete set of multi-scale stellar association data products for the full 38 PHANGS-HST galaxy sample will be published at a later date. 

The observed quantities provided in the DR4/CR2 catalogs include:
\begin{itemize}
\item persistent IDs to facilitate cross-identification between catalogs, and positional information (object IDs, RA, DEC, image x, y)
\item morphological classification (human classification, if available; machine learning classification for all sources)
\item NUV-U-B-V-I aperture photometry (corrected for aperture losses and foreground reddening; provided in Vega magnitudes and mJy; flags for non-detection and lack of HST coverage)
\item standard concentration indices measured in the V-band
\end{itemize}

A listing of these quantities is provided in Table~\ref{tab:cat_content}, while a full description can be found in the documentation accompanying the DR4/CR2 release at MAST. A discussion of the issues related to the completeness of the catalogs is provided in Section~\ref{sect:discussion}.
\begin{table*}
\centering
\caption{Content description for the PHANGS-HST DR4/CR2 observed property catalogs of clusters and compact associations.  Source positions were determined in the V-band at the detection stage, generally stemming from the \textsc{DOLPHOT} PSF-fitting photometry measurements and have not been optimized with post-facto centroiding or fitting of extended source models.  This can cause source positions to be shifted slightly ($\sim$~1~pixel) from the true location, but has negligible influence on our photometry due to use of a 4 pixel radius aperture. Upcoming structural fitting of C1+C2 clusters, for the purpose of measuring effective radii, will refine source positions.}
\label{tab:cat_content}
\begin{tabular}{lcl}
\hline\hline
\multicolumn{1}{c}{Column name} & \multicolumn{1}{c}{Unit} & \multicolumn{1}{c}{Description} \\ 
\hline
INDEX & int & Running index from 1 to N for each individual target \\ 
ID\_PHANGS\_CLUSTER & int & PHANGS cluster ID for each individual object classified as class 1,2 \\ 
 &  &  \,\,\, or 3, ordered by increasing Y pixel coordinate \\ 
ID\_PHANGS\_CANDIDATE & int & ID in the PHANGS-HST candidate catalog for each individual target, \\ 
 &  &  \,\,\, for cross-identification. \\ 
ID\_PHANGS\_ALLSOURCES & int & ID in the initial PHANGS-HST “all-source” detection catalog for each \\ 
 &  &  \,\,\, individual target, for cross-identification. \\ 
PHANGS\_X & pix & X coordinates on HST X-pixel grid (0...n-1). Scale = 0.03962 \\ 
 &  &  \,\,\, arcsec/pixel. \\ 
PHANGS\_Y & pix & Y coordinates on HST Y-pixel grid (0...n-1). Scale = 0.03962 \\ 
 &  &  \,\,\, arcsec/pixel. \\ 
PHANGS\_RA & deg & J2000 Right ascension, ICRS frame, calibrated against selected Gaia \\ 
 &  &  \,\,\, sources. \\ 
PHANGS\_DEC & deg & J2000 Declination, ICRS frame, calibrated against selected Gaia \\ 
 &  &  \,\,\, sources. \\ 
PHANGS\_CLUSTER\_CLASS\_HUMAN & int & Cluster class assigned through visual inspection. Integers 1 and 2 \\ 
 &  &  \,\,\, represent C1 and C2 compact clusters. 3 stands for C3 compact \\ 
 &  &  \,\,\, associations. Intengers > 3 are artefacts. \\ 
PHANGS\_CLUSTER\_CLASS\_ML\_VGG & int & Cluster class determined by VGG neural network. Integers 1 and 2 \\ 
 &  &  \,\,\, represent C1 and C2 compact clusters. 3 stands for C3 compact \\ 
 &  &  \,\,\, associations. Intengers > 3 are artefacts. \\ 
PHANGS\_CLUSTER\_CLASS\_ML\_VGG\_QUAL & float & Quality value for `cluster\_class\_ml' with values between 0.3 and 1, \\ 
 &  &  \,\,\, providing the frequency of the mode among the 10 randomly \\ 
 &  &  \,\,\, initialized models. \\ 
PHANGS\_[BAND]\_VEGA & mag & HST band apparent vega magnitude, MW foreground reddening and \\ 
 &  &  \,\,\, aperture corrected. Set to -9999 if source is not covered by HST \\ 
 &  &  \,\,\, filter. \\ 
PHANGS\_[BAND]\_VEGA\_ERR & mag & Uncertainty of `[BAND]\_VEGA' \\ 
PHANGS\_[BAND]\_mJy & mJy & HST band flux in mJy, MW foreground reddening and aperture corrected. \\ 
 &  &  \,\,\, Set to -9999 if source is not covered by HST filter. \\ 
PHANGS\_[BAND]\_mJy\_ERR & mJy & Uncertainty of `[BAND]\_mJy' \\ 
PHANGS\_NON\_DETECTION\_FLAG & int & Integer denoting the number of bands in which the photometry for the \\ 
 &  &  \,\,\, object was below the requested signal-to-noise ratio (S/N=1). 0 \\ 
 &  &  \,\,\, indicates all five bands had detections. A value of 1 and 2 means \\ 
 &  &  \,\,\, the object was detected in four and three bands, respectively. By \\ 
 &  &  \,\,\, design, this flag cannot be higher than 2. \\ 
PHANGS\_NO\_COVERAGE\_FLAG & int & Integer denoting the number of bands with no coverage for object. The \\ 
 &  &  \,\,\, specific bands can be identified as photometry columns are set to \\ 
 &  &  \,\,\, -9999. \\ 
PHANGS\_CI & float & Concentration index: difference in magnitudes measured in 1 pix and 3 \\ 
 &  &  \,\,\, pix radii apertures. \\ 
CC\_CLASS & str & Flag to identify in which region on the color-color diagram the \\ 
 &  &  \,\,\, object was associated with. Values are `YCL' (young cluster locus), \\ 
 &  &  \,\,\, `MAP' (middle aged plume) `OGCC' (old globular cluster clump) or \\ 
 &  &  \,\,\, `outside' (outside the main regions and therefore not classified). A \\ 
 &  &  \,\,\, detailed description is found in Section\,\ref{ssect:cc_regions}. \\ 
\hline
\end{tabular} 
\end{table*}
\section{Size and Depth of Cluster Samples}\label{sect:catalog_properties}
\begin{table*}
\begin{center}
\caption{Number count and absolute magnitude ($M_V$) catalog statistics. This table presents the number of star cluster candidates N$_{\rm Cand}$, the number of human inspected candidates N$_{\rm Insp}$, and the number of class 1, 2 and 3 objects (C1, C2, C3) resulting from the Human and ML morphological classifications in the catalogs for each of the 38 PHANGS-HST galaxies (39 fields - the sources in NGC 628 are reported in two separate catalogs). The minimum, median and maximum absolute V-band total magnitude (corrected for foreground MW reddening and aperture losses) are also given for the total C1$+$C2$+$C3 Human and ML samples. The last 3 rows provide the median, mean, and total numbers of objects summed over all 38 galaxies.}
\label{tab:numbers}
\begin{tabular}{lcccccccccccc}
\hline\hline
\multicolumn{1}{c}{Galaxy} & \multicolumn{2}{c}{Candidates} & \multicolumn{4}{c}{Human-classified} & \multicolumn{4}{c}{ML-classified} & \multicolumn{1}{c}{$M_V^{\rm Hum}$} & \multicolumn{1}{c}{$M_V^{\rm ML}$} \\ 
\hline
\multicolumn{1}{c}{} & \multicolumn{1}{c}{N$_{\rm Cand}$} & \multicolumn{1}{c}{N$_{\rm Insp}$} & \multicolumn{1}{c}{C1} & \multicolumn{1}{c}{C2} & \multicolumn{1}{c}{C3} & \multicolumn{1}{c}{C1+2+3} & \multicolumn{1}{c}{C1} & \multicolumn{1}{c}{C2} & \multicolumn{1}{c}{C3} & \multicolumn{1}{c}{C1+2+3} & \multicolumn{1}{c}{min$\vert$med$\vert$max} & \multicolumn{1}{c}{min$\vert$med$\vert$max} \\ 
\hline
\multicolumn{1}{c}{} & \multicolumn{1}{c}{} & \multicolumn{1}{c}{} & \multicolumn{4}{c}{} & \multicolumn{4}{c}{} & \multicolumn{1}{c}{mag} & \multicolumn{1}{c}{mag} \\ 
\hline
IC\,1954 & 1536 & 560 & 37 & 117 & 169 & 323 & 47 & 163 & 647 & 857 & -11.6$\vert$-7.3$\vert$-6.5 & -11.6$\vert$-6.9$\vert$-5.7 \\ 
IC\,5332 & 1432 & 628 & 78 & 152 & 147 & 377 & 35 & 147 & 416 & 598 & -9.4$\vert$-6.0$\vert$-5.3 & -9.4$\vert$-5.9$\vert$-5.1 \\ 
NGC\,628C & 7725 & 1308 & 263 & 225 & 188 & 676 & 534 & 1201 & 1953 & 3688 & -10.7$\vert$-7.6$\vert$-7.0 & -10.7$\vert$-6.2$\vert$-5.3 \\ 
NGC\,628E & 2321 & 283 & 51 & 40 & 22 & 113 & 165 & 357 & 540 & 1062 & -10.3$\vert$-7.5$\vert$-7.0 & -10.3$\vert$-5.8$\vert$-4.9 \\ 
NGC\,685 & 1568 & 704 & 111 & 194 & 172 & 477 & 63 & 168 & 672 & 903 & -12.2$\vert$-7.9$\vert$-7.1 & -12.2$\vert$-7.8$\vert$-6.9 \\ 
NGC\,1087 & 2636 & 976 & 278 & 226 & 174 & 678 & 185 & 384 & 1091 & 1660 & -11.9$\vert$-7.8$\vert$-7.0 & -11.9$\vert$-7.5$\vert$-6.3 \\ 
NGC\,1097 & 7139 & 1182 & 417 & 198 & 159 & 774 & 1037 & 772 & 1962 & 3771 & -13.1$\vert$-8.1$\vert$-7.2 & -13.1$\vert$-6.4$\vert$-4.7 \\ 
NGC\,1300 & 3602 & 892 & 169 & 149 & 179 & 497 & 830 & 824 & 680 & 2334 & -11.2$\vert$-8.0$\vert$-7.4 & -11.2$\vert$-6.8$\vert$-5.7 \\ 
NGC\,1317 & 401 & 180 & 16 & 18 & 34 & 68 & 18 & 32 & 128 & 178 & -11.3$\vert$-8.1$\vert$-6.9 & -11.3$\vert$-8.3$\vert$-6.7 \\ 
NGC\,1365 & 3291 & 1510 & 362 & 267 & 154 & 783 & 353 & 443 & 900 & 1696 & -15.1$\vert$-8.7$\vert$-7.5 & -15.1$\vert$-7.9$\vert$-6.8 \\ 
NGC\,1385 & 2531 & 958 & 269 & 260 & 208 & 737 & 204 & 348 & 1129 & 1681 & -13.1$\vert$-8.1$\vert$-7.2 & -13.1$\vert$-7.8$\vert$-6.5 \\ 
NGC\,1433 & 2083 & 646 & 90 & 104 & 99 & 293 & 148 & 233 & 463 & 844 & -11.5$\vert$-7.9$\vert$-7.3 & -11.5$\vert$-6.9$\vert$-6.1 \\ 
NGC\,1512 & 2675 & 925 & 188 & 120 & 116 & 424 & 220 & 349 & 648 & 1217 & -14.5$\vert$-9.6$\vert$-8.8 & -14.5$\vert$-8.5$\vert$-7.1 \\ 
NGC\,1559 & 8603 & 1592 & 419 & 303 & 218 & 940 & 657 & 839 & 3181 & 4677 & -13.9$\vert$-8.9$\vert$-7.9 & -12.9$\vert$-7.7$\vert$-6.1 \\ 
NGC\,1566 & 9045 & 1752 & 393 & 291 & 166 & 850 & 706 & 591 & 2619 & 3916 & -13.8$\vert$-8.4$\vert$-6.5 & -13.8$\vert$-7.4$\vert$-6.0 \\ 
NGC\,1672 & 8754 & 1419 & 238 & 134 & 121 & 493 & 930 & 1127 & 2855 & 4912 & -13.9$\vert$-9.3$\vert$-8.4 & -13.9$\vert$-7.1$\vert$-5.7 \\ 
NGC\,1792 & 4641 & 1215 & 265 & 301 & 108 & 674 & 255 & 501 & 1683 & 2439 & -12.3$\vert$-8.7$\vert$-7.1 & -12.3$\vert$-8.0$\vert$-6.6 \\ 
NGC\,2775 & 628 & 628 & 136 & 160 & 110 & 406 & 106 & 108 & 138 & 352 & -11.4$\vert$-8.2$\vert$-7.2 & -11.4$\vert$-8.2$\vert$-7.2 \\ 
NGC\,2835 & 3582 & 1567 & 223 & 346 & 324 & 893 & 110 & 369 & 1134 & 1613 & -10.7$\vert$-7.1$\vert$-6.4 & -10.7$\vert$-7.0$\vert$-6.1 \\ 
NGC\,2903 & 10837 & 1156 & 248 & 253 & 232 & 733 & 564 & 1126 & 3687 & 5377 & -13.3$\vert$-8.1$\vert$-7.4 & -13.3$\vert$-6.5$\vert$-5.1 \\ 
NGC\,3351 & 4766 & 1562 & 140 & 177 & 173 & 490 & 238 & 539 & 878 & 1655 & -13.3$\vert$-7.0$\vert$-5.9 & -13.3$\vert$-5.7$\vert$-4.6 \\ 
NGC\,3621 & 20347 & 1307 & 71 & 129 & 183 & 383 & 1148 & 1804 & 4895 & 7847 & -12.2$\vert$-7.8$\vert$-7.2 & -12.2$\vert$-5.4$\vert$-3.9 \\ 
NGC\,3627 & 10673 & 1522 & 462 & 312 & 184 & 958 & 1134 & 1694 & 3287 & 6115 & -12.9$\vert$-8.4$\vert$-7.8 & -12.9$\vert$-7.0$\vert$-5.4 \\ 
NGC\,4254 & 12284 & 1273 & 255 & 225 & 267 & 747 & 659 & 1554 & 4824 & 7037 & -12.8$\vert$-8.7$\vert$-8.1 & -12.8$\vert$-7.2$\vert$-5.5 \\ 
NGC\,4298 & 2272 & 547 & 173 & 103 & 79 & 355 & 161 & 333 & 760 & 1254 & -11.4$\vert$-7.5$\vert$-6.9 & -11.4$\vert$-6.6$\vert$-5.2 \\ 
NGC\,4303 & 9967 & 1192 & 264 & 293 & 140 & 697 & 439 & 1385 & 3813 & 5637 & -12.6$\vert$-9.4$\vert$-8.7 & -12.6$\vert$-7.9$\vert$-6.6 \\ 
NGC\,4321 & 6725 & 1381 & 436 & 279 & 235 & 950 & 521 & 965 & 2563 & 4049 & -14.2$\vert$-8.2$\vert$-7.4 & -12.6$\vert$-7.2$\vert$-5.9 \\ 
NGC\,4535 & 2648 & 972 & 202 & 202 & 127 & 531 & 196 & 310 & 833 & 1339 & -12.4$\vert$-7.8$\vert$-7.0 & -12.4$\vert$-7.4$\vert$-6.5 \\ 
NGC\,4536 & 3120 & 750 & 127 & 189 & 135 & 451 & 216 & 525 & 1106 & 1847 & -12.0$\vert$-7.7$\vert$-7.1 & -12.0$\vert$-6.9$\vert$-5.7 \\ 
NGC\,4548 & 788 & 414 & 96 & 99 & 76 & 271 & 100 & 106 & 242 & 448 & -10.7$\vert$-7.5$\vert$-6.6 & -10.7$\vert$-7.4$\vert$-6.4 \\ 
NGC\,4569 & 1309 & 726 & 212 & 213 & 100 & 525 & 228 & 276 & 322 & 826 & -11.2$\vert$-7.7$\vert$-7.0 & -11.2$\vert$-7.6$\vert$-6.7 \\ 
NGC\,4571 & 1085 & 465 & 61 & 101 & 100 & 262 & 44 & 102 & 377 & 523 & -10.0$\vert$-7.2$\vert$-6.4 & -9.9$\vert$-7.1$\vert$-6.2 \\ 
NGC\,4654 & 2812 & 1272 & 256 & 360 & 243 & 859 & 182 & 458 & 1079 & 1719 & -13.4$\vert$-8.6$\vert$-7.7 & -13.4$\vert$-8.3$\vert$-7.3 \\ 
NGC\,4689 & 1580 & 783 & 130 & 214 & 165 & 509 & 99 & 214 & 582 & 895 & -11.0$\vert$-7.3$\vert$-6.4 & -11.0$\vert$-7.2$\vert$-6.2 \\ 
NGC\,4826 & 1935 & 928 & 62 & 111 & 252 & 425 & 48 & 74 & 514 & 636 & -10.0$\vert$-5.7$\vert$-4.3 & -9.6$\vert$-5.6$\vert$-4.3 \\ 
NGC\,5068 & 6319 & 957 & 54 & 128 & 144 & 326 & 69 & 574 & 2286 & 2929 & -10.0$\vert$-6.8$\vert$-6.1 & -9.5$\vert$-5.0$\vert$-3.9 \\ 
NGC\,5248 & 3434 & 1154 & 211 & 324 & 194 & 729 & 232 & 506 & 1192 & 1930 & -13.2$\vert$-7.7$\vert$-6.9 & -12.0$\vert$-7.3$\vert$-6.2 \\ 
NGC\,6744 & 10276 & 1436 & 221 & 173 & 221 & 615 & 393 & 1122 & 3079 & 4594 & -10.3$\vert$-6.9$\vert$-6.4 & -10.3$\vert$-5.7$\vert$-4.4 \\ 
NGC\,7496 & 1390 & 618 & 105 & 158 & 110 & 373 & 72 & 211 & 452 & 735 & -13.6$\vert$-7.7$\vert$-6.9 & -12.3$\vert$-7.5$\vert$-6.4 \\ 
\hline
Median & 3120 & 972 & 202 & 194 & 165 & 509 & 216 & 443 & 1079 & 1681 & -15.1$\vert$ -8.1$\vert$ -4.3 & -15.1$\vert$ -7.0$\vert$ -3.9 \\ 
Mean & 4840 & 1008 & 199 & 196 & 159 & 555 & 342 & 585 & 1528 & 2456 & - & - \\ 
Total & 188760 & 39340 & 7789 & 7648 & 6228 & 21665 & 13346 & 22834 & 59610 & 95790 & - & - \\ 
\hline
\hline
\end{tabular} 
\end{center}
\end{table*}
In this section, we begin to characterize the star cluster and compact association populations within PHANGS-HST galaxies by examining the sizes and depths of the samples.  Table~\ref{tab:numbers} summarizes the numbers of cluster candidates, number that have human classifications, and number in each morphological class (based on human inspection and ML classification) for each galaxy.

\subsection{How many star clusters and compact associations are found?}\label{ssect:how_many_clusters}
\begin{figure}
\includegraphics[width=0.45\textwidth]{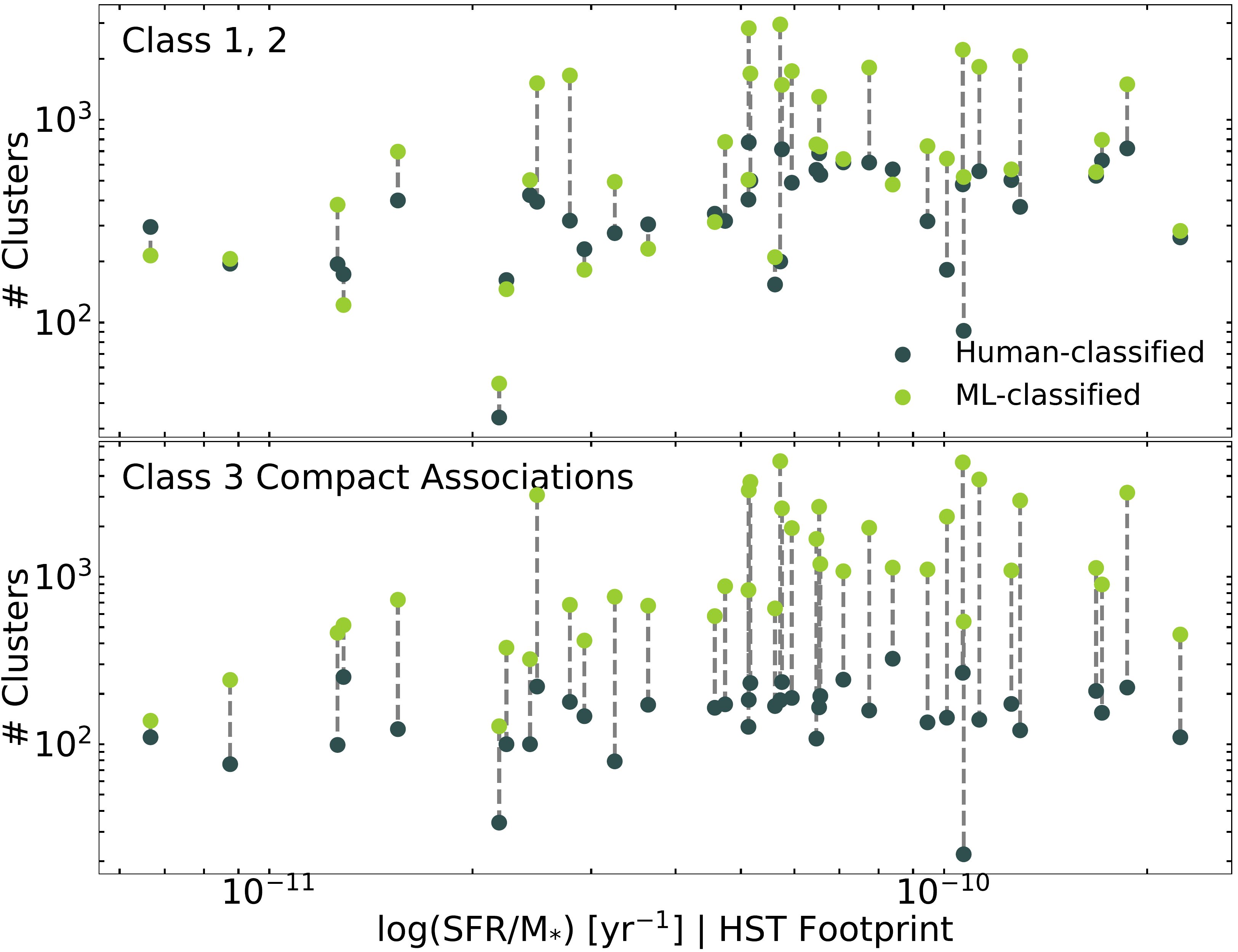}
 \caption{The number of star clusters (top panel) and compact associations (bottom panel) in each of the 38 PHANGS-HST galaxies, shown as a function of the specific star formation rate (sSFR), estimated inside the HST footprint. Sources which have been inspected and classified by a human (co-author BCW) are shown in dark green, while those which have been classified using a neural network model \citep{hannon_star_2023} are shown in light green.}
 \label{fig:n_cluster_ssfr}
\end{figure}

A variety of factors determine the number of star clusters and compact associations reported in the PHANGS-HST catalogs.
In addition to observational completeness (e.g., due to the depth of the imaging for individual targets, spatial resolution achieved, and selection function imprinted by our catalog production pipeline), the global physical properties of galaxies, in particular the star formation history, influence the properties of the cluster population. 

With these factors in mind, and to help visualize the variation in the sizes of the cluster samples across the PHANGS-HST survey, in Figure~\ref{fig:n_cluster_ssfr} we show the number of catalog sources as a function of the specific star formation rate (${\rm sSFR = SFR / M_{*}}$) evaluated inside the HST footprint. \footnote{DSS images with overlays of the HST footprint for each galaxy can be found at \url{https://archive.stsci.edu/hlsp/phangs/phangs-hst}.}  
The SFRs are based on a FUV$+$IR prescription, while the galaxy stellar masses are computed based on an IR flux and mass-to-light ratio \citep{leroy_z_2019,leroy_phangs-alma_2021}\footnote{Also see notes and references provided in Table 1 of \citet{lee_phangs-hst_2022}}.  We present clusters of class 1 and 2 together in the upper panel and compact associations (class 3) in the bottom panel. The human and the ML classified samples are shown separately, again to illustrate the differences in sample sizes.

The mean size of the human classified C1+C2+C3 sample per galaxy is $\sim560$, while for the ML classified C1+C2+C3 sample it is $\sim2500$ ($\sim4$ times larger).  Human classified ``inclusive" C1+C2+C3 samples span over a factor of ten in size from 68 in NGC\,1317 to 958 in NGC\,3627.  ML classified samples of the same variety span an even larger range from  178 in NGC\,1317 to 7847 in NGC\,3621.  This large variation in sample sizes is perhaps the most basic demonstration of the diversity of cluster populations in nearby spiral galaxies.

By construction, the C1+C2 ML classified sample is significantly larger than the human sample for the majority of galaxies. 
However, for IC\,5332, NGC\,685, 2775, 2835, 4571, 4689 and 4826 the human sample contains more C1+C2 clusters than the ML sample. Due to the relatively low number of cluster candidates in these galaxies, all available candidates were classified by co-author BCW. The higher number of C1+C2 human classifications are due to differences in the classification determination with the ML algorithm.

For the C3 compact associations, the ML samples are always significantly larger than the human samples (Figure~\ref{fig:n_cluster_ssfr} bottom panel). 
These large numbers are likely due to a combination of two factors.
First, we deploy our neural network models to classify the full candidate list, and the ML samples thus reach a fainter magnitude limit.  (Recall that our ML samples are a median of $\sim$1 mag deeper in the V-band as discussed in Section~\ref{ssec:classification}; we examine this further in the next section.)  Since the mass function of clusters and associations rises as $\rm{dN/dM} \propto M{^\beta}$, where $\beta\sim-2$ \citep[and references therein]{krumholz_star_2019}, there will be a factor of 100 increase in the number of sources for every additional decade of mass probed (or, up to a factor of 40 increase for every additional magnitude probed).  Second, the C3 compact associations in our catalogs tend to be young \citep[$\lesssim$10 Myr, e.g.,][see also Sec~\ref{ssect:cc_regions}]{lee_phangs-hst_2022}.  For a fixed magnitude limit, these young populations can be detected to much lower masses (between $\sim$0.5 to $\sim$2.5 dex lower, depending on the age of the comparison population) due to the high light-to-mass ratios of massive O and B stars, as illustrated in mass-age diagrams for star clusters \citep[e.g.,][Figure 13]{cook_star_2019}.  

In general, the number of clusters and associations found in each galaxy increases with the sSFR.  Further analysis of the variation in cluster populations with SFR is provided in Section~\ref{ssect:cc_sf}.

\subsection{V-band magnitude distributions}\label{ssect:v_mag}
\begin{figure*} 
\includegraphics[width=\textwidth]{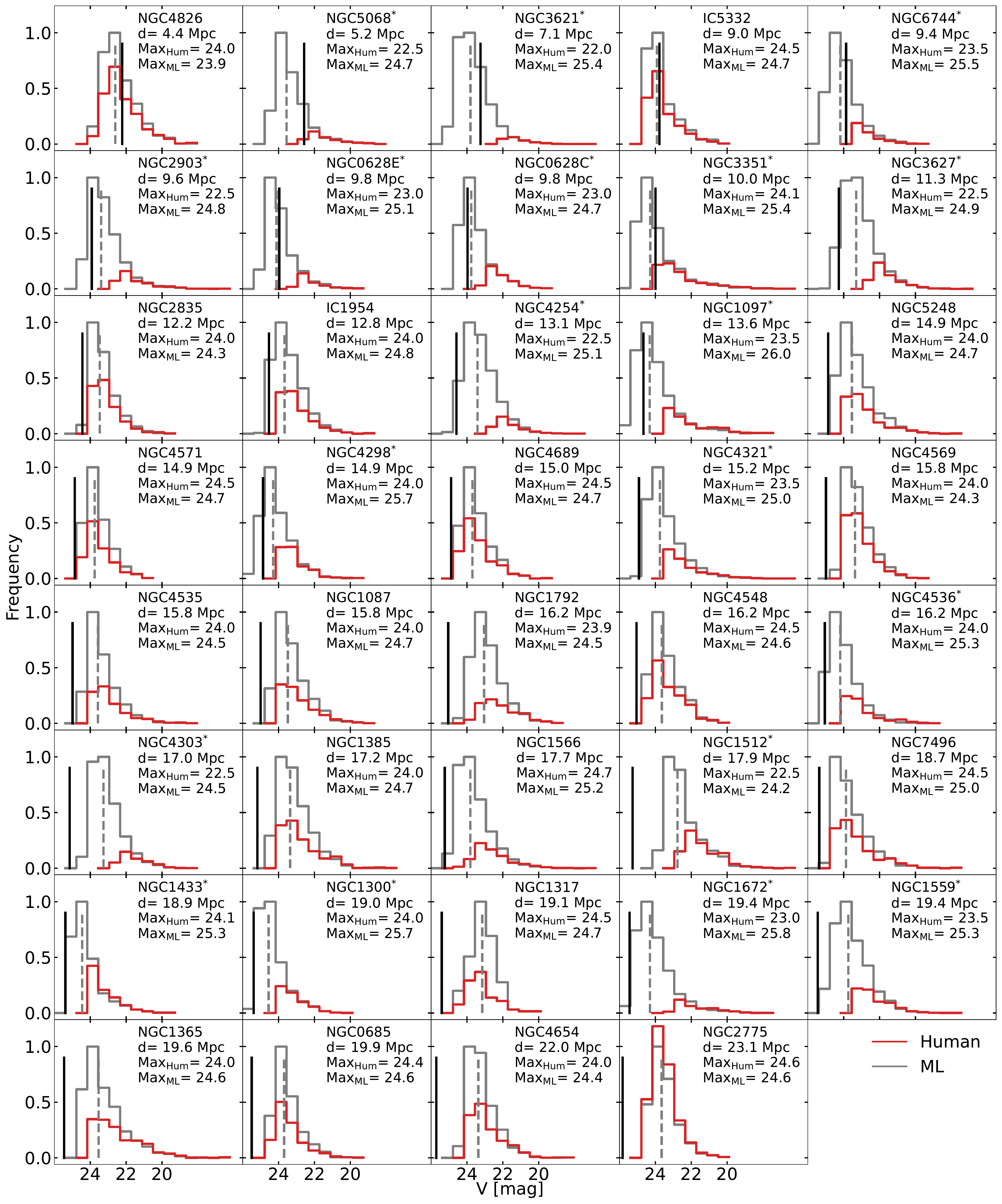}
 \caption{Probability distributions of apparent total V-band magnitude (i.e., corrected for aperture losses) for the cluster (class 1 + 2) and compact association (class 3) populations in all  38 PHANGS-HST galaxies. We show with red (grey) the Human (ML) classified catalogs. In order to compare their distribution we normalized the histograms to the highest bin of the ML sample. For each target, we display the distance and the faintest detected magnitude for the human and the ML classified clusters. A grey dashed line shows the median ML V-band magnitude and the solid black line the limit of ${\rm M_v=-6}$ used as the lower magnitude cut in \citet{adamo_legacy_2017}. We mark targets with a star, if the faintest human detected magnitude is brighter than the median ML detected magnitude.}
 \label{fig:v_mag_panel}
\end{figure*}
Table \ref{tab:numbers} shows the median, minimum, and maximum absolute V-band magnitude ($M_{\rm V}$) for the human and ML samples.  In the absence of a completeness analysis based on (computationally expensive) recovery simulations with artificial star clusters \citep[e.g.][]{mayya08, adamo_legacy_2017, messa_young_2018, linden_massive_2021, linden_star_2022, tang_cluster_2023}, these statistics provide an estimate of the depth of the cluster samples for each galaxy. In Figure~\ref{fig:v_mag_panel}, we show histograms of the apparent V-band magnitude ($m_{\rm V}$) for the clusters and associations in each of the galaxies in the PHANGS-HST sample. The panels are ordered by increasing galaxy distance, and the human and the ML samples are shown separately.

In 18 out of 38 galaxies, the human classified sample is shallower (by $\sim$2 mag) than the ML sample, which is a direct result of our strategy of only providing human classifications for the brightest clusters. For these galaxies (marked with a star next to their names in Figure~\ref{fig:v_mag_panel}), the faintest object in the human classified sample is brighter than the median magnitude of the ML classified sample. 
\begin{figure} 
\includegraphics[width=0.5\textwidth]{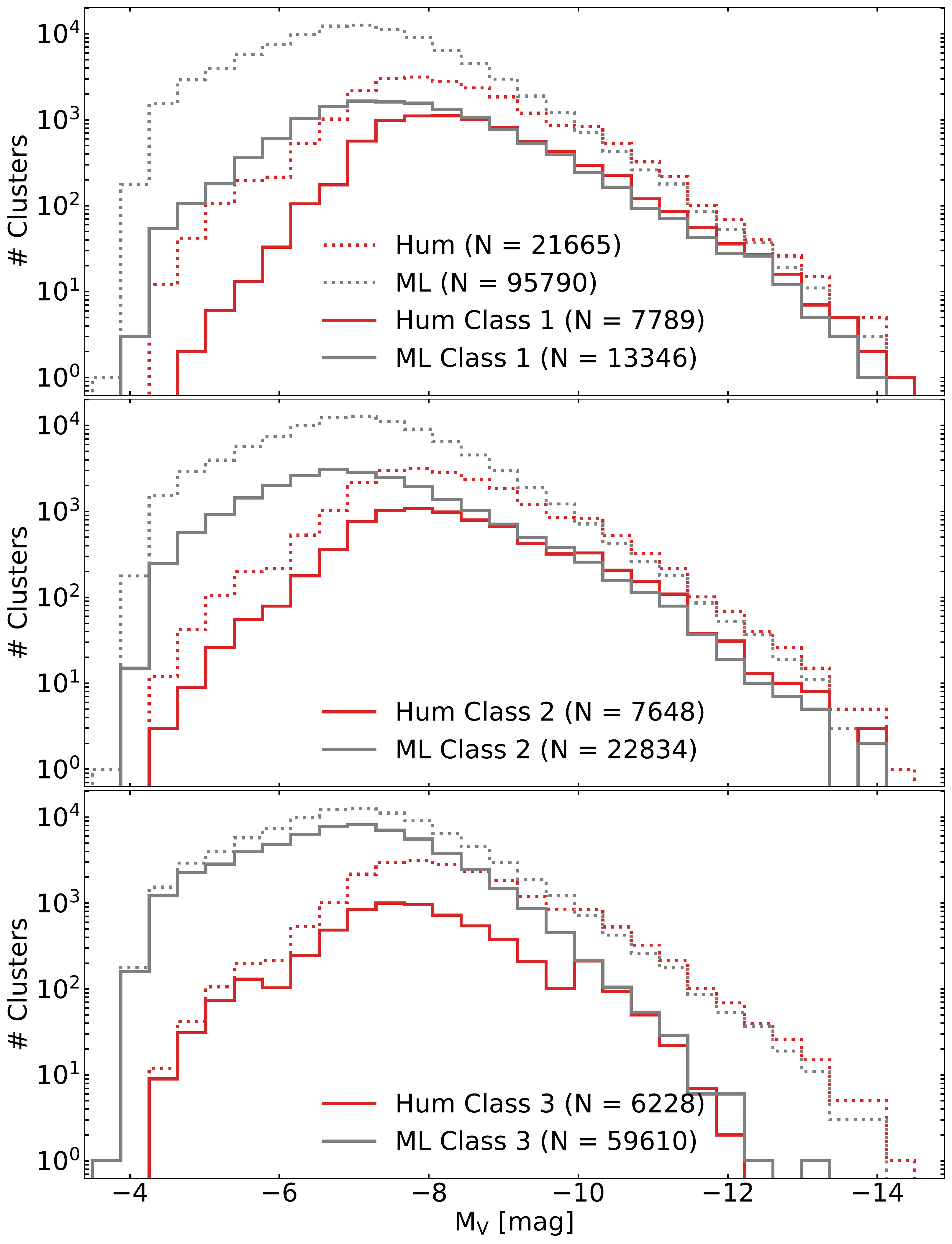}
 \caption{Distribution of the absolute V-band magnitude of the Human (red) and the ML (grey) samples for class 1 + 2 + compact associations shown as dotted lines in all three panels. To visualize individual cluster classes, we show their distributions with solid lines in each panel. The histograms are shown in logarithmic scale to visualize the zone where both samples have comparable sizes, as well as the differences when the machine learning sample size increases toward fainter magnitudes.}
 \label{fig:v_abs_mag}
\end{figure}

Figure~\ref{fig:v_abs_mag} shows histograms of the absolute V-band magnitude $M_{\rm V}$ for all C1 clusters, C2 clusters and C3 compact associations aggregated across the 38 galaxies, with the human and the ML samples shown separately. We also show the $M_{\rm V}$ distribution for each class individually. The distributions for the human and the ML samples are consistent for the brightest objects up to an absolute magnitude of $M_{\rm V} \sim -10$. After that the distributions diverge.
We note that there is a larger difference between human and ML classified objects at fainter magnitudes for C2 clusters and even more for C3 compact associations in comparison to C1 clusters. 
This is due to the fact that the ML sample is deeper than the human sample, and C1 clusters are on average older than C2 clusters, with C3 compact associations representing the youngest objects (see Section~\ref{ssect:cc_overview}). As just discussed in Section~\ref{ssect:how_many_clusters} a larger number of C2 clusters and C3 compact associations will be detected at fainter magnitudes due to a combination of lower mass-to-light ratio at young ages and the shape of the cluster mass function.
For the aggregate Human and the ML samples, the median absolute V-band magnitude is $-8.1$ and $-7.0$ , and their 99-percentiles are $-5.5$ and $-4.5$, respectively. Thus, when combined across the 38 galaxies, the ML sample is about 1 magnitude deeper in the V-band than the human sample.

We note that at the bright end, the aggregate ML sample has 409 fewer C1+C2+C3 objects than the human sample for $M_{\rm V} < -10$\,mag, and this is generally consistent with the accuracy of the ML classifier \citep{hannon_star_2023}.  In cases where a human classification exists, it is preferred for most science applications relative to the ML classification.

\begin{figure*}
\includegraphics[width=\textwidth]{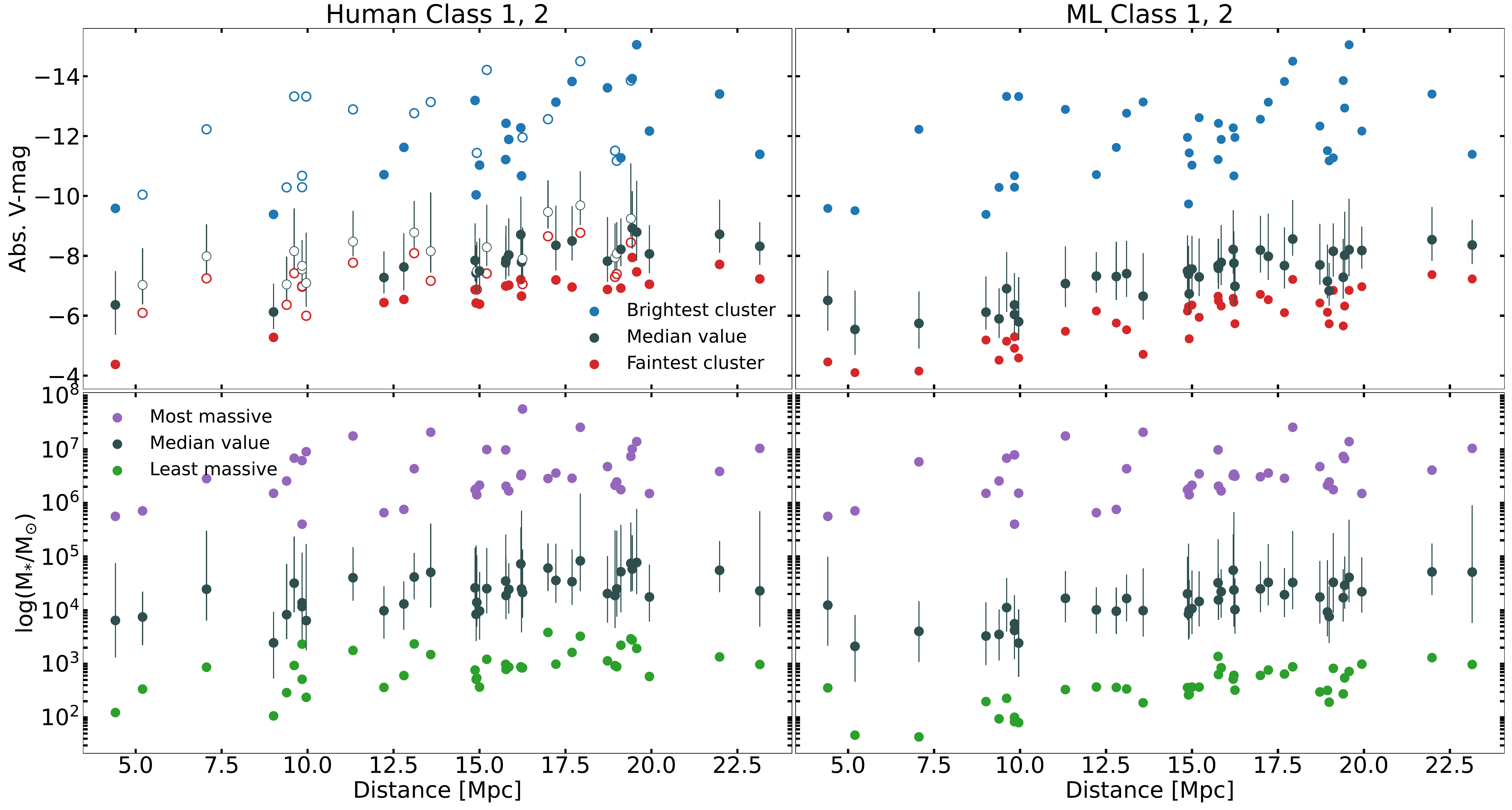}
 \caption{Absolute V-band magnitude (top panels) and stellar mass (bottom panels) as function of galaxy distance. The left (resp. right) panels are for human (resp. ML) classified star clusters (class 1 + 2) for each galaxy. Grey dots with error-bars denote the median value and the 16-84 percentile range, red (resp. blue) dots represent the brightest (resp. faintest) V-band magnitude. The most (resp. least) massive clusters are shown with violet (resp. green) dots.  
 In the top left panel, we use open circles, if the maximal human detected magnitude is brighter than the median ML detected magnitude (See Figure\,\ref{fig:v_mag_panel}).} 
 \label{fig:mag_mstar}
\end{figure*}

\begin{figure*}
\includegraphics[width=\textwidth]{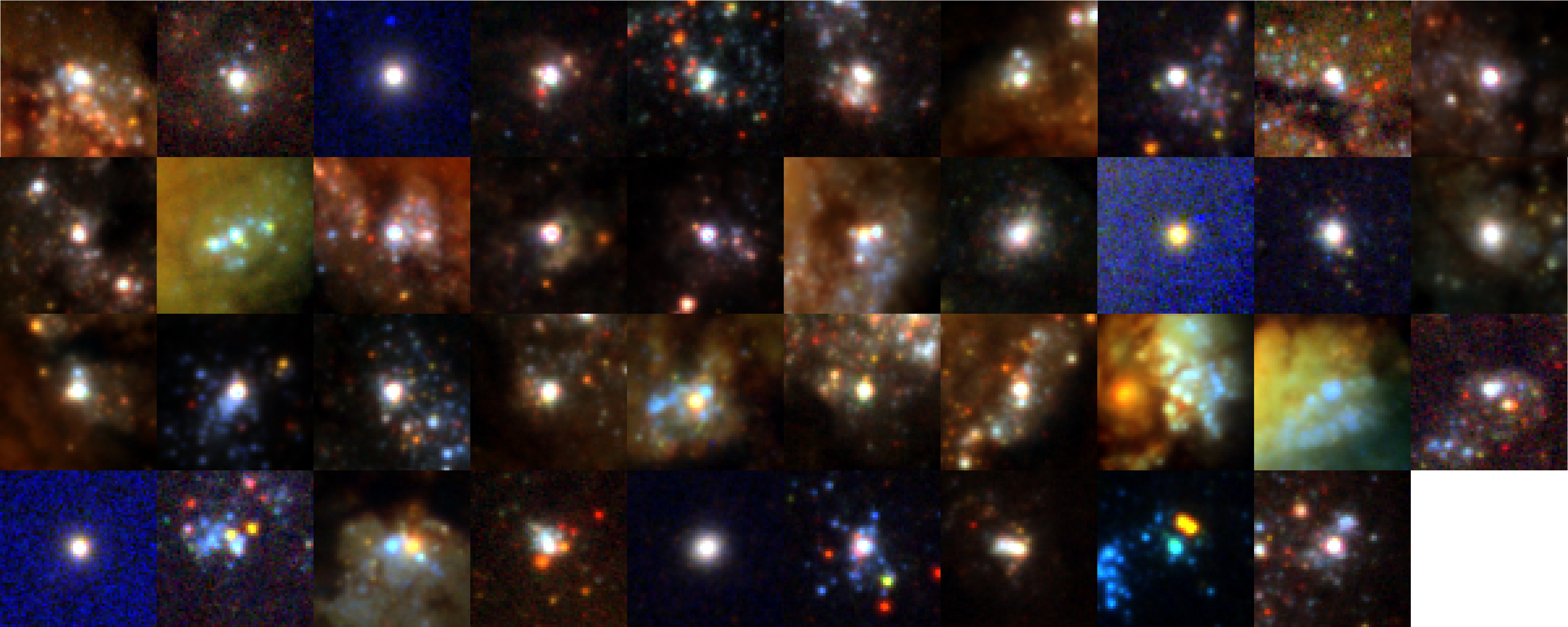}
 \caption{The brightest (absolute V-band magnitude, uncorrected for internal extinction) cluster in each PHANGS-HST target. Color images are constructed from I-, V-, U-band data, and each cutout spans 2.38 arcsec (corresponding to $\sim$50-270~pc, depending on the distance to each galaxy). The clusters are arranged in from left-to-right, top-to-bottom in the order of Table~\ref{tab:numbers}.} 
 \label{fig:ML12_brightestMV}
\end{figure*}

The detection limit depends primarily on the distance of the target since the exposure times for all new HST observations (i.e., as opposed to recycled archival data) were generally uniform (Table 1).  In Figure~\ref{fig:mag_mstar}, we plot the brightest, median, and faintest absolute V-band magnitude, and corresponding quantities for the stellar masses for the C1+C2 samples, in each galaxy as a function of the galaxy distance. The stellar mass is estimated through SED fitting of the 5 filter UV-optical PHANGS-HST photometry as described in \citet{thilker23sed}.
In the upper left panel of Figure~\ref{fig:mag_mstar}, the galaxies where the human classified sample is far shallower are indicated with open circles, consistent with the annotation provided in the Figure~\ref{fig:v_mag_panel} histograms. In Fig.~\ref{fig:ML12_brightestMV} we present a montage showing the brightest cluster in each of our targets.  These luminous clusters are almost all very young (1--3~Myr), though a few middle-age objects and one globular cluster (in NGC\,2775) are also in the sample.

Our catalogs will of course include a population of fainter star clusters in the galaxies which are closer to us which are not detectable in the more distant targets.
The median absolute V-band magnitude is $-6.6$~mag for C1+C2 ML clusters below a distance of 14~Mpc. At distances $>14~{\rm Mpc}$, the median absolute V-band magnitude is $-7.7$~mag. The medians for the human classified samples are $-7.9$~mag for galaxies at distance $<14~{\rm Mpc}$, and $-8.4$~mag for those that are further away. 
In terms of stellar mass, we find median stellar masses of $\log(M_*/M_{\odot}) = 3.9$ and $4.3$ for ML and human clusters, respectively, at distances $<14~{\rm Mpc}$. For the more distant clusters ($>14~{\rm Mpc}$) we find median $\log(M_*/M_{\odot}) = 4.3$ and $4.6$. 


\section{Color-Color Diagrams: The PHANGS-HST 38 galaxy aggregate distribution}\label{sect:color_color}
The SED of a single-age stellar population (or simple stellar population - SSP) evolves over time such that young populations ($\sim 10 {\rm Myr}$) are dominated by blue light from massive stars (e.g. brighter in the NUV or U-band), while old stellar populations ($\sim 1 {\rm Gyr}$), are dominated by red light from lower mass main sequence and evolved intermediate mass stellar populations (e.g. brighter in the I-band).  
Hence, the distributions of star clusters in color-color diagrams have long been studied to gain insight into the properties and evolution of the cluster population 
\citep[e.g.,][]{van_den_bergh_ubv_1968,searle_classification_1980,
girardi_age_1995, larsen_young_1999, chandar_luminosity_2010, adamo_legacy_2017}, 
as well as to test SSP models \citep[e.g.,][]{bruzual_stellar_2003,vazquez_optimization_2005,maraston_evolutionary_1998}.

Our large sample of $\sim$100,000 star clusters and associations combined across the 38 galaxies of the PHANGS-HST sample reveals that the distribution in the U-B vs. V-I color-color diagram can be described in terms of three main features: a young cluster locus (YCL), a middle-age plume (MAP), and an old globular cluster clump (OGC). Here, we examine variations in these features for 
\begin{itemize}
\item different color combinations (NUV-B-V-I and B-V-I as well as standard U-B-V-I), 
\item the three morphological classes of clusters and compact associations, 
\item the machine-learning and human classified samples
\item low and high mass samples
\item the individual 38 galaxies in the survey
\end{itemize}

\subsection{Comparison of the C1, C2, C3 morphological classes and different color combinations}\label{ssect:cc_overview}
\begin{figure*} 
\includegraphics[width=\textwidth]{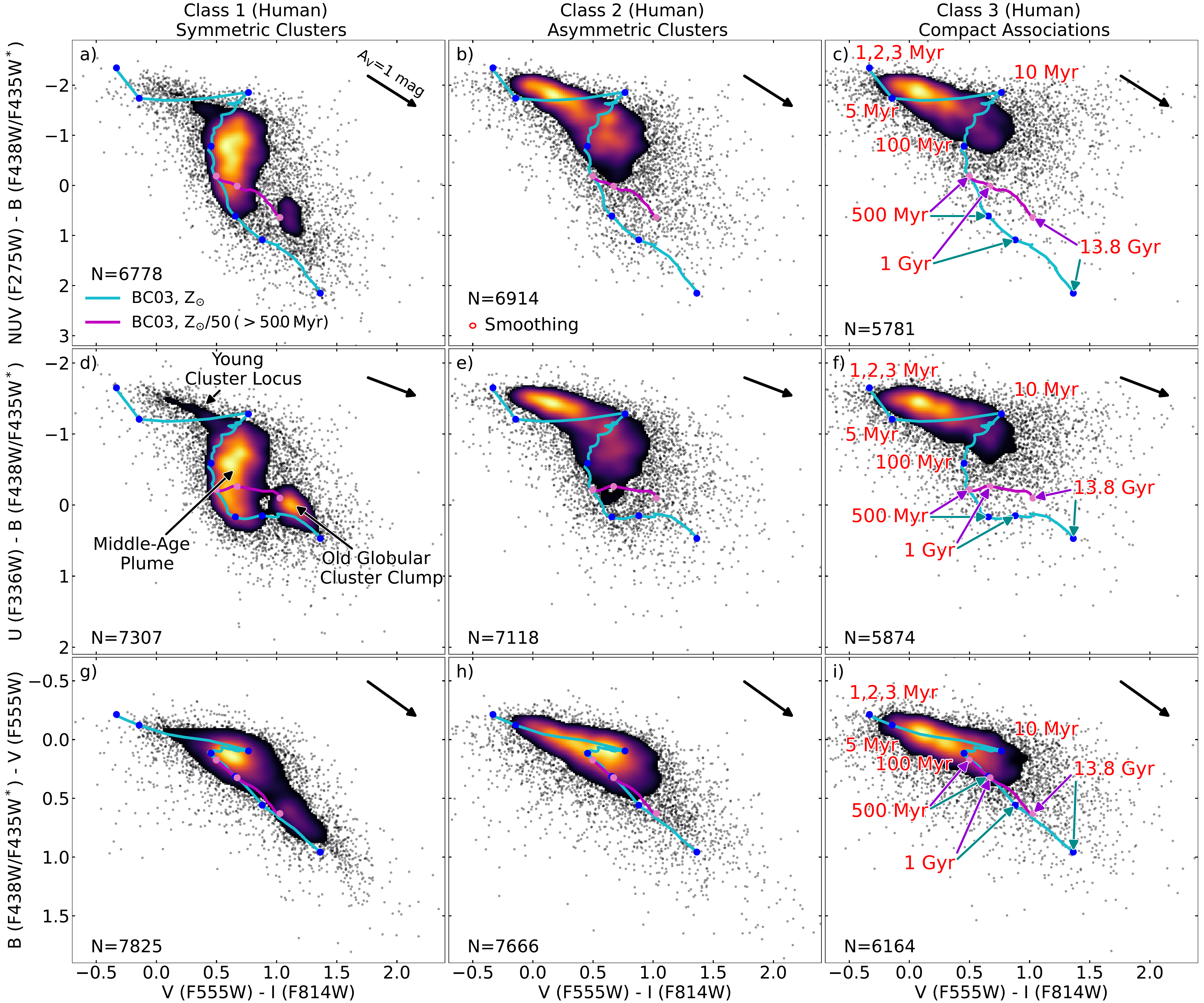}
 \caption{Color-color diagrams for the PHANGS-HST human classified sample, with each morphological class shown separately: C1 single-peaked symmetric clusters (left column); C2 single-peaked asymmetric clusters (middle column); and C3 multi-peaked compact associations (right column). In all panels V-I is plotted along the horizontal axis, and three other colors are shown along the vertical axis: NUV-B (top row), U-B (middle row) and B-V (bottom row).  We only show data points for clusters which are detected with at least a ${\rm S/N > 3}$ in the plotted bands. 
 Individual clusters are represented by black dots whereas in crowded regions we show a Gaussian-smoothed heat map indicating the relative density. 
 The size of the smoothing kernel is shown by a red circle on the top middle panel.
 A cyan track denotes the \citetalias{bruzual_stellar_2003} SSP model for ${\rm Z_{\odot}}$ metallicity at ages from 1\,Myr till 13.7\,Gyr. The portion of the SSP track ${\rm Z_{\odot}/50}$ metallicity from 0.5-13.7\,Gyr is also shown with a magenta track. Key ages are indicated on the right column and are marked with blue and pink dots on each track.  A reddening vector (top right of each panel) corresponds to ${\rm A_v = 1.0 mag}$. In panel d), we indicate names for relevant loci in the color-color space.}
 \label{fig:cc_compare}
\end{figure*}
We begin by presenting color-color diagrams formed from NUV-B-V-I, U-B-V-I and B-V-I photometry for clusters and compact associations in the three human-determined morphological classes (Figure~\ref{fig:cc_compare}).  As in our previous papers \citep[e.g.,][]{turner_phangs-hst_2021,lee_phangs-hst_2022,deger_bright_2022}, we examine the color-color diagram in the context of  \citetalias{bruzual_stellar_2003} SSP model tracks with no addition of nebular emission, and the dust reddening vector.  We show SSP models of $Z_{\odot}$ and $Z_{\odot}/50$ metallicity since it has been well-established by past studies including PHANGS-MUSE that the spiral galaxies, both in our sample and more generally, have nebular metallicities around $Z_{\odot}$ \citep[e.g.,][]{zkh94, skillman_virgo_96, moustakas10, groves_phangs-muse_2023, scheuermann_stellar_2023}, and because our catalogs include objects with a full range of ages, including old globular clusters which are metal poor. The $Z_{\odot}/50$ metallicity BC03 models (based on the Padova 1994 tracks) correspond to $[$Fe/H$]=1.65$ which should generally cover the range of globular cluster metallicities for spiral galaxies \citep[][and references therein]{BS06}. 

Examination of Figure~\ref{fig:cc_compare}, where the human-classified C1, C2, and C3 samples are shown in separate panels, provides insight into how the three morphological classes map onto cluster physical properties.  

The C1 single-peaked symmetric clusters are predominantly older than $\sim$10 Myr (Figure\,\ref{fig:cc_compare} left panels).  Both the middle-age plume and old globular cluster clump are evident in the NUV-B vs V-I, and U-B vs V-I diagrams of the C1 population (Figure\,\ref{fig:cc_compare} a and d). 
Although there are younger C1 clusters which define a sharp diagonal locus roughly parallel to the reddening vector in the U-V vs B-I diagram (Figure\,\ref{fig:cc_compare} d), these objects are in the minority of the C1 population.  

In contrast, the populations of C2 single-peaked asymmetric clusters and C3 multi-peaked compact associations are predominately young, and both show a prominent, clearly defined young cluster locus, which again appears to be roughly parallel to the reddening vector.  The C2 sample YCL exhibits an extension into the middle-age plume to $\sim$500 Myr (Figure\,\ref{fig:cc_compare} b and e).  The shape of the left side of the extension, which follows the BC03 SSP track, suggests that this distribution contains middle-age clusters, and are not solely reddened young clusters.  The C3 YCL human-classified (bright) sample does not have an obvious extension into the middle-age plume.

In the B-V vs V-I diagrams, the three main features are blended and far less distinct (Figure\,\ref{fig:cc_compare} bottom row); this reaffirms the need for NUV and U band photometry for cluster age dating \citep{smith_young_2007}. 
After 100 Myr, not only is the reddening vector parallel to the B-V vs V-I SSP track, but the solar and sub-solar metallicity SSP models trace a similar path (Figure\,\ref{fig:cc_compare} i).  The NUV band (F275W) is the shortest wavelength filter available on the HST WFC3 camera that avoids the 2175 \AA\ dust feature, while the U and B bands straddle the 4000 \AA\ break.  The combination of the NUV-U-B-V-I filters serve to break the age-extinction-metallicity degeneracy, as illustrated by the untangling of the SSP tracks in the NUV-B vs V-I (top row) and U-B vs V-I (middle row) planes, and by the separation of metal-rich and metal-poor tracks, as reflected in the segregation of the old globular cluster clump from the middle-age plume. 

\begin{figure*} 
\includegraphics[width=\textwidth]{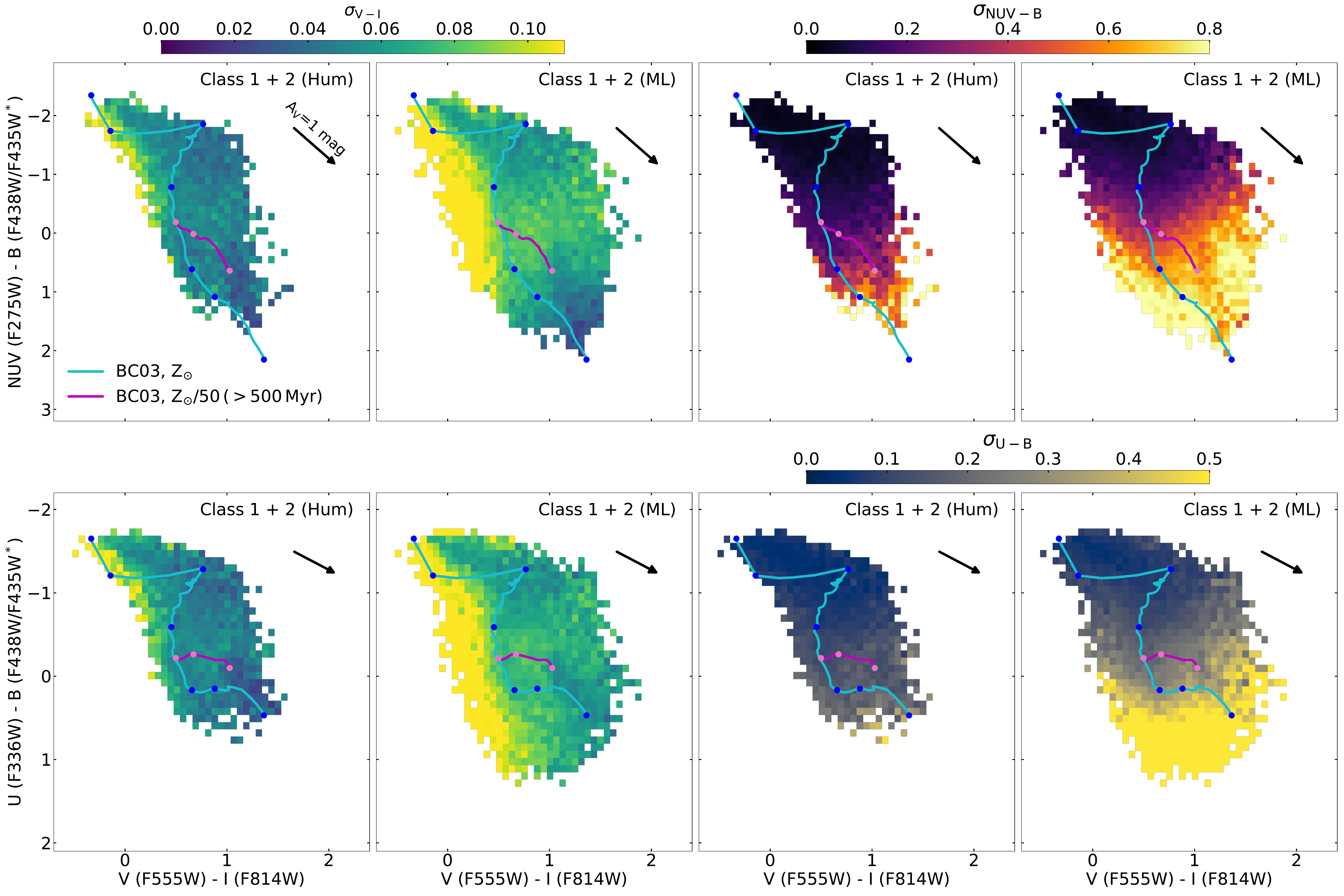}
 \caption{Mean color uncertainties for the NUV-B vs V-I (top row) and U-B vs V-I (bottom row) diagrams. We present class 1+2 clusters for ML (left two panels) and human classifications (right two panels) separately. The maps show the mean uncertainty in each bin, and only bins with at least 5 clusters are displayed. }
 \label{fig:color_color_uncert}
\end{figure*}

Hereafter, we choose to focus on the U-B vs V-I color-color diagram.  While the separation between the middle-age plume and the old globular cluster clump is larger in the NUV-B vs V-I plane, the NUV detection rate and signal-to-noise for old clusters (which are significantly dimmer in the blue) are lower relative to the U-band (Figure~\ref{fig:color_color_uncert}) despite the factor of $\sim2$ larger NUV exposure time (Table 1).

\subsection{Comparison of human and machine-learning classified samples}\label{ssect:cc_compare}
As discussed earlier, by construction, an important difference between the human and ML classified catalogs is the depth of the samples.
\citet{whitmore_star_2021} looked for other possible systematic differences between the ML and human classified samples, and assessed the performance of the ML classifications by examining the UVBI color-color diagram of five individual galaxies processed with the first generation of our CNN models \citep{wei_deep_2020}\footnote{DR3/CR1 at \url{https://archive.stsci.edu/hlsp/phangs/phangs-cat}}.
Here, we compare the samples aggregated over all 38 galaxies, and classified using the current version of our CNN model \citep{hannon_star_2023}.\footnote{DR3/CR2 at \url{https://archive.stsci.edu/hlsp/phangs/phangs-cat}} 

\begin{figure*} 
\includegraphics[width=\textwidth]{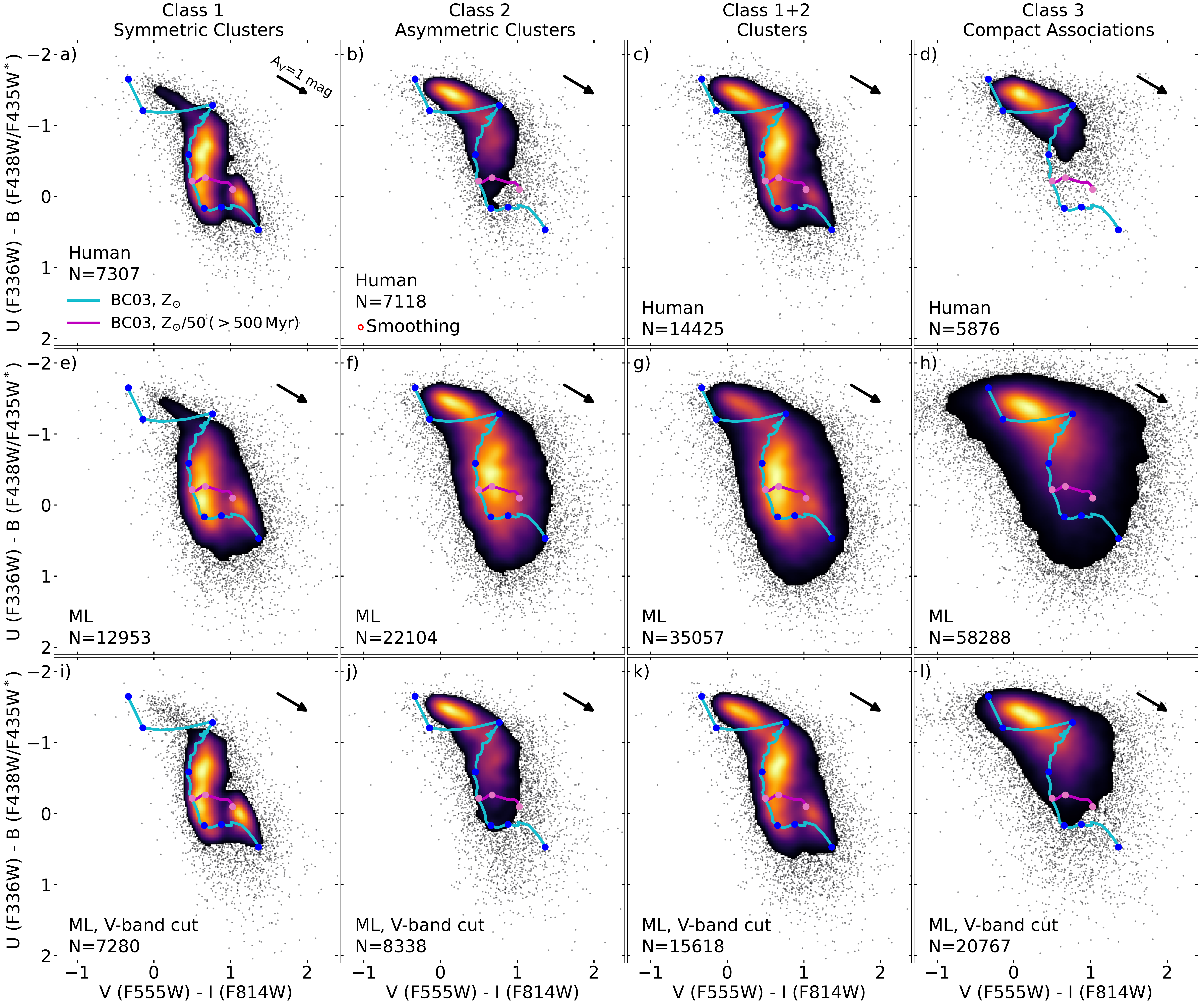}
 \caption{Color-color diagrams for the Human cluster sample (top row) and the ML cluster sample (middle and bottom rows). In the middle row we show all ML classified clusters, whereas the bottom row only shows ML classified clusters up to the same V-band magnitude for each target as detected for the human sample. The individual V-band cuts are estimated with the maximal detected magnitude as presented in Figure~\ref{fig:v_mag_panel}. Cluster classes 1, 2, 1+2 and class 3 compact associations are shown individually in each column from left to right, respectively. Clusters are represented by black dots and in crowded regions by a Gaussian-smoothed heat map indicating the relative density.}
 \label{fig:colo_colo_first_view}
\end{figure*}

In Figure~\ref{fig:colo_colo_first_view} we compare the U-B vs V-I diagram for each cluster class for the Human (top row) and the ML samples (middle and bottom rows).
In the bottom row a V-band magnitude cut corresponding to the depth of the human classified sample (as indicated in Figure~\ref{fig:v_mag_panel}) is applied to the ML sample for each individual galaxy. Qualitatively, it appears that this magnitude cut results in the same color-color features seen in the human-classified sample, which provides evidence for the robustness of the ML classifications for the brighter sources. 

For the C1 clusters, the old globular cluster clump shows a slightly broader distribution for the full ML sample (compare Figure~\ref{fig:colo_colo_first_view} a and e). This slightly broader distribution is mostly due to the fact that fainter globular clusters in the ML sample are detected in the U and B bands, but have low signal-to-noise.
These fainter clusters in the ML samples also appear to shift the peak of the middle-age plume towards older ages (compare Figure~\ref{fig:colo_colo_first_view} a and e).   For the C2 clusters, the increase of fainter sources in the ML sample results in a prominent middle-age plume, which were under-represented in the human classified sample, but does not result in a distinct old globular cluster clump (compare Figure~\ref{fig:colo_colo_first_view} b and f).

Comparison of the human and ML classified C3 compact associations, shows a significantly broader distribution for the ML sample stretching over the entire color-color diagram (compare Figure~\ref{fig:colo_colo_first_view} d and h).    The broadening of the distribution is not surprising given that the ML C3 sample (1) is dominated by young populations and will probe to lower masses relative to the C1/C2 samples (as discussed in Section 3.1), and (2) will thus have the lowest mean S/N values.   We find about 4 times as many ML C3s when applying the human-classified catalog V-band magnitude limit. For the human classified sample, C3s are the smallest category (N=6235, 28\%), however, for the ML classified sample, it is by far as the largest category (N=59684, 62\%).
The low mass ML C3 associations (${\rm <10^4 M_{\odot} }$) will also be affected by stochasticity in sampling of the stellar initial mass function \citep[e.g.][]{fouesneau_accounting_2010,popescu_age_2012,de_meulenaer_deriving_2013,krumholz_star_2015, OD2022}, which leads to large scatter in their luminosities and colors relative to the predictions of the BC03 SSP model track, which assumes a fully sampled IMF.

\subsection{Comparison of high and low mass clusters}\label{ssect:cc_mag_m_star}
\begin{figure*} 
\includegraphics[width=\textwidth]{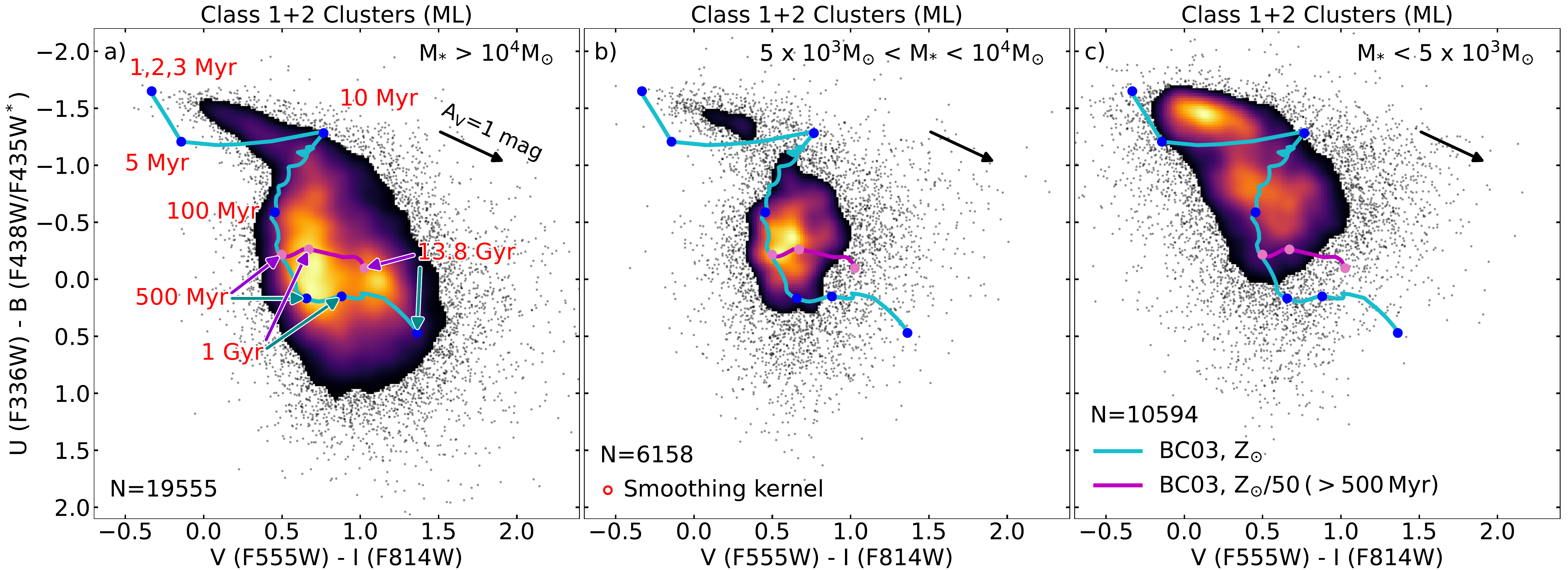}
 \caption{Color-color diagram of ML classified class 1 and 2 clusters in three bins of stellar mass. The most massive clusters with ${\rm M_{*} > 10^4 M_{\odot}}$ are shown in the left panel, intermediate masses of ${\rm 5\times10^3 M_{\odot} > M_{*} < 10^4 M_{\odot}}$ are in the middle panel, and low mass clusters of ${\rm M_{*} < 5\times10^3 M_{\odot}}$ are in the right panel. Similar to Figure\,\ref{fig:cc_compare}, we use a density heat map to illustrate the distribution of clusters.}
 \label{fig:cc_mass_cut}
\end{figure*}

To further explore the impact of stellar IMF stochasticity on the observed properties of low mass clusters, in Figure~\ref{fig:cc_mass_cut} we present the color-color diagram for the C1+C2 aggregate sample in three different mass bins.

The differences in the predominance of the YCL, MAP, and OGC in the three mass bins is primarily due to the dependence of the mass limit with age. As discussed at the end of Section 3.1, for a fixed magnitude limit, due to evolution of the mass-to-light ratio, the YCL ($<10^7$ Myr) can be detected to masses 100 times lower than the OGC ($>$1 Gyr), as illustrated in mass-age diagrams for star clusters \citep[e.g.,][]{cook_star_2019}.
However, the effects of IMF stochasticity are clear when comparing the YCL across the three mass bins.  The YCL is narrow, well-defined, and roughly parallel to the reddening vector in the highest mass bin.  In the lowest mass bin, the distribution is much broader and similar to the stochastic synthesis model predictions shown in Figure 2 of \citet{fouesneau_analyzing_2012}.

\subsection{Quantitative characterization}\label{ssect:cc_regions}
\begin{figure*} 
\includegraphics[width=\textwidth]{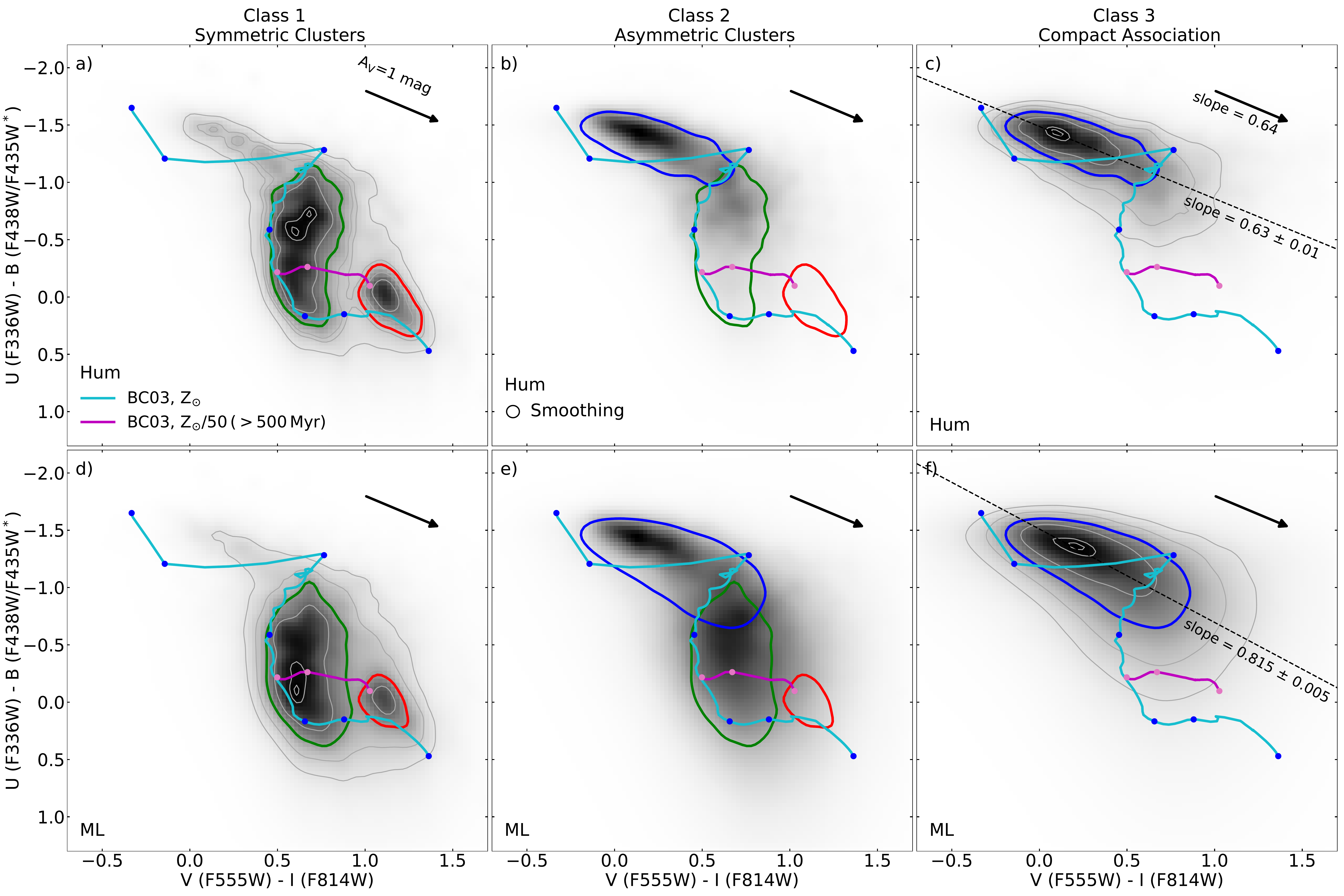}
 \caption{Characteristic regions in U-B vs V-I color-color diagrams of C1 and C2 clusters and C3 compact associations.
 We show human and ML classified samples in the top and bottom row, respectively.
 We compute the color-color maps by stacking each cluster as a normalized Gaussian function on a grid using the color uncertainties as standard deviations. 
 We then identify the YCL (blue) and the MAP (green) as the contour lines encircling 50\,\% of the highest point for C1 clusters and C3 compact associations, respectively. We then find the largest contour line which only encircles the OGC (red), separating this region from the MAP. We show the hulls of all three regions for C2 clusters.
 In order to compare the slope of the reddening vector and the sequence of dust-reddened objects in the YCL, we fit a linear function to all C3 compact association which are inside the blue segmented area.
 }
 \label{fig:color_color_regions}
\end{figure*}
\begin{figure*} 
\includegraphics[width=\textwidth]{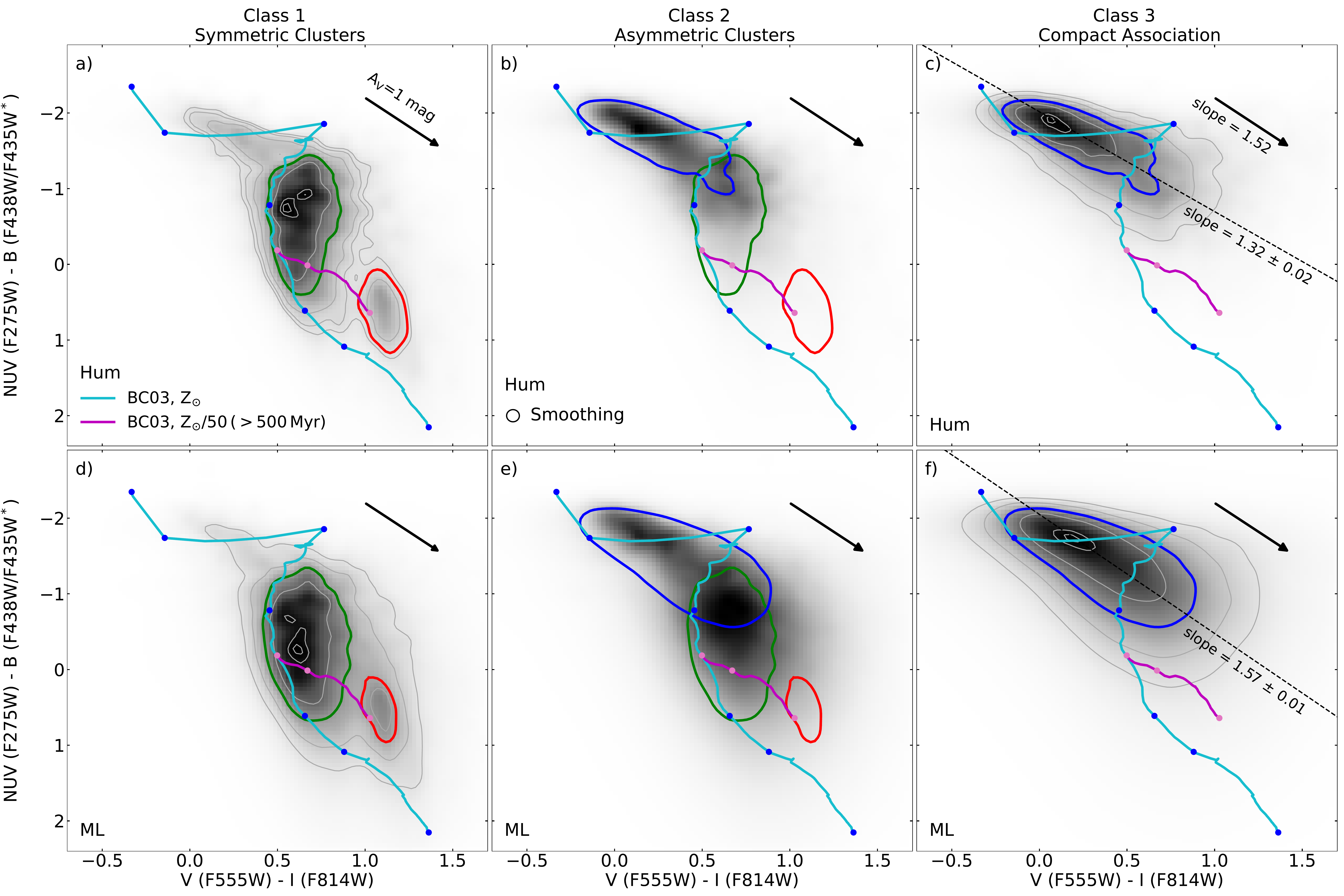}
 \caption{Same as Figure\,\ref{fig:color_color_regions} but with NUV-B colors on the \textit{y}-axis.}
 \label{fig:color_color_regions_nuvb}
\end{figure*}

We now proceed to a quantitative characterization of the three principal features to facilitate further analysis. In particular, in Section~\ref{sect:spatialdist} we will examine the spatial distribution of the populations associated with these features. 

Our first step is to produce an uncertainty weighted color-color diagram. In Figure~\ref{fig:color_color_regions}, each cluster is represented as a normalized Gaussian function with the color uncertainties adopted as standard deviations. Using this approach, clusters with low S/N color measurements are blurred out and do not provide high signal at their specific location in the diagram. On the other hand, more luminous clusters with more precise color-color measurements will dominate the distribution at their positions in these maps.  Figure~\ref{fig:color_color_uncert} shows that the color uncertainties are highest in regions that cannot be reached through reddening of the \citetalias{bruzual_stellar_2003} models.
Uncertainties in the V-I color are highest (left four panels) for clusters on the blue side of the \citetalias{bruzual_stellar_2003} model for middle age clusters (100 to 500\,Myr), and are particular prominent for the ML sample.   U-B color uncertainties (bottom right panels) are highest redward of the \citetalias{bruzual_stellar_2003} model of old clusters (500\,Myr to 13.8\,Gyr). 
By incorporating the uncertainty in the color-color diagrams in Figure~\ref{fig:color_color_regions}, these region with large photometric uncertainties are down-weighted and are less prominent as a result.  

We provide definitions of the Young Cluster Locus, Middle-Age Plume, and Old Globular Cluster Clump by selecting contour lines enclosing the respective regions.
We define the MAP and YCL with contour-lines enclosing $50\,\%$ of all C1 clusters and C3 compact associations, respectively. 
To define the OGC, we select the largest contour lines of C1 clusters which separates it from the MAP. We perform this analysis for the human and ML classified samples separately, as well as for the NUV-B vs V-I diagram.  The results are presented in  Figures~\ref{fig:color_color_regions} and Figure~\ref{fig:color_color_regions_nuvb}. Files providing these contours are at \url{https://archive.stsci.edu/hlsp/phangs/phangs-cat}.

Earlier in this section, we noted that the YCL appears roughly parallel to the reddening vector. The reddening vector corresponding to the \citet{cardelli_relationship_1989} reddening curve has a slope of $0.64$ in the U-B vs V-I diagram.
To probe the orientation of the YCL with respect to the reddening vector, we fit a straight line to the C3 compact associations which are inside the $50\,\%$ contour, and find a slope of $0.63\pm0.01$ and $0.814\pm0.005$ for the human and ML classifications, respectively. 
The general consistency for the human classified C3 compact associations suggests that the shape of the C3 locus is indeed the result of the dust reddening of young clusters (for the ML sample this is affected by the increased scatter due to IMF stochasticity). This exercise illustrates the potential of using color-color diagrams to test reddening laws using carefully selected young, dusty clusters and compact associations. 

As discussed in Section~\ref{ssect:cc_compare},the human and the ML classified samples result in MAP distributions with the same overall shape, but with a peak shifted toward redder (U-B) by $\sim$0.5 (i.e., implying older ages) for the ML sample which appears to be due to its increased depth. Figures~\ref{fig:color_color_regions} and \ref{fig:color_color_regions_nuvb} show that the maximum of the MAP distribution for C1 clusters is located near an age of ${\rm \sim 100~Myr}$ for human classified sample, whereas it is closer to an age of ${\rm \sim 400~Myr}$ for the ML sample.  There does not appear to be as clear of a difference in the peaks of the human and ML classified samples for the C2 clusters.  When using the parametrization for these regions  one should keep in mind that depending whether the human or ML sample is used, populations of slightly different ages are represented.

\begin{figure*}[h]
\includegraphics[width=\textwidth]{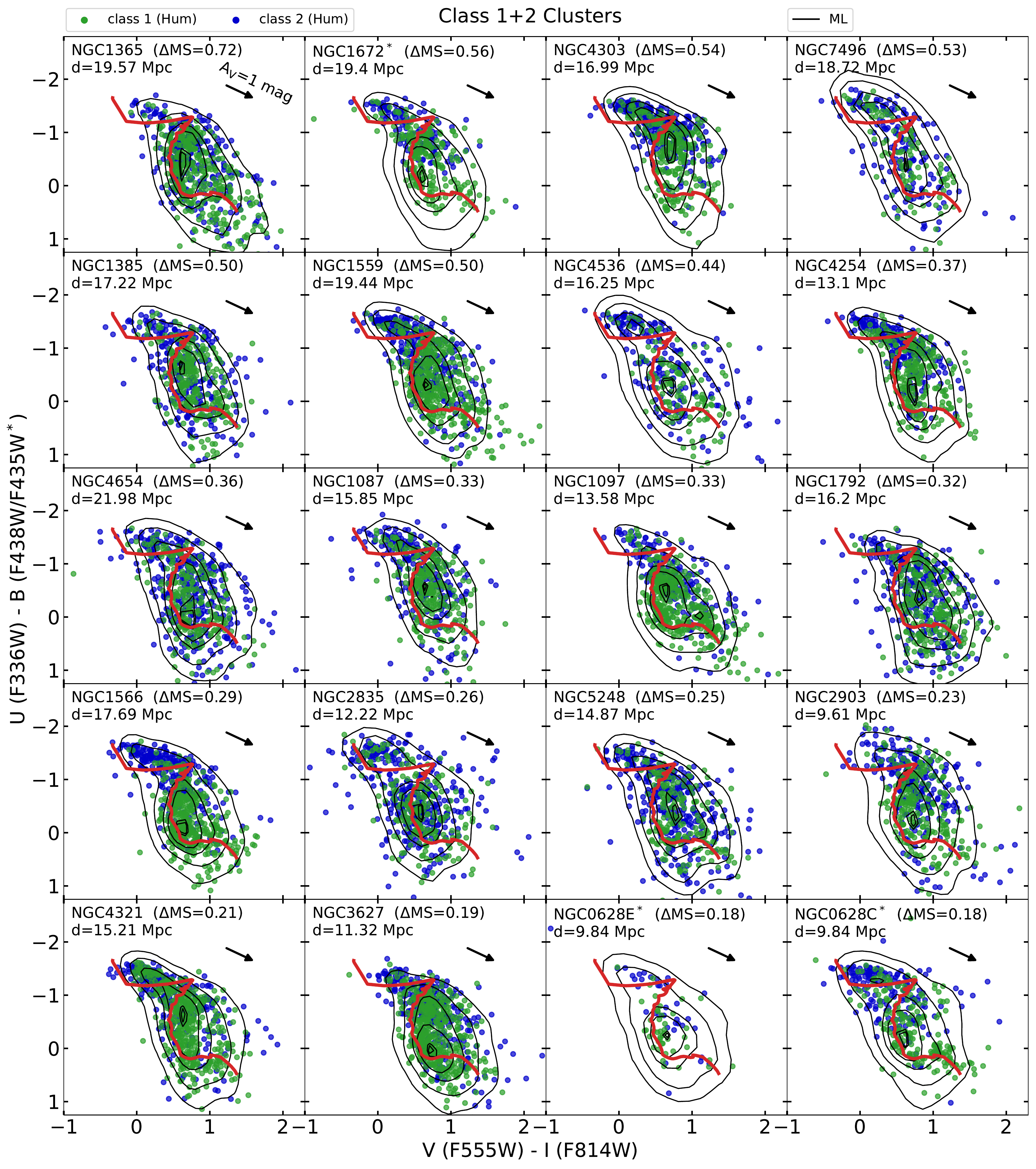}
 \caption{UB-VI color-color diagrams for each individual PHANGS-HST galaxies. We present ML classified class 1 and 2 clusters with black contours. With green and blue points, we over-plot human classified class 1 and 2 clusters, respectively. For reference, we show the solar metallicity track with a red line of the \citetalias{bruzual_stellar_2003}-model. To indicate the direction of color-color shift due to reddening, we show a black arrow in the top left which indicates a reddening of ${\rm A_{V} = 1}$. To study the color-color distribution of each galaxy with respect to the position of the Main Sequence (MS) of star-forming galaxies (see Figure~\ref{fig:ms}), we sort the diagrams in decreasing order of $\Delta$MS values.}
 \label{fig:ub_vi_1}
\end{figure*}
\begin{figure*}[h]
\includegraphics[width=\textwidth]{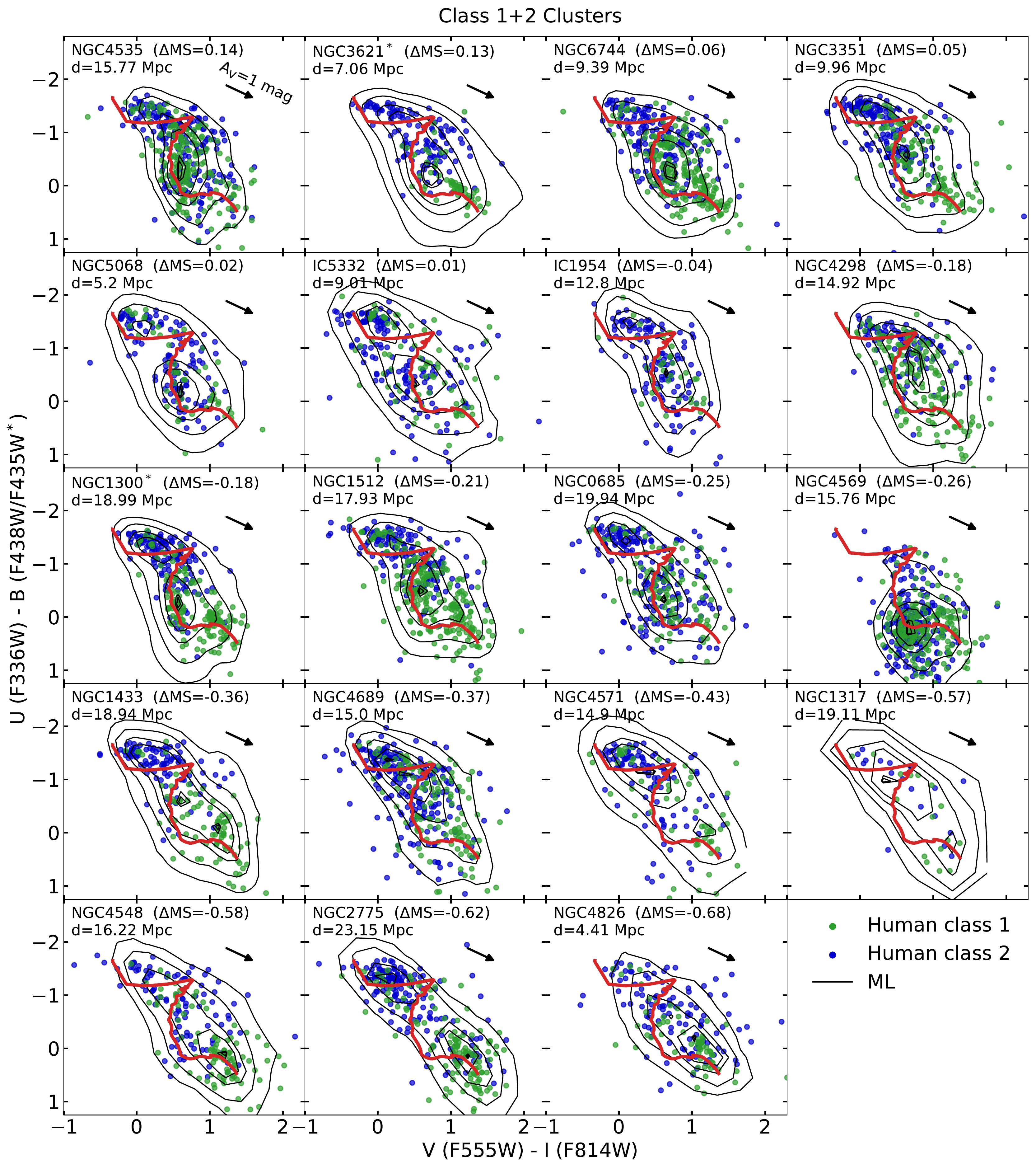}
 \centering{Continuation of Figure~\ref{fig:ub_vi_1}.}
\end{figure*}

\section{Color-Color Diagrams: Individual galaxies}\label{ssect:cc_sf}

Until this point, our analyses of the color-color diagrams have followed the approach of \citet{lee23ubvi} and have been based on the cluster population aggregated across the full sample of PHANGS-HST galaxies.  Here, we return to the more conventional approach of studying color-color diagrams for individual galaxies.

To provide a framework for analysis of the star cluster color-color distributions in the 38 individual galaxies (Figure~\ref{fig:ub_vi_1}), we consider the global star formation rate (SFR) and stellar mass (M$_*$) of the galaxies, but now in the context of the star-forming galaxy main sequence \citep[e.g.,][]{salim_uv_2007, noeske_star_2007, lee_star_2007,peng_mass_2010}.  As in Section 3.1, SFRs are based on an FUV$+$IR prescription, while the galaxy stellar masses are computed based on an IR flux and mass-to-light ratio.

\begin{figure*}
\includegraphics[width=\textwidth]{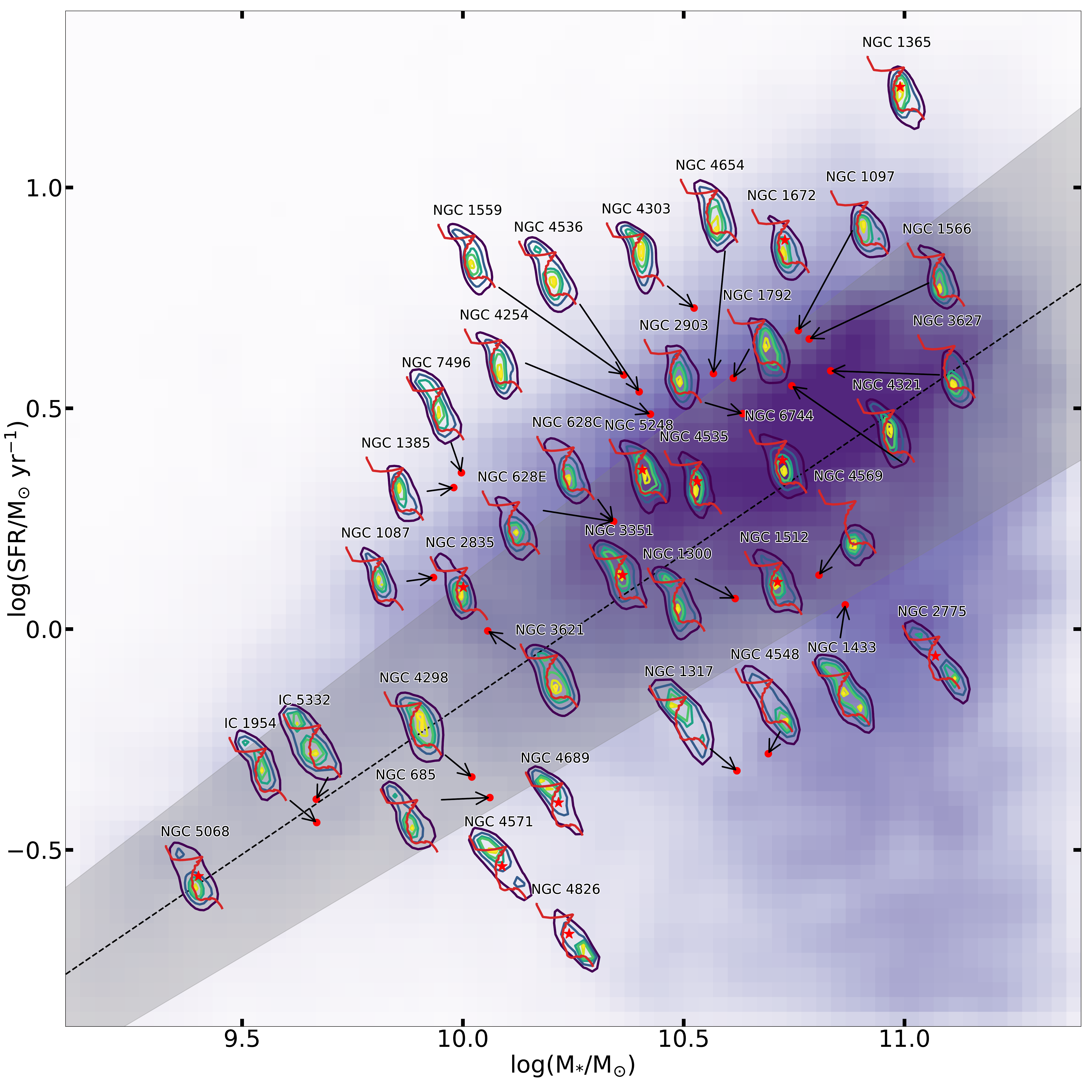}
 \caption{The main sequence (MS) of star-forming galaxies. We represent each galaxy of the PHANGS-HST sample as a U-B vs. V-I color-color diagram (Figure~\ref{fig:ub_vi_1}) at the position on the MS of their host galaxy. The color-color diagrams are presented by contours computed for the ML catalog of C1 and C2 clusters. As a reference, we show for each diagram the  \citetalias{bruzual_stellar_2003}-model track in red. For crowded regions, we shift the color-color diagrams and denote their position on the MS with a red point and an arrow. For those galaxies which are not in a crowded region we mark their position on the MS with a red star, situated in the center of the color-color diagrams. The purple background represents the density of SDSS galaxies of $z < 0.2$ with M$_*$ and SFR values computed by \citet{salim_galex-sdss-wise_2016}. The dashed line is the predicted MS at $z = 0$ defined by \citet{leroy_phangs-alma_2021} and the grey area shows the standard deviation computed by \citet{catinella_xgass_2018}. 
 The essence of this figure is the connection between star formation activity and the cluster population of all PHANGS-HST galaxies. As discussed in the text, the star formation rates are sensitive to timescales of $< 100\,{\rm Myr}$ and therefore the relative fractions of MAP clusters correlate with the relative position on the MS as shown in Figure~\ref{fig:ms_stats}.}
 \label{fig:ms}
\end{figure*}

To visualize trends in the star cluster color-color distributions with galactic star formation properties, in  Figure~\ref{fig:ms} we plot the contours of individual color-color diagrams at the parent galaxy’s position in the star formation rate (SFR)-stellar mass (M$_*$) diagram.  We compute $\Delta$MS, the offset of the galaxy's position in the SFR-M$_*$ diagram relative to the galaxy main sequence.   We order the individual color-color diagrams in Figure~\ref{fig:ub_vi_1} by $\Delta$MS, from the most intensely star-forming galaxies furthest above the MS to those below the MS.  Table~\ref{tab:DeltaMS} provides $\Delta$MS and M$_*$ for each galaxy.  In these plots we show only C1 and C2 clusters, which have a higher likelihood of being gravitationally bound.  

\begin{figure*}
\includegraphics[width=\textwidth]{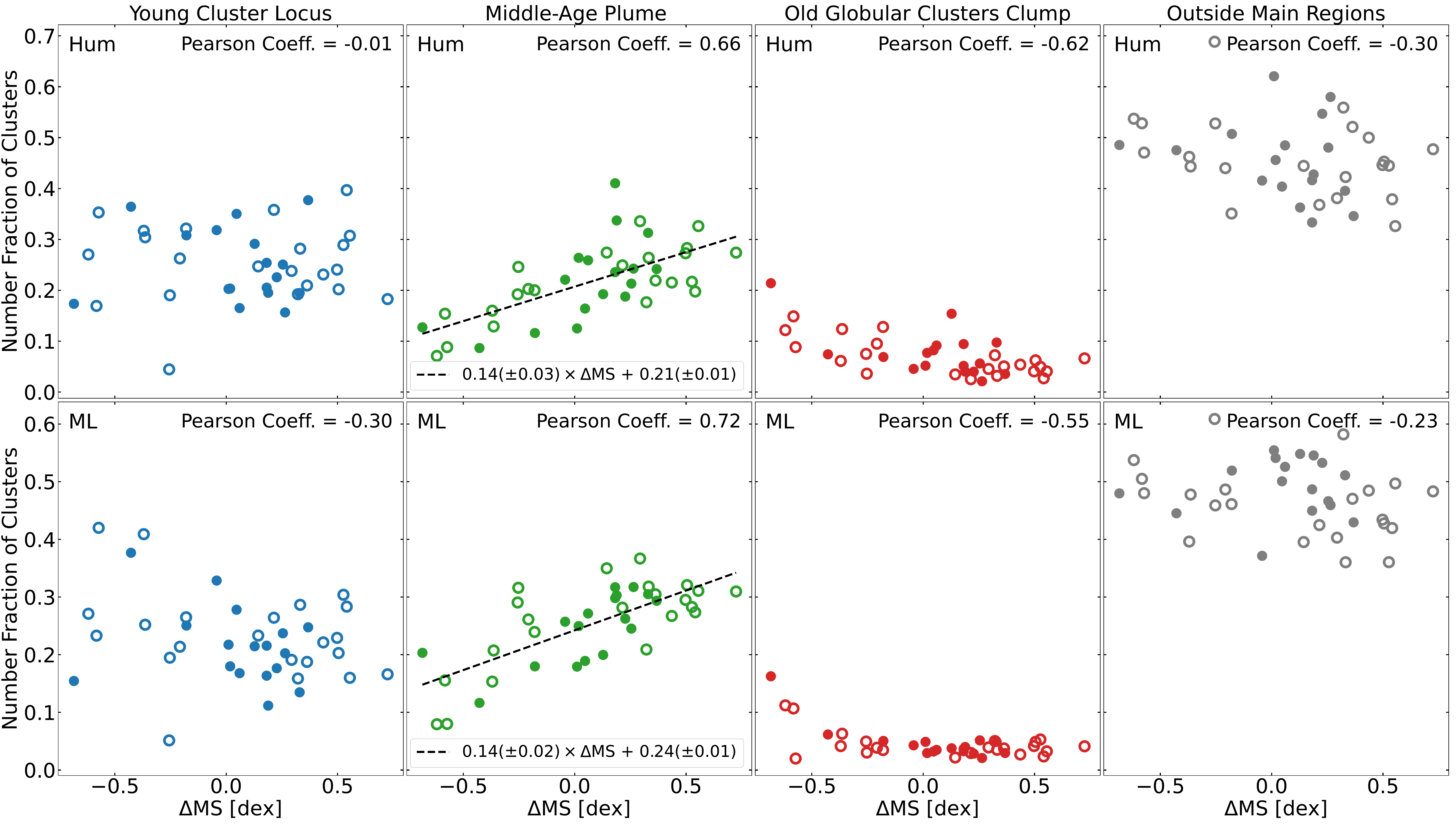}
 \caption{Number fraction of C1 and C2 clusters of each galaxy associated with the main characteristic regions in color-color diagrams found in Section~\ref{ssect:cc_regions} as a function of $\Delta$MS. We show the YCL, the MAP and the OGC in blue, green and red, respectively. In gray, we show clusters outside the main regions. We distinguish galaxies at a distance of smaller and larger than $15\,{\rm Mpc}$ with full and open circles respectively. For each panel we show the Pearson correlation coefficient in the top right. Since the MAP shows a strong correlation which we explain in the text, we fitted a linear function to the data points and provide the fit parameters.}
 \label{fig:ms_stats}
\end{figure*}

To quantify changes in the relative distribution of clusters and associations among the three principal features of the color-color diagram, we compute the relative number fractions in the YCL, MAP, and OGC for each individual galaxy and examine them as a function of $\Delta$MS (Figure~\ref{fig:ms_stats}). No attempt was made to correct for the variation in the depth of the YCL, MAP, and OGC populations due to evolution in the mass-to-light ratio with age prior to computing these fractions.  Thus, the absolute values of the number fractions themselves may not be physically meaningful.  However, the general 
relative trends in Figure~\ref{fig:ms_stats} should still provide insights into differences in the global processes which drive, regulate, and extinguish star and cluster formation across the galaxy sample.  We also show that the differences in depth between the cluster samples for the different galaxies (e.g., due to distance) does not seem to affect the results.

\begin{table*}
\centering
\begin{center}
\caption{Tabular representation of dependence between $\Delta$MS and galaxy morphological properties. PHANGS-HST galaxies are sorted in order of decreasing MS deviation, and the NGC/IC number is shown in the following columns whenever the specified column property is applicable to a particular galaxy.  The galaxy stellar mass is also provided.}
\label{tab:DeltaMS}
\begin{threeparttable}
\begin{tabular}{lcccccccc}
\hline\hline
\multicolumn{1}{c}{$\Delta$MS} & \multicolumn{1}{c}{log M$_*$} & \multicolumn{7}{c}{Morphological features} \\ 
\hline
\multicolumn{1}{c}{[dex]} & \multicolumn{1}{c}{[M$_{\odot}$]} & \multicolumn{1}{c}{Bar-driven SF \tnote{a}} & 
\multicolumn{1}{c}{Central-Ring} & 
\multicolumn{1}{c}{SF-end-of-Bar \tnote{b}} & 
\multicolumn{1}{c}{Global-Arms \tnote{c}} & 
\multicolumn{1}{c}{Bulge \tnote{d}} & 
\multicolumn{1}{c}{Flocculent \tnote{e}} & 
\multicolumn{1}{c}{Quiescent \tnote{f}} \\ 
\hline
0.72 & 10.99 & N1365 & N1365 & N1365 & N1365 &  &  & \\
0.56 & 10.73 & N1672 & N1672 & N1672 &  &  &  & \\
0.54 & 10.52 & N4303 &  & N4303 &  &  &  & \\
0.53 & 10.00 & N7496 &  & N7496 &  &  &  & \\
0.50 & 9.98 & N1385 &  &  & N1385 &  & N1385 & \\
0.50 & 10.36 & N1559 &  &  & N1559 &  &  & \\
0.44 & 10.40 &  &  & N4536 &  &  &  & \\
0.37 & 10.42 &  &  &  &  &  &  & \\
0.36 & 10.57 & N4654 &  & N4654 & N4654 &  &  & \\
0.33 & 9.93 &  &  &  &  &  & N1087 & \\
0.33 & 10.76 & N1097 & N1097 & N1097 &  &  &  & \\
0.32 & 10.61 &  &  &  &  & N1792 &  & \\
0.29 & 10.78 &  &  &  & N1566 & N1566 &  & \\
0.26 & 10.00 &  &  &  & N2835 &  &  & \\
0.25 & 10.41 &  &  &  & N5248 &  &  & \\
0.23 & 10.63 & N2903 &  & N2903 &  &  &  & \\
0.21 & 10.75 &  & N4321 &  & N4321 &  &  & \\
0.19 & 10.83 & N3627 &  & N3627 & N3627 & N3627 &  & \\
0.18 & 10.34 &  &  &  & N628 & N628 &  & \\
0.14 & 10.53 & N4535 &  &  & N4535 &  &  & \\
0.13 & 10.06 &  &  &  &  & N3621 & N3621 & \\
0.06 & 10.72 &  &  &  &  & N6744 & N6744 & \\
0.05 & 10.36 & N3351 & N3351 &  &  & N3351 & N3351 & \\
0.02 & 9.40 &  &  &  &  &  & N5068 & \\
0.01 & 9.67 &  &  &  &  &  & I5332 & \\
-0.04 & 9.67 &  &  &  & I1954 &  &  & \\
-0.18 & 10.02 &  &  &  &  &  & N4298 & \\
-0.18 & 10.62 & N1300 & N1300 & N1300 & N1300 & N1300 &  & \\
-0.21 & 10.71 & N1512 & N1512 & N1512 &  & N1512 & N1512 & \\
-0.25 & 10.06 &  &  &  &  &  & N685 & \\
-0.26 & 10.81 & N4569 &  & N4569 &  &  &  & N4569\\
-0.36 & 10.87 &  &  &  &  &  & N1433 & \\
-0.37 & 10.22 &  &  &  &  &  & N4689 & \\
-0.43 & 10.09 &  &  &  &  &  & N4571 & \\
-0.57 & 10.62 &  &  &  &  &  & N1317 & \\
-0.58 & 10.69 &  &  & N4548 &  & N4548 &  & \\
-0.62 & 11.07 &  &  &  &  & N2775 & N2775 & N2775\\
-0.68 & 10.24 &  &  &  &  & N4826 &  & N4826\\
\hline
\end{tabular} 
\begin{tablenotes}
\item[a] \textbf{Bar-driven SF}: i.e., short bars (like NGC 4536 and NGC 685) and stellar bars 
with minimal star formation
(e.g., NGC 6744, and NGC 4548) are not included, since they do not appear  to be generating much star formation. 
\item[b] \textbf{SF-end-of-Bar}: A clear enhancement of star formation at the end of the bar (like NGC 1300) compared to downstream.
\item[c] \textbf{Global-Arms}: Relatively continuous  star formation along the spiral arm for at least 180 degrees (like NGC 1566 and and NGC 4535).
\item[d] \textbf{Bulge}: Evidence of a old (red), roughly spherical or slightly flattened central component without extensive star formation (e.g., NGC 3351, NGC 2775). Generally associated with the presence of old globular clusters.  
\item[e] \textbf{Flocculent}: Rather than global arms, star formation is in short, irregular regions of star formation. See \citet{EE87}.
\item[f] \textbf{Quiescent}: Large regions without active star formation. Often associated with galaxies that have had their gas removed by ram-pressure stripping (e.g., NGC 4689 - \citep{kenney_co_1986}.
\end{tablenotes}
\end{threeparttable}
\end{center}
\end{table*}

\subsection{$\Delta$MS \& the Young Cluster Locus (YCL)}\label{sec:ycl}
Figure~\ref{fig:ms_stats} shows no correlation between $\Delta$MS and the relative number fraction of clusters associated with the YCL. There are at least two reasons for the lack of correlation.  First, the dust-corrected FUV star formation indicator traces galaxy SFRs over $\sim$100 Myr timescales, while the YCL population is $\lesssim$ 10 Myr.  Nevertheless, SFR tracers over these two timescales have been shown to correlate \citep[e.g.,][and references therein]{salim_uv_2007, lee_comparison_2009}.  A more important issue involves the impact of dust on the observed colors of young clusters.   An absent or weak YCL does not necessarily signify the lack of recent cluster formation.  In fact, NGC\,1365 and 1672, neither of which have a prominent YCL, have the largest $\Delta$MS and are host to the most extreme central starbursts in the sample \citep{brandt_rosat_1996, querejeta_stellar_2021, whitmore_phangs-jwst_2023}.  These high sSFR galaxies have significant dust, which shifts the YCL feature along the reddening vector into the middle age plume \citep{thilker23sed} and even into the old globular cluster clump \citep{hollyhead_studying_2015}. On the other hand, galaxies with low $\Delta$MS values would be expected to have a lack of recent cluster formation, and a weak YCL.  Examples of this are NGC~4826, and NGC~4569, which has the most peculiar color-color distribution of the sample.  In this context it is notable that NGC 4569 is the brightest late-type galaxy in the Virgo cluster.  It experienced a ram pressure stripping event about 300 Myr ago \citep{vollmer_ngc_2004,crowl_stellar_2008,boselli_spectacular_2016} which drained the galaxy's gas reservoir and quenched its star formation.  This event is reflected in the nearly complete absence of the YCL and unusual MAP in NGC~4569's cluster color-color diagram.

PHANGS-HST galaxies with prominent YCLs relative to the other color-color diagram features are
NGC~7496, 1559, 4536, 1566, 1300, 685 and 2775. It is notable that in their YCL regions, we mostly find C2 clusters, indicating that their asymmetric shape is associated with young age.  

\subsection{$\Delta$MS \& Middle Age Plume (MAP)}\label{sec:map}
The MAP feature is visible for most of our galaxies and for some galaxies this feature is by far the most prominent one. 
Figure~\ref{fig:ms} shows that galaxies with more positive $\Delta$MS values have more distinct MAP features. In fact, galaxies below the MS tend to lack this feature, as in NGC~4569, 4689, 4571, 1317, 4548, 2775 and 4826. 
This trend is apparent in Figure~\ref{fig:ms_stats} through a clear correlation between the number fraction of clusters situated in the MAP and the $\Delta$MS value. 

A linear fit to this correlation yields the same slope of $0.14$ for both human and ML classified cluster samples. This behaviour may be expected since the SFR values are based on the UV emission and thus is an average of the star formation history over a few hundred Myr, and the MAP holds the largest fraction of such clusters.

It may be surprising that the correlations resulting from the human and the ML classified samples are the same given that the MAP distribution shows different peaks in color-color diagrams with the two samples.  As discussed in Sections~\ref{ssect:cc_compare} and  ~\ref{ssect:cc_regions}, the two peaks are separated by (U-B)$\sim$0.5 which implies an age difference of a few hundred Myr.  Despite this, there is no significant difference between the correlations in Figure~\ref{fig:ms_stats}.  This could be due to the fact that the star (and cluster) formation rate should be relatively constant over a dynamical timescale for the galaxy, which happens to also be several hundred Myrs for spiral galaxies.
We can estimate the the dynamical timescales as $\tau_{\rm dyn} \approx r / v_0$, where $r$ is the galaxy radius and $v_0$ is the asymptotic velocity of the modeled CO-rotation curves \citep{lang_phangs_2020}. The average for the PHANGS-HST galaxy sample is $\overline{\tau_{\rm dyn}} = 760~{Myr}$ with the smallest measurement for NGC1559 of $\tau_{\rm dyn}=335~{Myr}$. 
If the dynamical timescales of the galaxies in the sample were shorter (e.g., for dwarf galaxies), difference in depths of the samples would more likely affect the results.

To further investigate cluster sample completeness issues that may influence the relative fraction of clusters in the MAP, we tested for correlations with the galaxy distance (Figure~\ref{fig:ms_stats_dist}). There is no correlation with the distance. 
There is also no correlation with the median absolute V-band magnitude ${\rm M_V}$ of the cluster sample. 
The lack of correlation between the cluster sample depth and the fraction of MAP clusters is most likely explained by the fact that we are computing the relative fraction of these cluster groups and not the total numbers. 
This suggests that the relative fractions are not sensitive to the variation in depth, which is described in in Section~\ref{ssect:v_mag}, spans over $\sim 1~{\rm mag}$ in the V-band.

\subsection{$\Delta$MS \& Old Globular Cluster Clump (OGC)}
The OGC feature in the color-color diagram contains the oldest star cluster populations in each galaxy. A larger relative number of globular clusters may indicate intense star formation in the early evolutionary phase of the galaxy \citep{BS06}, whether in-situ or ex-situ \citep[and references therein]{CG19}, but also means that the clusters have not been disrupted and have persisted through time. 
In particular NGC 4826, 6744, 3621, 628c, 1097, 1512, 1433, 1300 and 2775 host a significant population of old globular clusters, which are almost exclusively classified as class 1.  There is no correlation with $\Delta$MS.

\section{An atlas of star cluster spatial distributions}\label{sect:spatialdist}

A careful examination of Figure~\ref{fig:ms} in combination with our HST imaging reveals a number of trends between the positions of the galaxies in the diagram and galaxy morphology.  This motivates examination of the properties of the cluster populations in relation to both $\Delta$MS and galaxy morphology.  For this and other science applications \citep[e.g., calculation of correlation functions, constraints on star formation timescales, and comparison with simulations, e.g.,][]{Gouliermis14, grasha_spatial_2015, grasha_hierarchical_2017, grasha_spatial_2019, turner22}, it essential to examine the 2D spatial distribution of clusters in each galaxy.

Here, we provide an atlas of star cluster maps for the full PHANGS-HST 38 galaxy sample.  In Figure~\ref{fig:spatial_dist} to \ref{fig:spatial_dist_9}, we present the spatial distributions of the clusters associated with the three principal features of the color-color diagram -- the old globular cluster clump (OGC), middle age plume (MAP), and young cluster locus (YCL).  A color composite HST image is included, and ALMA CO(2-1) intensity contours are overlaid on the maps of the YCL. Following the analysis of the previous section, we show the maps in decreasing order of $\Delta$MS values.   

\begin{figure} 
\includegraphics[width=0.48\textwidth]{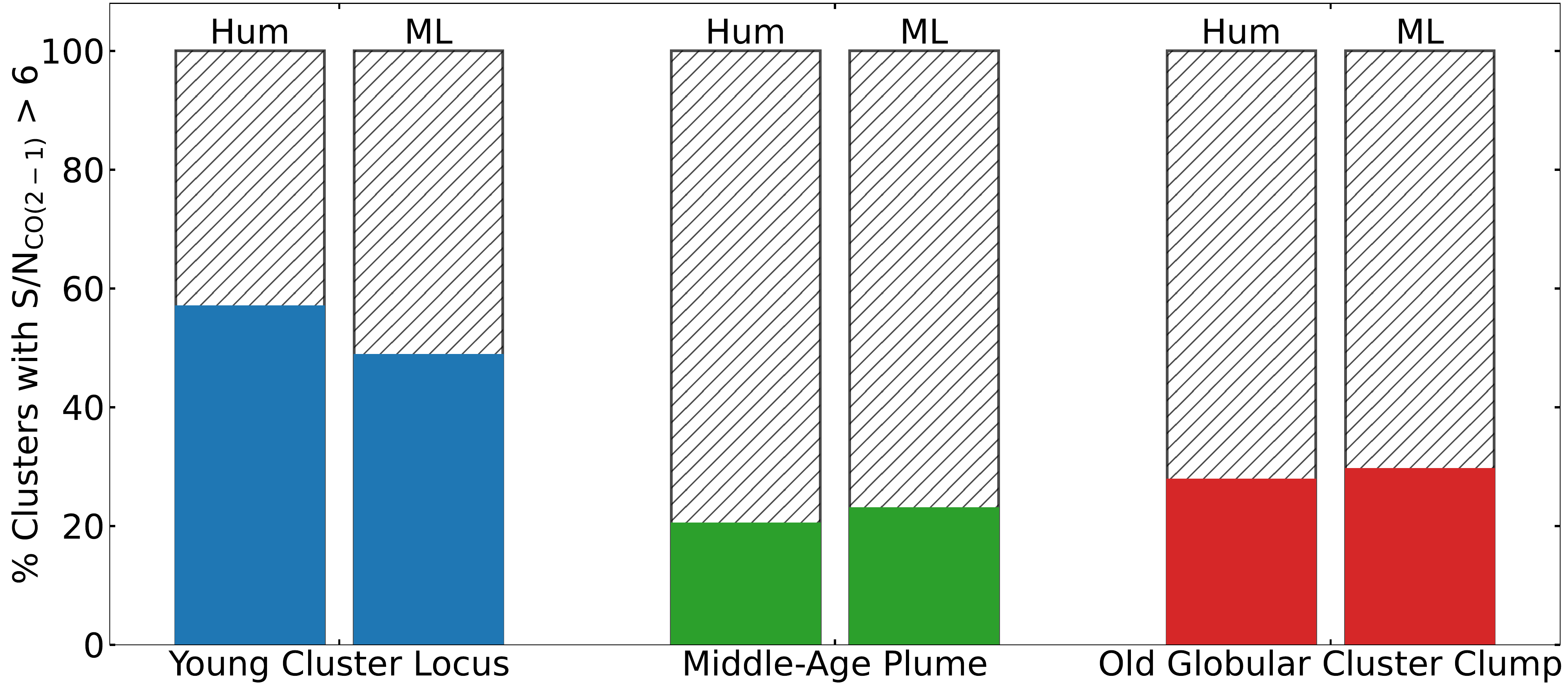}
\caption{Histograms representing the percentage of C1 + C2 compact clusters associated with molecular clouds. We show for each characteristic region (YCL, MAP and OGCC) the percentages of clusters which are co-spatial with ALMA CO(2-1) molecular gas detection with a S/N$>6$. The human-classified and machine learning classified samples are shown separately.}
 \label{fig:dist_gmc}
\end{figure}

A broad examination of the overall atlas shows that objects associated with the YCL are generally found in areas with CO, as expected. On average, we find that YCL objects are coincident with CO twice as often than objects associated with the MAP or OGC (Figure~\ref{fig:dist_gmc}).   As also expected, YCL objects closely trace the spiral structure and central dynamical rings, and reflect the structure of the ISM from which they are born.  These structures then disperse with age -- the spatial organization is broader for the MAP objects, and is closest to a random distribution for objects associated with OGC.

\section{Galaxy morphology}
To facilitate a multi-scale examination of trends across the 38 PHANGS-HST galaxies, we combine information about key galaxy morphological features with galaxy $M_*$ and $\Delta$MS in Table~\ref{tab:DeltaMS}. The classifications in Table~\ref{tab:DeltaMS} are based on visual examination of a BVI image by co-author BCW. 

We have checked how well our visual classifications agree with prior reference studies in the literature for bars, global spiral structure, and flocculent star formation. For example, we find that all 15 galaxies in which we have identified bar-driven SF (i.e., either in the bar, in a central star-forming ring at the inner end of the bar, or at the outer end of the bar) are indeed classified as barred (11 / 15 as SB and 4 / 15 as SAB) by \citet{buta15}. 

We performed a similar check on our classification of spiral structure, as determined by \citet{EE87}. Here we find that 8 of the 9 galaxies in which we have identified global spiral structure, and that are within the sample defined by \citet{EE87}, are consistent with the their determinations. Similarly, 9 of the 11 galaxies in common characterized as flocculent agree. We conclude that our classifications are in reasonably good agreement with previously established determinations.

Starting at the top of Figure~\ref{fig:ms} and Table~\ref{tab:DeltaMS}, we note that several of the galaxies with the largest positive residuals are galaxies with star forming bars, such as NGC 1365, NGC 1672, NGC 4303, NGC 7496, NGC 1385, and NGC 1559. On the other hand, most of the galaxies with the largest negative residuals are flocculent and quiescent galaxies, like NGC 4826, NGC 2775, NGC 4548, NGC 1317, NGC 4571, and NGC 4698. Other properties that tend to be correlated with positive $\Delta$MS are the presence of star formation at the end of the bars and the presence of global spiral arms. Galaxies with bulges tend to have negative $\Delta$MS as expected.

\section{Relation of cluster population properties to $\Delta$MS and Galaxy Morphology} \label{sect:deltams}

\subsection{Bars and central rings}

As just mentioned, many of the galaxies with the largest $\Delta$MS are those with bars that appear to be driving star formation.  
The presence of a strong bar is known to effectively funnel gas into the galaxy's central regions \citep[e.g.][]{athanassoula_existence_1992,sellwood_dynamics_1993,kuno_nobeyama_2007, sormani_fuelling_2023, schinnerer_phangs-jwst_2023}.  This process creates high gas densities, leads to more efficient star formation, and often promotes cluster formation. 

An examination of the star cluster color-color diagrams for such galaxies in Figure~\ref{fig:ub_vi_1} shows they all have prominent middle age plumes, as expected based on Figure~\ref{fig:ms_stats}. NGC 1365, the galaxy with the highest SFR in the sample (16.90 M$_{\odot}$ yr$^{-1}$), is exceptional, and this activity results from the combination of a bar which drives a central star-forming ring, and strong spiral arm structure.
It not only has a particularly prominent middle age plume, but also has the richest population of massive young clusters of any known galaxy within 30 Mpc, with $\sim$30 star clusters more massive than 10$^6$M$_{\odot}$ and younger than 10 Myr \citep{whitmore_phangs-jwst_2023}.
 
The cluster spatial distribution maps (Figures~\ref{fig:spatial_dist}-\ref{fig:spatial_dist_9}) reveal star formation hot-spots where young clusters dominate, many of which are related to the presence of a bar. Beyond NGC~1365, central star-forming rings are found in an additional 6 galaxies, and all but one of these galaxies also exhibit 
a clear bar morphology (Table~\ref{tab:DeltaMS}).  The presence of the ring is reflected in the distribution of young star clusters.
Concentrations of young clusters also appear at the connection points between bars and spiral arms as observed in NGC\,1365, 7496, 1097, 1300, and 1512. The enhanced star formation at these parts of galaxies are explained by the increase of density due to the elliptical orbits in bars \citep[e.g.][]{nguyen_luong_w43_2011,beuther_interactions_2017,tress_simulations_2020,sormani_simulations_2020,levy_morpho-kinematic_2022}. 
Interestingly, these cluster hot-spots are dominated by highly dust-reddened ($>1.5 A_V$) young ($<{\rm 10\,Myr}$) clusters, which are located in the middle-age plume or globular cluster region rather than the young cluster locus \citep{whitmore_improving_2023, thilker23sed}. This means that these high density regions have large amounts of dust which have a major impact on our HST UV-optical observations, and long-wavelength JWST and ALMA observations become essential for studying the earliest stages of dust and embedded star and cluster formation \citep[e.g.][]{johnson_physical_2015, leroy_phangs-alma_2021, emig_super_2020,rico-villas_super_2020, costa_toward_2021, levy_outflows_2021, levy_morpho-kinematic_2022, schinnerer_phangs-jwst_2023, whitmore_phangs-jwst_2023, linden_goals-jwst_2023, sun_hidden_2024}.

Another common feature of galaxies with bar-driven star formation is that middle-age clusters are found near the young cluster hot-spots, as well as throughout the bar (e.g., NGC~1672, NGC~2903, NGC~1097).  Comparison with the distribution of the old globular clusters, which are more uniformly distributed, makes it clear that the middle-age clusters still reflect the dynamical features of their galaxy.

Some galaxies show a string of middle-age clusters parallel to the bar (e.g., NGC~1097). This population seems to be a relic from a star-formation episode after which the star clusters remained on a similar orbit. In fact, this scenario is described by simulations in \citet{dobbs_age_2010} and their Figure 2 reflects a situation where $\sim50$\,Myr old clusters are orbiting parallel to the bar. \citet{sormani_simulations_2020} suggested that such clusters are formed near the central ring and then collectively moved out into the galaxy. Considering the relative position above the MS of these galaxies, we can infer that such a past star formation episode contributes to the enhanced SFR value.

\subsection{Flocculent star formation} \label{sect:flocculent}

Galaxies with flocculent morphologies dominate the galaxies below the main sequence (i.e., with negative $\Delta$MS; see Table~\ref{tab:DeltaMS}). As already discussed in Section~\ref{sec:map} and shown in Figure~\ref{fig:ms_stats}, galaxies with negative $\Delta$MS tend to have peculiar color-color diagrams (Figure~\ref{fig:ms}) which lack a distinct MAP feature, indicating a major departure from steady-state star-formation due to interactions with their external environments. 

Examination of individual cases shows the connection between the MAP deficiency, and galaxy morphology. In particular, NGC 2775 is of Type a SA(r)ab with an intermediate sized bulge, a flocculent disc, and a color-color distribution that appears strongly bimodal.  It has the second lowest $\Delta$MS in the sample and flocculent structure so striking that its HST imaging has captured broad attention\footnote{\url{https://esahubble.org/images/potw2026a/}}. Almost all class 1 clusters are situated in the bulge and class 2 clusters in the disc (Figure~\ref{fig:spatial_dist_9}). The bimodal distribution originates from the combination of a relatively dust free old central region with no recent star formation \citep{hogg_hot_2001}, and flocculent star formation thought to be seeded by accreted gas \citep[i.e., from the nearby companion NGC\,2777,][]{arp_properties_1991} which led to a disk rejuvenation event. 

Two other flocculent galaxies NGC\,4571 \citep{kennicutt_evolution_1983} and NGC\,4689 \citep{elmegreen_arm_2002} exhibit a strong YCL feature. They are adjacent in Figure~\ref{fig:ms} below the main sequence.  NGC\,4689 is a member of the Virgo cluster. The galaxies are not able to sustain their star formation as they are presumed to have lost most of their gas due to their environment \citep{kenney_co_1986}, resulting in a weak middle-age plume.

Our multi-scale observational analysis is consistent with a two-component disk model which predicts a dearth of intermediate age stars in the disk of a flocculent galaxy \citep{ET93,SM22}.  In this model, flocculent patterns arise through gravitational instabilities in a low-mass cool disk component embedded in a massive halo. \citet{SM22} suggest that a two-component disk could arise naturally with the abrupt accretion of gas following a period of gas starvation. Flocculent instabilities would then give rise to star formation in short arm segments. 
All of these flocculent galaxies below the MS show an evenly distributed cluster population with no significant hot-spots of clusters. 

\section{Discussion}\label{sect:discussion}

With the completion of the largest HST census to-date of star clusters and compact associations, we are beginning to realize the scientific potential of PHANGS-HST to build upon the previous generation of star cluster studies \citep[e.g.,][and references therein]{portegies_zwart_young_2010, renaud_star_2018, krumholz_star_2019, adamo20}, and break new ground in the multi-scale characterization of their observational properties.  

The nature and size of our dataset allow us to bring together once-separate techniques for the characterization of galaxies (galaxy morphology and location relative to the galaxy main sequence) and clusters (color-color diagrams and 2D spatial distributions) for a diverse sample of spiral galaxies.  We provide a broad overview of the demographics of the objects in our catalogs, which demonstrates that tremendous insight can be gained from the observed properties of clusters alone, irrespective of the exact choice of model SSP track, and even in the absence of their transformation into physical quantities.  

In particular, we show how the PHANGS-HST cluster sample greatly expands utility of the color-color diagram.  In particular, the UBVI CCD reveals that the three standard morphological classes of clusters and associations map to distinct combinations of YCL, MAP, and OCG features, and hence to distinct age distributions.  It provides a model-independent graphical representation of both the star formation history of individual galaxies as traced by clusters, as well as the cosmic cluster formation history of disk galaxies.  When coupled with population synthesis model tracks and dust reddening laws, the UBVI CCD is important for not only testing SSP models \citep[e.g.,][]{larsen_young_1999, bruzual_stellar_2003, vazquez_optimization_2005, maraston_evolutionary_1998}, but also for exposing the uncertainties in the translation of photometric colors into ages, and specific degeneracies between age and reddening.  The much broader distribution of low luminosity/low mass systems in the UBVI CCD confirms how photometric colors do not map uniquely to a given age for this population, even if the reddening and metallicity are known, due to stochasticity in the presence of massive stars and short-lived stellar evolutionary phases \citep[e.g.][]{fouesneau_accounting_2010, silva-villa_star_2011,de_meulenaer_deriving_2013,krumholz_star_2015, OD2022}.

 Of course, when comparing the photometric properties to model predictions it is important to understand the accuracy of the model and its limitations. Throughout the paper, a BC03 SSP solar metallicity model is shown to provide context for discussion of the distribution of the cluster population in color-color diagrams, but there are apparent inconsistencies between this track and the observed color distribution as noted in our previous papers \citep[e.g.,][]{turner_phangs-hst_2021, deger_bright_2022}.  For example, the color evolution of the model between 3 and 5 Myr is too blue in V-I and/or too red in U-B by a few tenths relative to the observed YCL (even accounting for the impact of dust along the reddening vector).  The sharp turn to the red at 5 Myr in NUV-BVI and UBVI does not seem to be reproduced by the shape and position of the YCL.   There appears to be an inconsistency between the relative position of YCL and MAP and tracks in the BVI compared with that in the UBVI diagrams.  The track does not incorporate nebular emission which would produce a red ``hook" for ages $<$3 Myr, which would be important for some fraction of the youngest clusters.  These complications are one motivation for the focus on the observed properties of our sample in this paper, which are far more likely to stand the test of time.

Clearly a great deal of work lies ahead to use this sample to test and constrain SSP models \citep[e.g.,][and references therein]{wofford_comprehensive_2016}, and this will be the focus of upcoming work.  Quantitative determination of ages, reddenings, and stellar masses through SED fitting assuming the BC03 SSP model track is presented in Paper II. Proper quantitative study of the timescales and processes governing the star and cluster formation cycle requires robust determination of these physical properties, a clear understanding of underlying model uncertainties, together with proper determination of  ompleteness limits of the catalogs.  In the remainder of this section, we  discuss issues related to sample completeness both to outline future work and to provide advice to users of the catalog.

Characterizing the completeness of star cluster samples is known to be a messy business. While completeness will depend on the distance of the galaxy (which changes by a factor of 4 from 5-23 Mpc in PHANGS-HST), it also is affected by:
\begin{itemize}
    \item local background in the galaxy, which can be highly variable.  For example, cluster candidates are not detected in the bright central regions of some galaxies (e.g., NGC\,1566, 3627, 1317 and 4548; Figure~\ref{fig:spatial_dist}-\ref{fig:spatial_dist_9}). Completeness will also be a function of the density of resolved sources (crowding).
    \item dust, which can also be highly variable across a galaxy.  Incompleteness will be higher for the youngest clusters ($\lesssim5$ Myr), which are still clearing the natal gas and dust from the environments in which they are born.  The earliest stages of star and cluster formation will be entirely dust enshrouded and unobservable in the optical.   The PHANGS-JWST dataset will be critical in this regard for completing the cluster census at young ages, and this was a key science driver for the survey \citep[][and references therein]{lee_phangs-jwst_2023}.
    \item the size of the cluster, and the underlying cluster size distribution.  Incompleteness is likely higher for the most compact clusters, which may not be distinguishable from a point source \citep[e.g.,][]{ryon_effective_2017, brown_radii_2021}.  
    \item the details of the source detection algorithm and candidate selection criteria.  Two issues are particularly important to note in this context.
    \item the age of the cluster, due to the evolution of the mass-to-light ratio,
    \begin{itemize}
    \item As discussed in \citet{lee_phangs-hst_2022} and Section~\ref{ssect:cat_content}, the PHANGS-HST pipeline is optimized to identify single-peaked clusters, which leads to a high level of incompleteness for multi-peaked stellar associations (class 3).  The majority of star formation occurs in stellar associations \citep[][and references therein]{lada03,ward18, ward20, wright20}.  Whether the C3 compact associations provided in this catalog should be used will thus be heavily dependent on the science goal of the analysis. A separate pipeline for stellar associations, based on a watershed algorithm, provides a far more complete inventory of young stellar populations across multiple physical scales \citep{larson_multiscale_2023}. Multi-scale stellar association data products for the full 38 PHANGS-HST galaxy sample will be published at a later date.  
    \item Even when pipelines are specifically developed for single-peaked clusters, differences in the adopted detection algorithm and morphological selection criteria \citep[which has generally been based on some form of concentration index, e.g.,][]{chandar_luminosity_2010, adamo_legacy_2017} can lead to significant differences in the populations captured.  As discussed in \citet{thilker_phangs-hst_2022}, LEGUS \citep{calzetti_legacy_2015} has produced cluster catalogs for four of the seven galaxies in common with PHANGS (NGC~628, NGC~1433, NGC~1566, NGC~3351)\footnote{\url{https://archive.stsci.edu/prepds/legus/dataproducts-public.html}}, and there is an overlap of 50-75\% of human verified C1 and C2 clusters in the union of the LEGUS$+$PHANGS-HST catalogs.  Understanding the differences in the catalogs, and comparison of results based on the union of the two catalogs with those based on the separate catalogs from each survey will be important subjects for future investigation.
    \end{itemize}
    \item unknown-unknowns, e.g., systematics in the neural network morphological classifications, particularly for the fainter sources in the sample for which human classifications were not generally performed. 
    
\end{itemize}

In the future, analysis of artificial star clusters added to the HST imaging can be performed to quantify catalog completeness \citep[e.g.,][]{adamo_legacy_2017, tang_cluster_2023}.  In the meantime:
\begin{itemize}
\item In Section~\ref{sect:catalog_properties}, we provide basic statistics for the size and depths of the catalogs for both individual galaxies and the total sample aggregated across all 38 galaxies.  These data can be used to estimate the completeness limit of the catalogs, by locating the turn-over point in the luminosity (or mass functions) as has been done in prior work \citep[e.g.,][]{mayya_hst_2008, ryon_effective_2017, cook_star_2019, cuevas-otahola_cluster_2023}.
\item analysis can be conducted using different sub-samples of the catalog, selected based on a completeness-dependent parameter, and the results compared.  For example, sub-samples can be defined with different magnitude limits, galaxy distances, from different regions of the galaxies (e.g., excluding the inner crowded, high background parts of the galaxy).  Comparative analysis using C1 vs C2 vs C1+C2 samples, as suggested in \citet{whitmore_star_2021} and demonstrated in several figures in the current paper (e.g., Figures~\ref{fig:cc_compare} and \ref{fig:colo_colo_first_view}) can also be performed.
\end{itemize}

Finally, due to the black-box nature of the neural-network models, comparative analysis with human-classified and machine-classified catalogs should be performed.
It would be hoped that the agreement between human and ML classification would be so robust that the we can rely entirely on the ML catalog once it is built. While the current state of the art is quite promising (especially for C1+C2), we are not yet at a stage where  ML classification can be used blindly - care  must be taken.
Machine learning classifications will continue to improve, but the subject is still at an early stage of development. See \citet{wei_deep_2020, whitmore_star_2021, perez_starcnet_2021, hannon_star_2023} for additional discussion and other examples of how well the ML classifications perform for specific science applications.

\section{Summary}\label{sect:summary}
We present the largest catalog of star clusters and associations to-date for nearby galaxies.  For the 38 spiral galaxies of the PHANGS-HST survey, which span distances between 5 to 23 Mpc, our catalog provides aperture-corrected photometry in the NUV-U-B-V-I filters for:
\begin{itemize}
    \item a total of $\sim20,000$ star clusters and compact associations, with a median of $\sim500$ sources per galaxy, which have been visually inspected and morphologically classified by a human \citep[co-author BCW,][]{whitmore_star_2021}.  This subset of the catalog is comprised of $\sim$8000 class 1 and $\sim$8000 class 2 clusters, and $\sim$6000 compact associations (class 3).  The median $m_V$ of this human-classified sample is $\sim-8$ mag (Vega).
    \item a larger sample of $\sim100,000$, with a median of $\sim1700$ sources per galaxy, which have passed neural network classification \citep{hannon_star_2023}.  This sample is comprised of $\sim$13000 class 1 and $\sim$23000 class 2 clusters, and $\sim$60000 compact associations (class 3).  The neural network models were trained on the human-classified sample, and deployed on the entire cluster candidate list of $\sim$190,000 sources. This yields a sample of clusters and associations $\sim$1 V-band magnitude deeper than the human classified sample.
\end{itemize}

We provide a broad overview of the observed properties of the photometric catalogs.  A summary of our findings is as follows.

Regarding UV-optical color-color diagrams for star clusters and associations:
\begin{enumerate}  
\item given the typical depth of HST Treasury surveys of nearby galaxies with the WFC3 camera, the U-B-V-I color-color diagram provides the greatest diagnostic power (relative to B-V-I and NUV-B-V-I) for distinguishing between different age populations and separating its three principal features --  the young cluster locus (YCL, $\lesssim$ 10 Myr), the middle-age plume (MAP, 1 Gyr$\lesssim$t$\lesssim$ 100 Myr), and the old globular cluster clump (OGC, t$\gtrsim$ 1 Gyr) (Section~\ref{ssect:cc_overview}).  We provide contour based definitions for each feature (Section~\ref{ssect:cc_regions}). 
\item We study the observed properties of the cluster population on the color-color diagram combined across all 38 spiral galaxies in the PHANGS-HST survey.  This shows that the C1, C2, C3 morphological classes each have distinct color-color diagrams, and hence map to distinct age distributions.  C1 clusters have a prominent MAP and OGC, and weak, but narrow YCL.  C2 clusters have a clear YCL and MAP, but no OGC.  C3 compact associations have a strong YCL, and no significant MAP or OGC. In particular, the large sample demonstrates that the properties of C1 and C2 clusters are distinguishable. (Section~\ref{ssect:cc_overview})  
\item The differences in the YCL, MAP, and OGC features indicate that age distributions skew younger as the degree of cluster asymmetry and central concentration increases from C1 to C3, and are consistent with the expectation that the process of cluster dissolution should yield some correlation between age and morphology \citep[e.g.,][and references therein]{adamo_legacy_2017, whitmore_star_2021, cook23}. (Section~\ref{ssect:cc_overview})
\item  The distribution of clusters in the color-color diagram is qualitatively similar for human and ML classified clusters when both samples have similar magnitude limits.  This provides evidence for the robustness of the ML classifications.  (Section~\ref{ssect:cc_compare})
\item The distribution of low mass young clusters ($<$5000 M$_\odot$) on the color-color diagram show increased scatter which is generally consistent the impact of stochastic sampling of the stellar IMF.  (Section~\ref{ssect:cc_mag_m_star})
\end{enumerate}

We bring together various techniques -- the characterization of galaxies (galaxy morphology and location relative to the galaxy main sequence) and cluster populations (color-color diagrams and 2D spatial distributions) -- to explore the dataset in a multi-scale context and demonstrate that the UBVI color-color diagram is a highly valuable, model-independent, observational diagnostic of the star and cluster formation history and evolutionary status of the galaxy.
\begin{enumerate}
\item As expected YCL populations closely trace spiral structure.  They are coincident with CO twice as often than objects associated with the MAP or OGC, and reflect the structure of the ISM from which they were born. These structures then disperse with age as has been found previously -- the spatial organization is broader for the MAP objects, and is closest to a random distribution for objects associated with OGC.  (Section~\ref{sect:spatialdist})

\item There is no correlation between $\Delta$MS and the fraction of clusters in the YCL. The absence of a strong YCL feature at above the MS is generally due to dust reddening and does not necessarily imply the absence of cluster formation.  Above the MS, strong bars, a number of which are associated with central star forming rings, appear to be driving high star formation densities and promote cluster formation. Clusters trace the star forming rings, concentrations of clusters appear at the bar ends, and these populations tend to be highly dust-reddened.  At low $\Delta$MS the relative fractions of the cluster populations in each of the features reflects a complex star formation history due to the external environment of the galaxy (e.g., Virgo cluster) and interactions with neighboring galaxies.  At low $\Delta$MS, many galaxies have flocculent morphologies and evidence of a recent gas accretion (``rejuvenation'') event which is fueling low levels of star and cluster formation.  (Section~\ref{sect:deltams})
\item  There is a strong linear correlation between a galaxy's offset from the MS and the fraction of its cluster population in the middle-age plume.  In contrast to the YCL, dust is not a confounding factor as the width of the MAP indicates low amounts of reddening.  Above the MS, the presence of a strong MAP feature indicates the elevated star and cluster formation activity must have a duration on the order of 100 Myr.  Below the MS, galaxies appear to have a deficient MAP feature, which is consistent with a two-component disk model where flocculent patterns arise through gravitational instabilities in a low-mass cool disk component embedded in
a massive halo which has recently accreted gas after a period of quiescence.    (Section~\ref{sect:flocculent})
\end{enumerate}

This presentation of the PHANGS-HST star cluster and association catalogs of observed photometric properties provides a foundation for a broad range of science. Previous studies of star formation and feedback timescales, and cluster formation and evolution, which were performed with more limited samples can now be expanded with this large sample of $\sim$100,000 star clusters and compact associations to probe the interplay of the small-scale physics of gas and star formation with galactic structure and galaxy evolution.  These catalogs are an essential complement for JWST studies of the earliest phases of dust embedded star and cluster formation, and for extending the study of the observed cluster properties into the infrared.  In Paper II, we discuss the derivation of cluster masses, ages, and reddenings based on improved SED fitting methods for UV-optical photometry, and present the companion catalog of physical properties.

\section*{Acknowledgements}
We dedicate this paper to the memory of Julie Whitmore (January 13, 1954 - August 23, 2023).

JCL is enormously grateful to Michele Judd, Janet Seid, and the W.M. Keck Institute for Space Studies (KISS) at Caltech for its sustained support of collaboration meetings where key work for this paper was performed, and for providing a quiet space for JCL to think.  The PHANGS-HST survey benefited from early discussions between JCL, AL, and other current PHANGS team members which date back to the 2014 KISS workshop, ``Bridging the Gap: Observations and Theory of Star Formation Meet on Large and Small Scales.''

R.C.L. acknowledges support for this work provided by a National Science Foundation (NSF) Astronomy and Astrophysics Postdoctoral Fellowship under award AST-2102625.

MB gratefully acknowledges support from the ANID BASAL project FB210003 and from the FONDECYT regular grant 1211000.

This work was supported by the French government through the France 2030 investment plan managed by the National Research Agency (ANR), as part of the Initiative of Excellence of Université Côte d’Azur under reference number ANR-15-IDEX-01.

KK gratefully acknowledges funding from the Deutsche Forschungsgemeinschaft (DFG, German Research Foundation) in the form of an Emmy Noether Research Group (grant number KR4598/2-1, PI Kreckel) and the European Research Council’s starting grant ERC StG-101077573 (“ISM-METALS"). 

QT acknowledges generous support from the Adelphic Educational Fund Wesleyan Summer Grants for a summer internship in Baltimore and thanks the Johns Hopkins University Department of Physics and Astronomy and the Space Telescope Science Institute for hosting him.

KG is supported by the Australian Research Council through the Discovery Early Career Researcher Award (DECRA) Fellowship (project number DE220100766) funded by the Australian Government, and by the Australian Research Council Centre of Excellence for All Sky Astrophysics in 3 Dimensions (ASTRO~3D), through project number CE170100013.

We thank summer undergraduate intern Lucius Brown (Yale) for his help with artifact identification in the HST images.

This work is based on observations made with the NASA/ESA Hubble Space Telescope, obtained at the Space Telescope Science Institute, which is operated by the Association of Universities for Research in Astronomy, Inc., under NASA contract NAS 5-26555. These observations are associated with program \#15654.

\section*{Data Availability}
The PHANGS--HST star cluster catalogs will be made publicly available through the Mikulski Archive for Space Telescopes (MAST) Janurary 2024 at \url{https://archive.stsci.edu/hlsp/phangs/phangs-cat}.
{DOLPHOT \citep[v2.0][]{dolphin_dolphot_2016}, CIGALE \citep{burgarella_star_2005,noll_analysis_2009,boquien_cigale_2019}}

\bibliography{bibliography, bibliography_add}{}
\bibliographystyle{aasjournal}


\appendix
\section{Additional figures}\label{append:add_fig}
\begin{figure*}
\includegraphics[width=\textwidth]{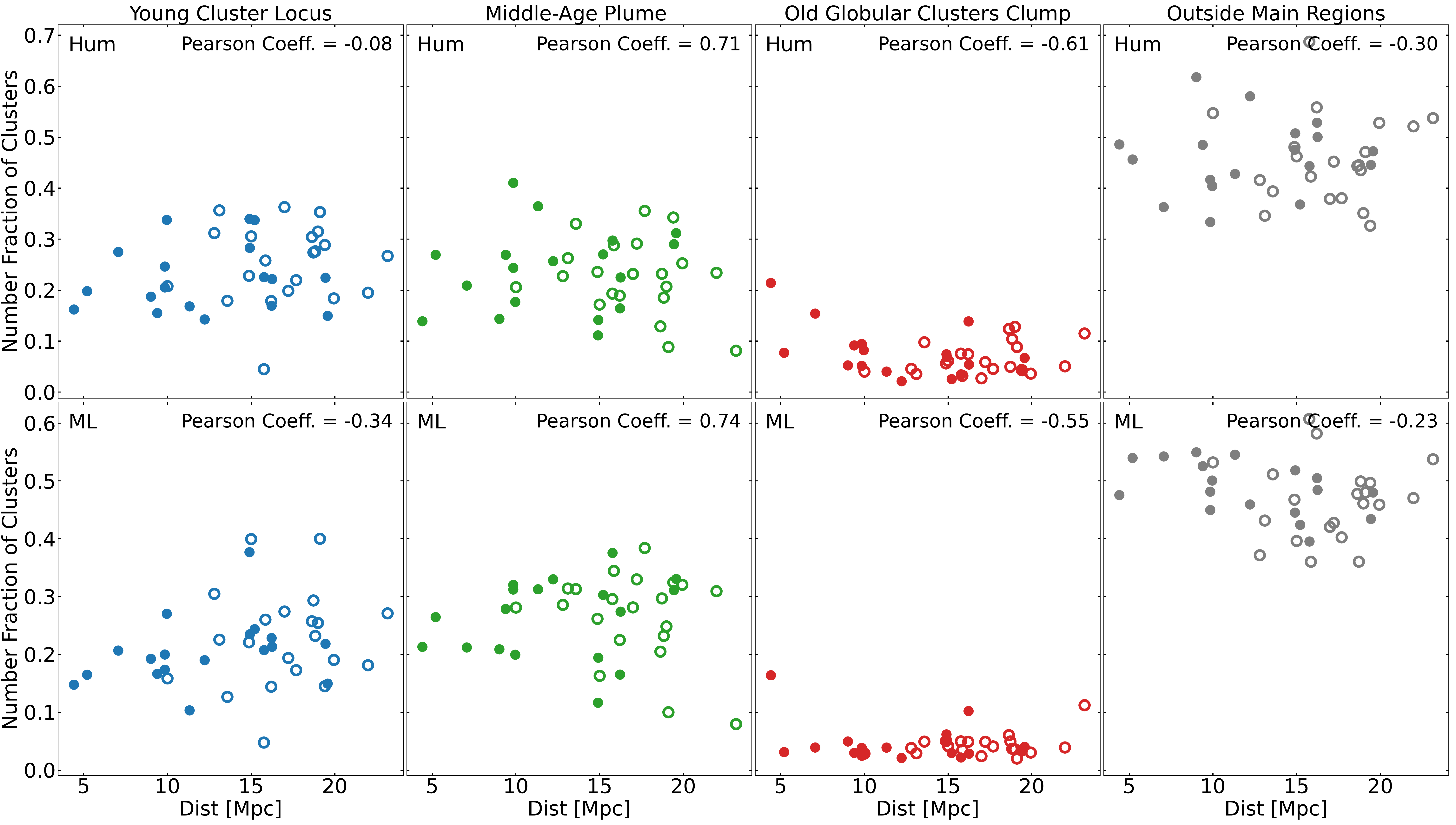}
 \caption{Number fraction of C1 and C2 clusters of each galaxy associated with the main characteristic regions in color-color diagrams found in Section~\ref{ssect:cc_regions} as a function of galaxy distance. We show the YCL, the MAP and the OGC in blue, green and red, respectively. In gray, we show clusters outside the main regions. We distinguish distance measurements which are estimated from stellar markers such as Tip of the Red giant Branch (TRGB) or from Cepheid variable stars are marked with full circles whereas other distant measurements which are less precise are marked by empty circles. A complete discussion on each individual distance measurement is provided in \citet{anand_distances_2020} and \citet{anand_distances_2021}. For each panel we show the Pearson correlation coefficient in the top right.}
 \label{fig:ms_stats_dist}
\end{figure*}

In this section we show additional figures. Figure~\ref{fig:ms_stats_dist} presents the correlation plot between host galaxy distances and relative fraction of characteristic regions in color-color diagrams as discussed in Section~\ref{ssect:cc_sf}. Figures~\ref{fig:spatial_dist} to \ref{fig:spatial_dist_9}present the spatial distribution plots discussed in Section~\ref{sect:spatialdist}.
\begin{figure*} 
\includegraphics[width=\textwidth]{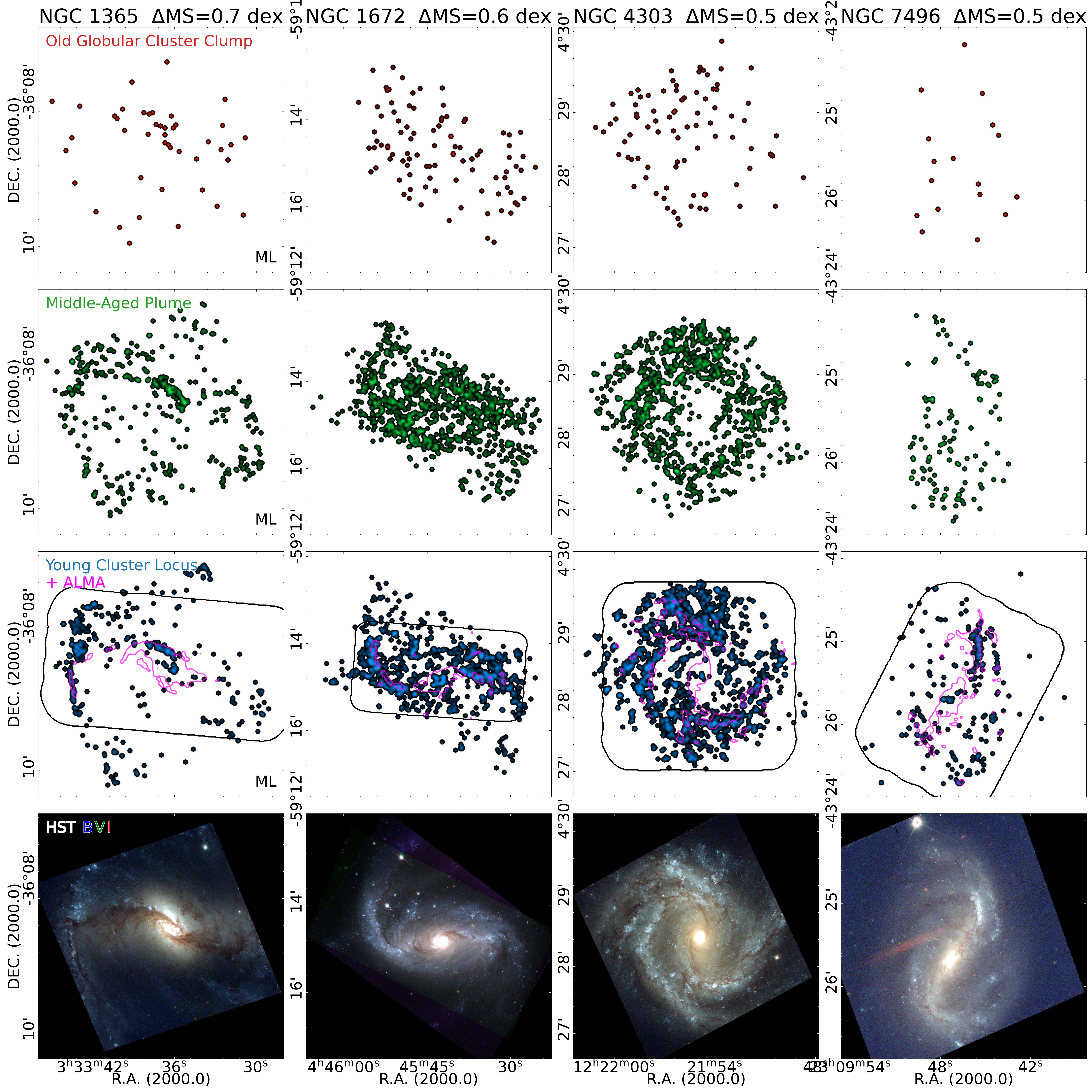}
 \caption{Spatial distributions for ML-classified star clusters of class 1 and 2 as categorized  (Sect.\,\ref{ssect:cc_regions}) into three groups: OGC (red, top panels), MAP (green, upper middle panels) and YCL (blue, lower middle panels); plus color-composite images created from the HST U-B-V bands (bottom panels).  The cluster spatial distribution maps are produced by binning the cluster positions onto a pixel grid which is subsequently convolved with a Gaussian and normalized to unity. 
 In order to highlight the relation between young clusters and molecular gas, with magenta lines, we overlay the ALMA CO(2-1) intensity contours of the 95 percentile on the YCL distribution maps.
 In this figure and Figures\,\ref{fig:spatial_dist_1}--\ref{fig:spatial_dist_9}, we show the spatial cluster distribution for all PHANGS--HST galaxies sorted by decreasing $\Delta$MS values (see Figure\,\ref{fig:ms}).}
 \label{fig:spatial_dist}
\end{figure*}
\begin{figure*} 
\includegraphics[width=\textwidth]{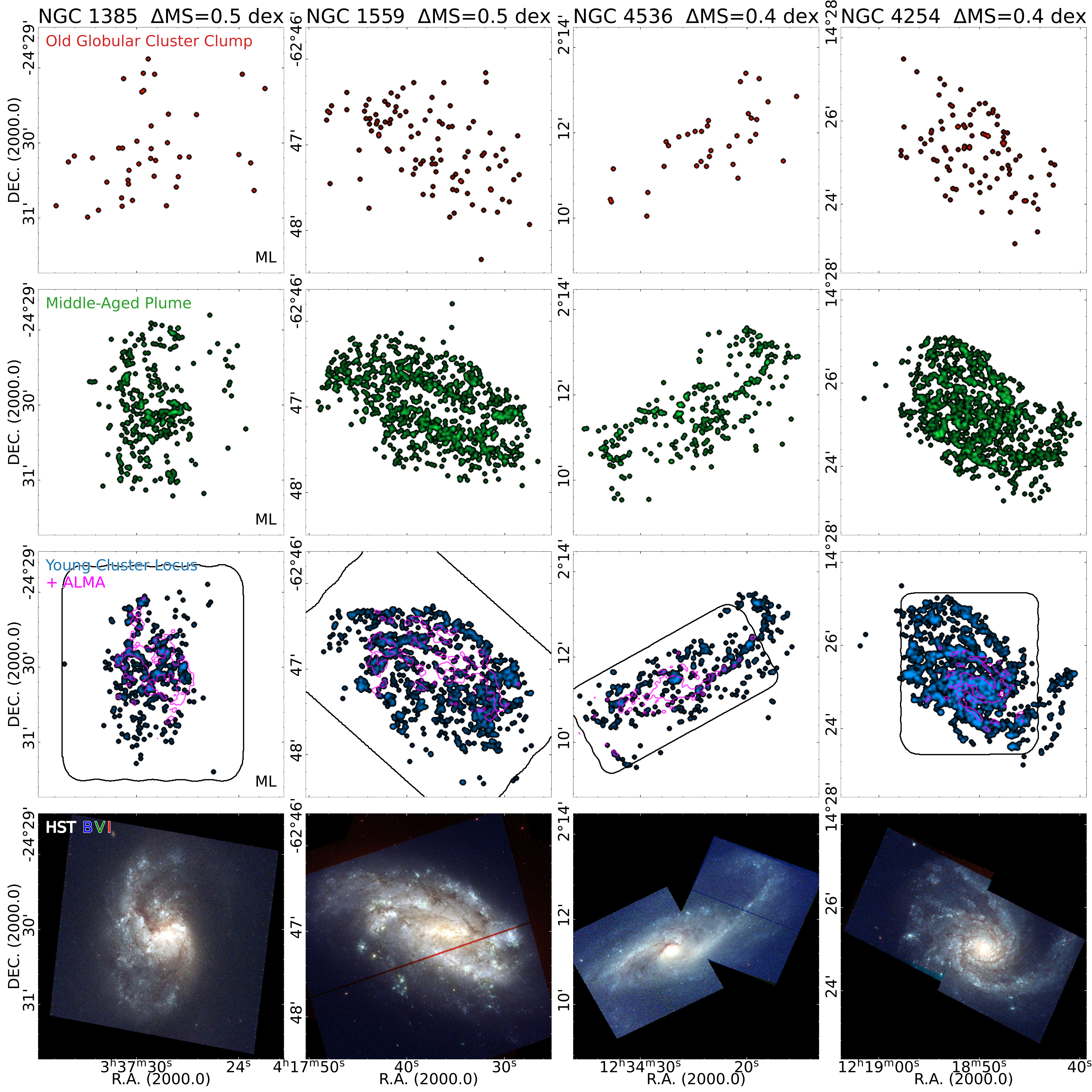}
 \caption{Continuation of Figure\,\ref{fig:spatial_dist}}
 \label{fig:spatial_dist_1}
\end{figure*}
\begin{figure*} 
\includegraphics[width=\textwidth]{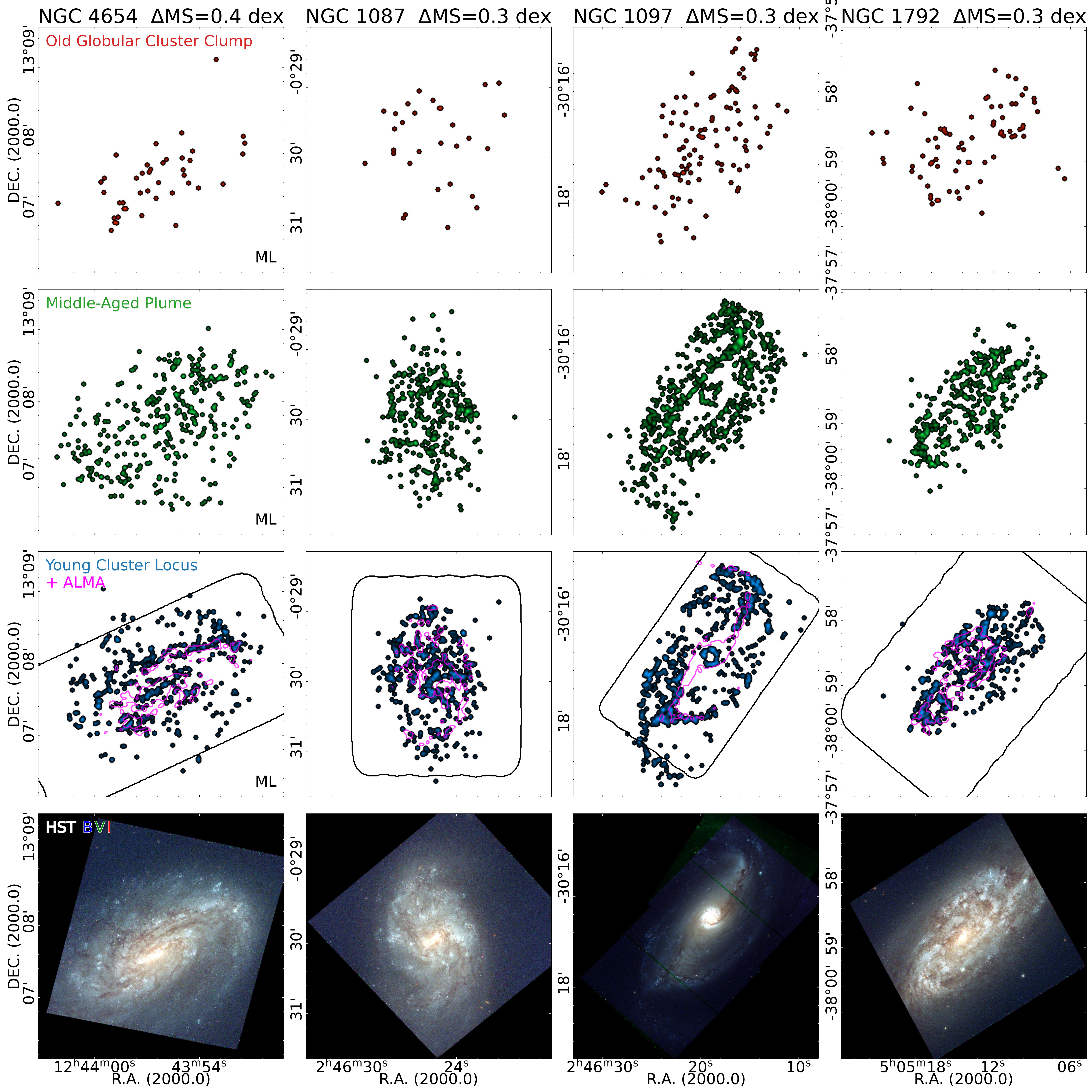}
 \caption{Continuation of Figure\,\ref{fig:spatial_dist}}
 \label{fig:spatial_dist_2}
\end{figure*}
\begin{figure*} 
\includegraphics[width=\textwidth]{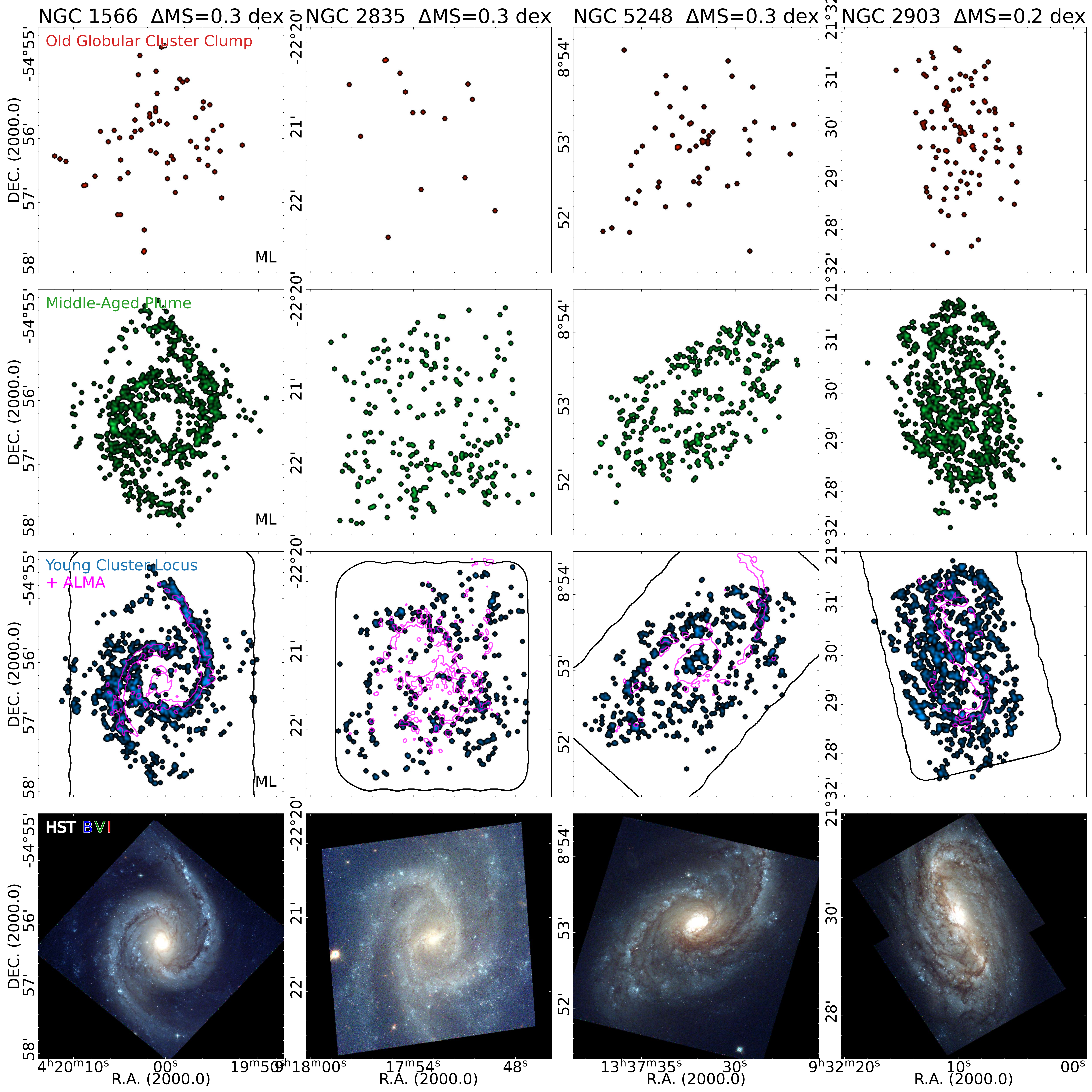}
 \caption{Continuation of Figure\,\ref{fig:spatial_dist}}
 \label{fig:spatial_dist_3}
\end{figure*}
\begin{figure*} 
\includegraphics[width=\textwidth]{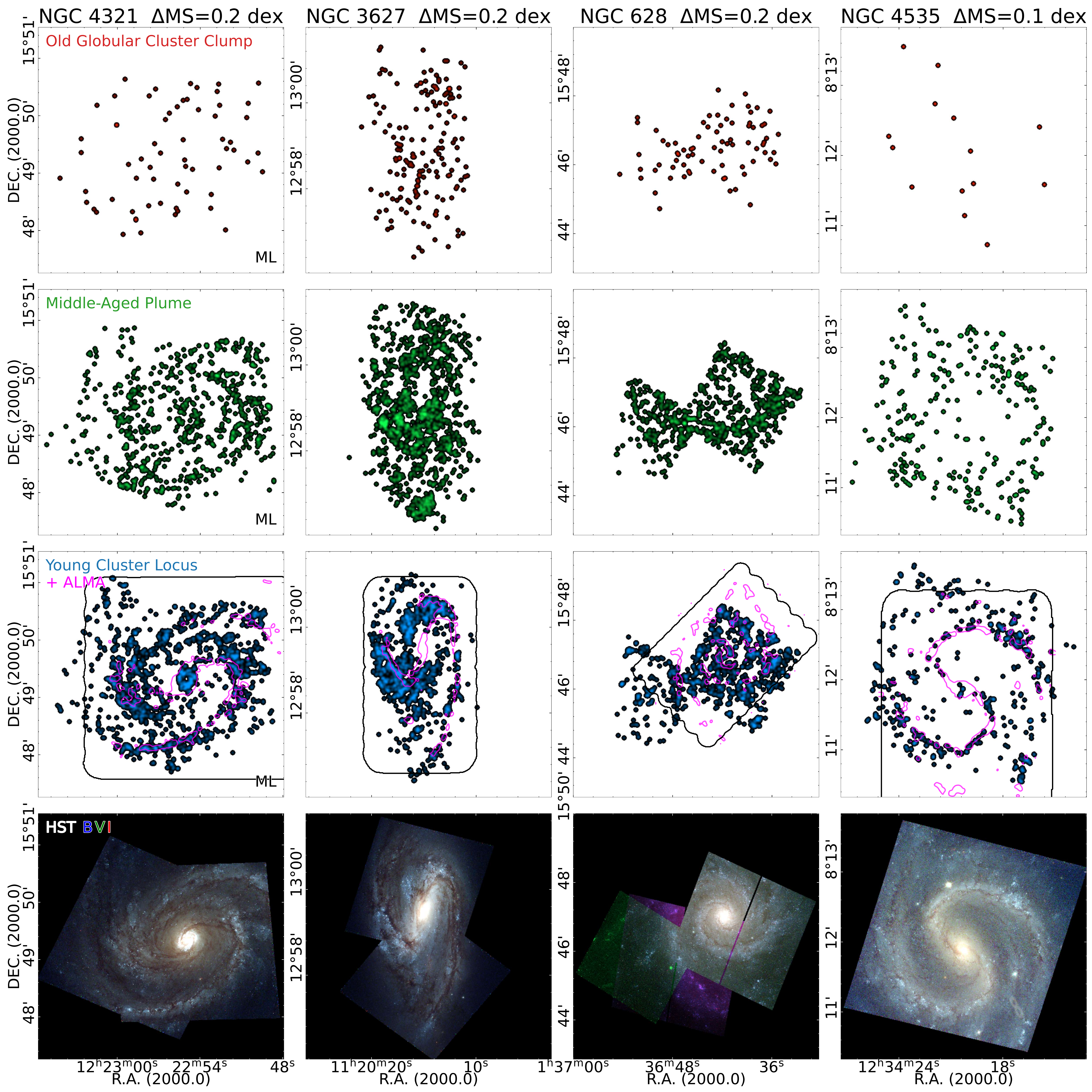}
 \caption{Continuation of Figure\,\ref{fig:spatial_dist}}
 \label{fig:spatial_dist_4}
\end{figure*}
\begin{figure*} 
\includegraphics[width=\textwidth]{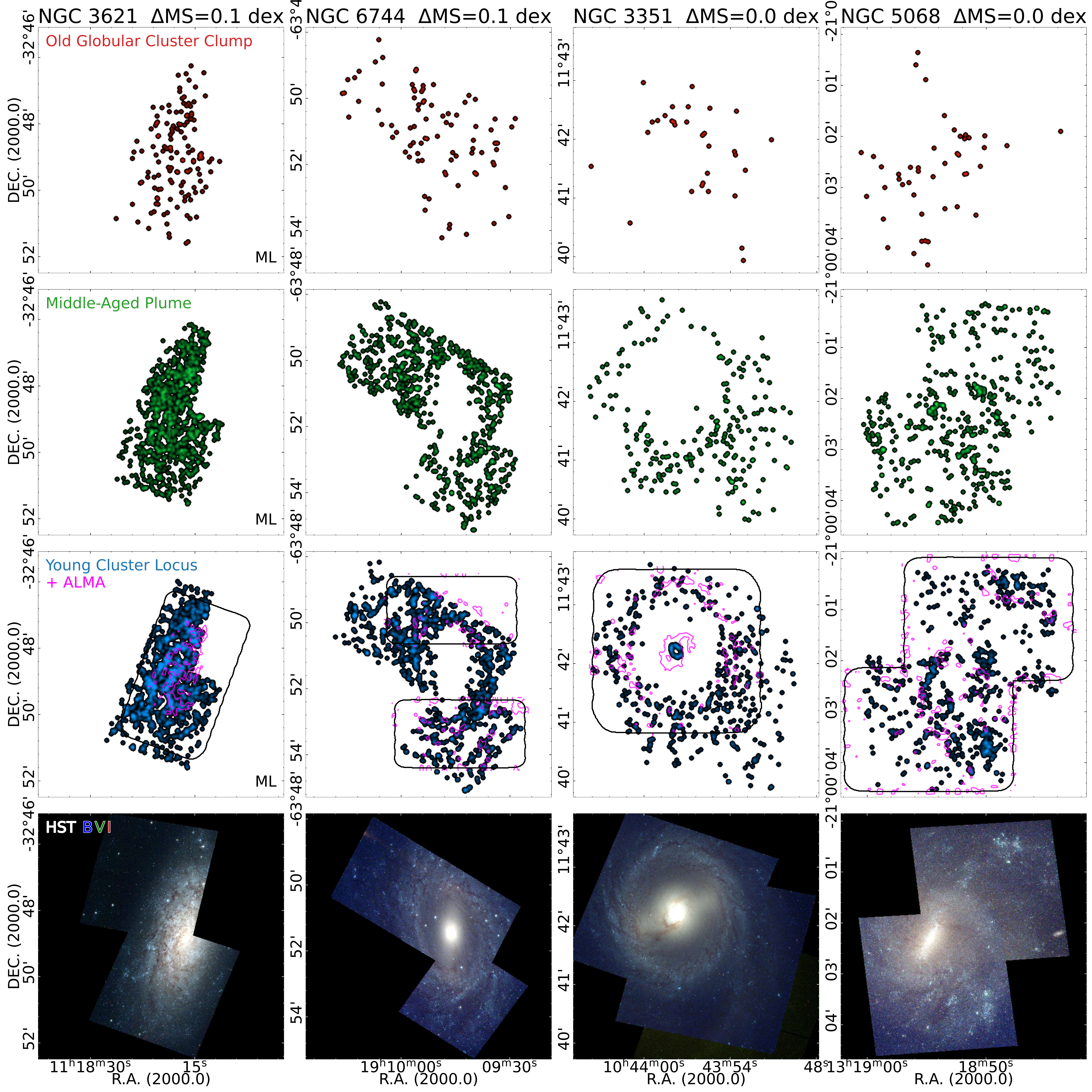}
 \caption{Continuation of Figure\,\ref{fig:spatial_dist}}
 \label{fig:spatial_dist_5}
\end{figure*}
\begin{figure*} 
\includegraphics[width=\textwidth]{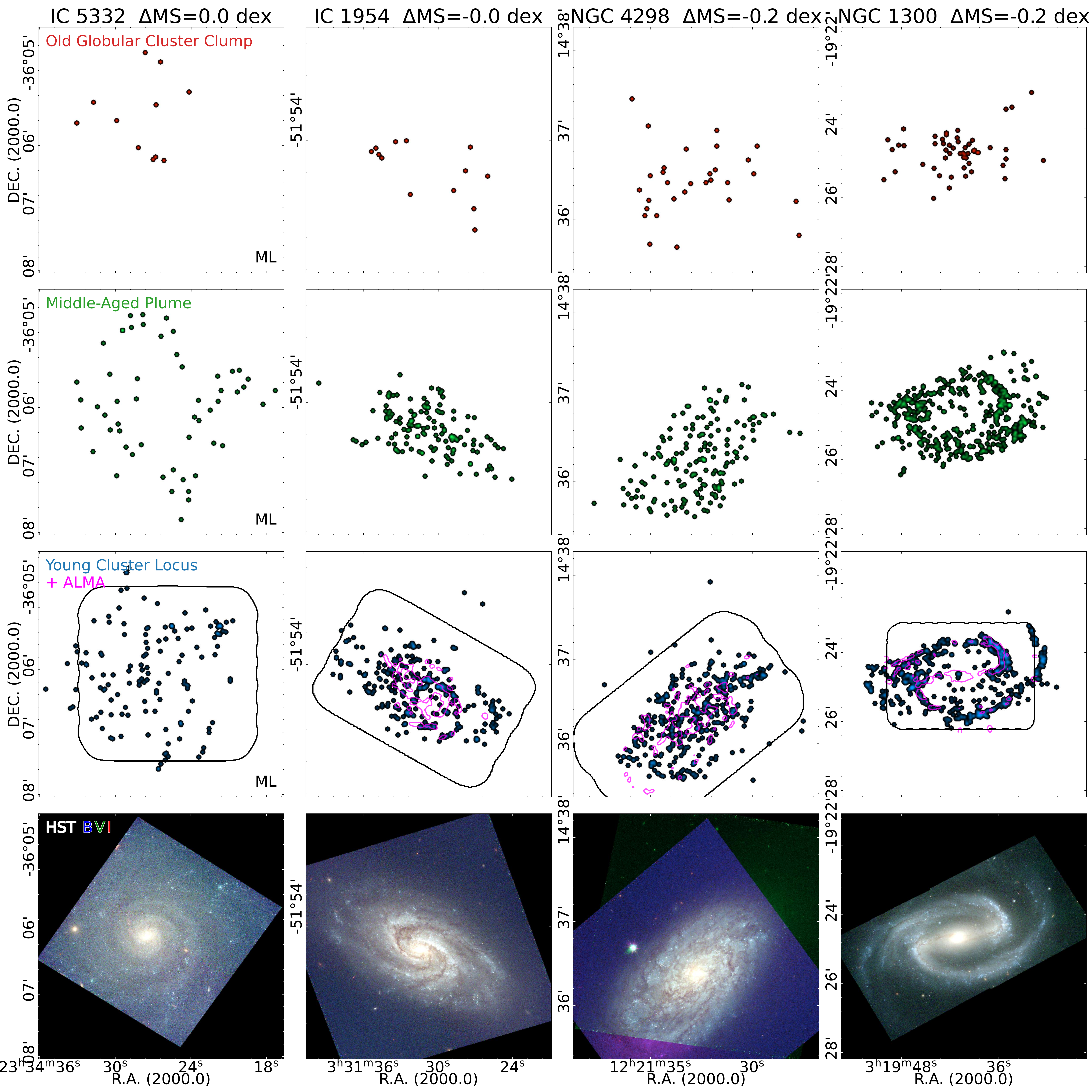}
 \caption{Continuation of Figure\,\ref{fig:spatial_dist}. We note that for IC\,5332 there is no ALMA CO(2-1) detection.}
 \label{fig:spatial_dist_6}
\end{figure*}
\begin{figure*} 
\includegraphics[width=\textwidth]{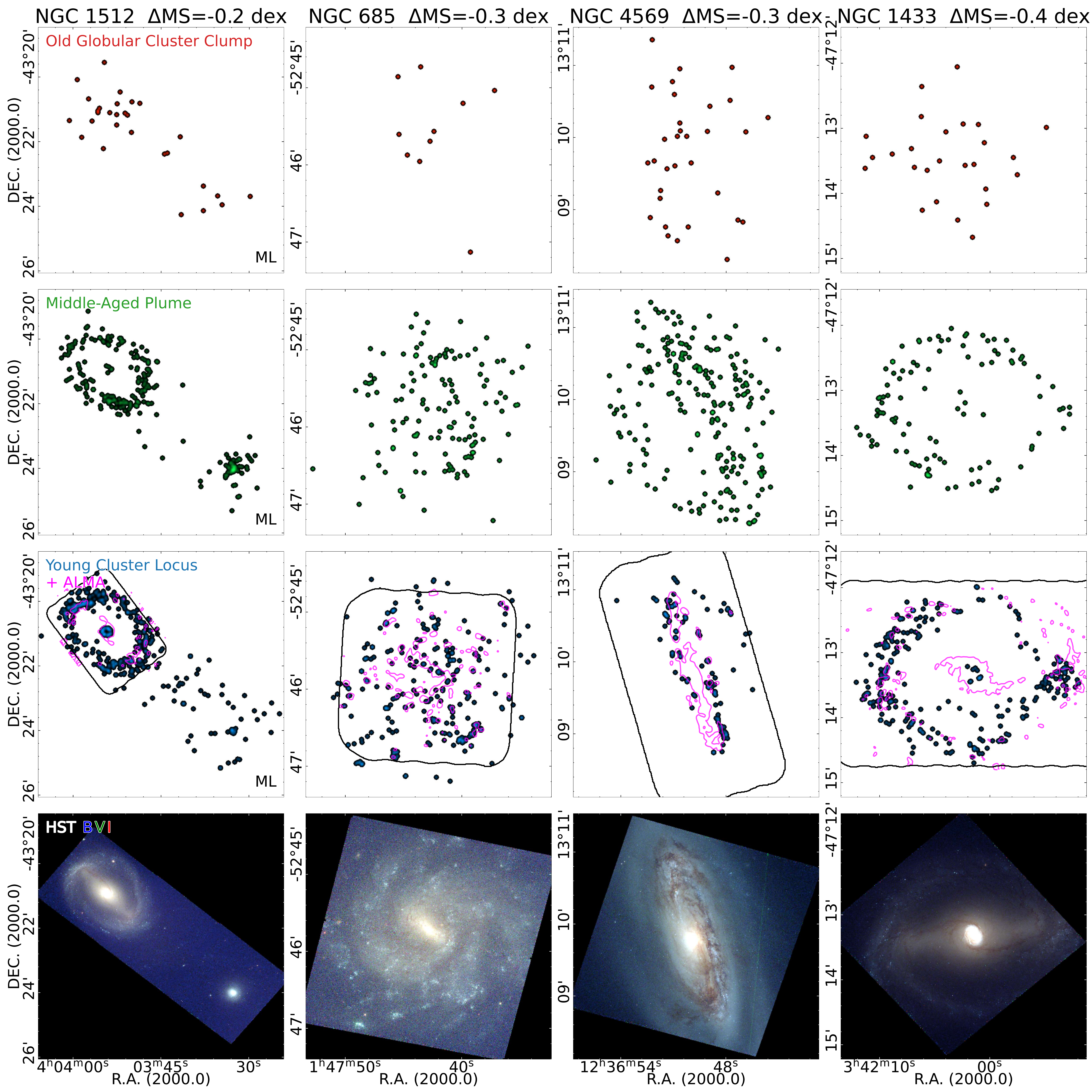}
 \caption{Continuation of Figure\,\ref{fig:spatial_dist}}
 \label{fig:spatial_dist_7}
\end{figure*}
\begin{figure*} 
\includegraphics[width=\textwidth]{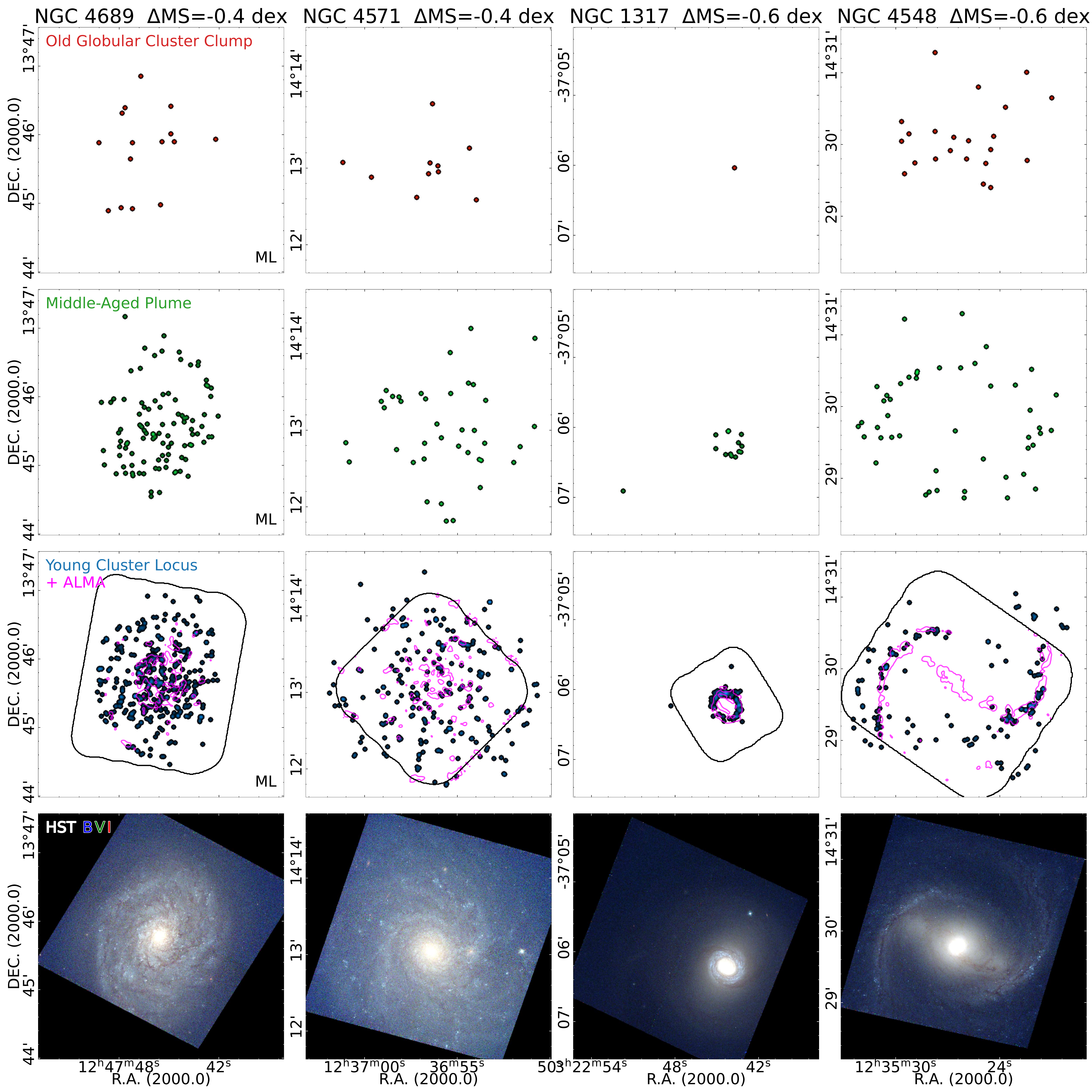}
 \caption{Continuation of Figure\,\ref{fig:spatial_dist}}
 \label{fig:spatial_dist_8}
\end{figure*}
\begin{figure*} 
\includegraphics[width=\textwidth]{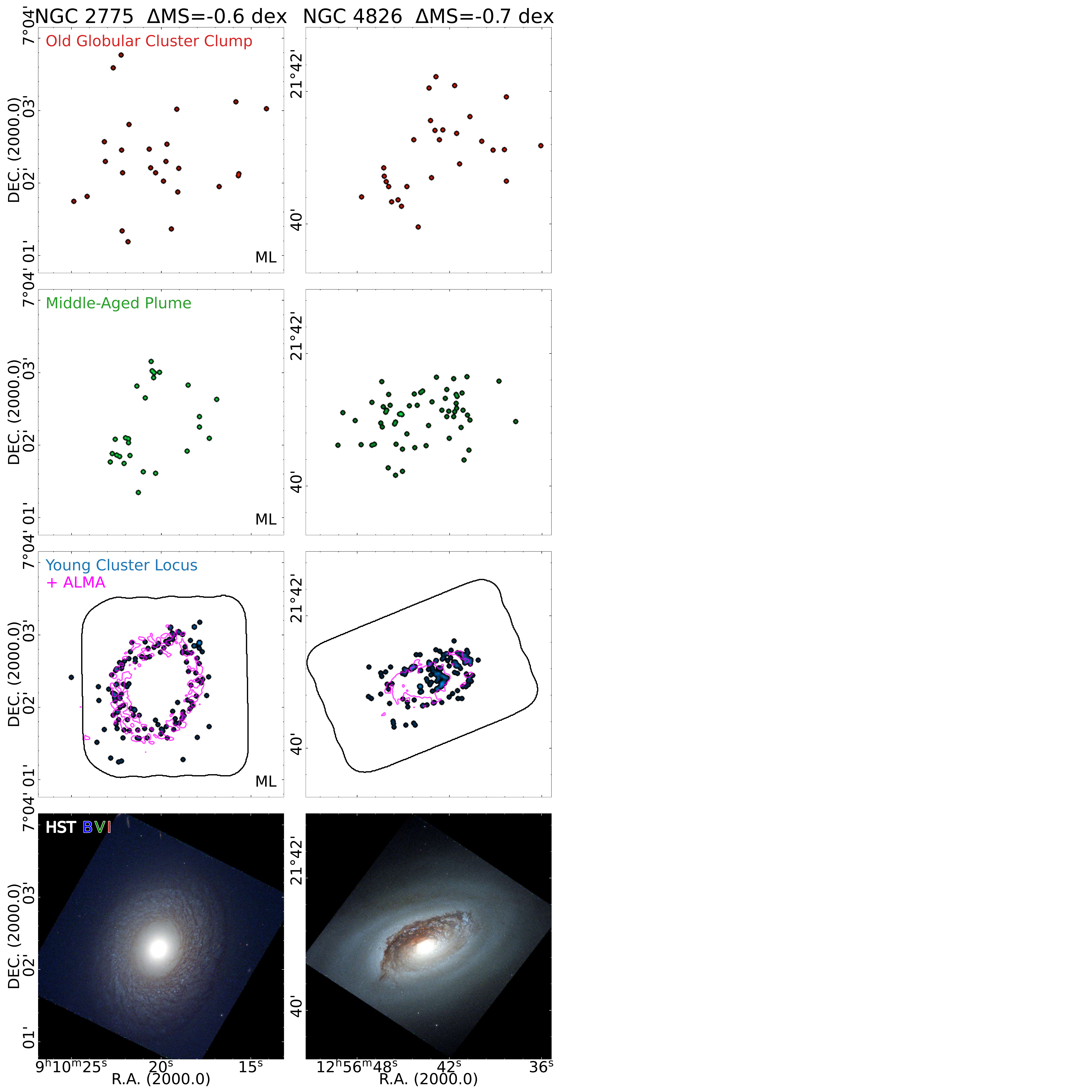}
 \caption{Continuation of Figure\,\ref{fig:spatial_dist}}
 \label{fig:spatial_dist_9}
\end{figure*}
%

\end{document}